\documentclass[aps,showpacs,prc,eqsecnum]{revtex4}
\usepackage{color,epsfig}

\def\lixo#1{}

\begin{document}

\title{Low-energy constants and relativity in\\
peripheral nucleon-nucleon scattering}

\author{ R. Higa$^{}$\footnote{Email address: higa@jlab.org} }

\affiliation{ Jefferson Laboratory, 12000 Jefferson Avenue, 
Newport News, VA 23606, USA }

\begin{abstract}
In our recent works we derived a chiral $O(q^4)$ two-pion exchange 
nucleon-nucleon potential (TPEP) formulated in a relativistic baryon 
(RB) framework, expressed in terms of the so called low energy 
constants (LECs) and functions representing covariant loop integrations. 
We showed that the expansion of these functions in powers of the inverse 
of the nucleon mass reproduces most of the terms of the TPEP derived 
from the heavy baryon (HB) formalism, but such an expansion is ill 
defined and does not converge at large distances. In the present work we 
perform a study of the phase shifts in nucleon-nucleon ($NN$) scattering 
for peripheric waves ($L\geq 3$), which are sensitive to the tail of the 
potential. We assess quantitatively the differences between the RB and 
HB results, as well as variations due to different values of the LECs. 
By demanding consistency between the LECs used in $\pi N$ and $NN$ 
scattering we show how $NN$ peripheral phase shifts could constrain 
these values. We demonstrate that this idea, first proposed by the 
Nijmegen group, favors a smaller value for the LEC $c_3$ than the 
existing ones, when considering 
the TPEP up to order $q^4$ in the chiral expansion.
\end{abstract}

\date{\today}

\maketitle


\section{Introduction} \label{secI}

One of the most intriguing mysteries in the history of modern 
physics is trying to describe and understand nuclear forces in 
terms of a basic, elementary theory of strong interactions. Despite 
more than fifty years of intense research in this area only in the 
last decade such a connection could be established, between what is 
believed to be the theory of strong interactions, QCD, and the rich 
phenomenology of low-energy nuclear physics. Chiral symmetry plays 
an essencial role in this game. Weinberg \cite{wein79} first pointed 
out that it can be used to obtain not only the same predictions of 
current algebra \cite{c-algeb}, but also corrections to them. 
Later these ideas were systematically elaborated by Gasser and Leutwyler 
\cite{chpt}, evolving to what is known as chiral perturbation theory 
(ChPT). It relies on the fact that the QCD lagrangian exhibits chiral 
$SU(N_f)\times SU(N_f)$ symmetry, $N_f$ being the number of light quarks 
($u$, $d$, and, eventually, $s$ quarks), in the limit where their 
masses, $m_q$, go to zero. This underlying theory is then described 
effectively including the relevant degrees of freedom (pions and 
nucleons for the case we are interested in, $N_f=2$) in the most 
generic way constrained only by the 
allowed symmetries, and caracterized by an expansion parameter $q$ 
representing either the momenta involved or the symmetry breaking terms 
$m_q\propto \mu^2$, $\mu$ being the Goldstone boson (pion) mass. The 
effective theory is then evaluated by means 
of feynman diagrams, strictly obeying a given power 
counting scheme. These ideas were successfuly applied to many processes 
in the mesonic sector, forming a compelling evidence that ChPT indeed 
represents QCD in the low-energy regime \cite{ecker,bkmeis95,scherer02}. 

The power counting is an essential ingredient in ChPT, allowing us 
to evaluate corrections in a consistent and controlled way. In 
the mesonic sector it follows naturally, once loop contributions are 
regularized using dimensional regularization. On the other hand, with 
baryons an additional energy scale (the baryon mass) is introduced, 
which does not vanish in the chiral limit and makes the theory much 
more complicated, with chiral loops (using dimensional regularization) 
spoiling the power counting \cite{gss}. For many years it was believed 
that a relativistic treatment of baryons in ChPT was not consistent with 
power counting. The first and widely used formalism which restores the 
latter consists in applying a sort of non-relativistic expansion at the 
level of the lagrangian, around the limit of infinitely heavy baryon (HBChPT), 
proposed by Jenkins and Manohar in the early 90s \cite{jenkman}. Almost 
ten years later, Becher and Leutwyler \cite{bl99,bl01} showed that it 
is possible 
to formulate baryonic ChPT preserving explicit Lorentz invariance and power 
counting (RBChPT). Based on previous works of Ellis and Tang \cite{EllisT}, 
the so called infrared regularization allows one to not only remove the 
power counting violating terms from loop integrals, but also recover 
the results from HBChPT when subjected to an expansion in powers of the 
inverse of the baryon mass. From the latter, Becher and Leutwyler showed 
that the heavy baryon approach fails to preserve the correct analytic 
structure of the theory in a specific low energy domain. In $\pi N$ 
scattering, this region corresponds to the Mandelstam variable $t$ close 
to $4 \mu^2$. This motivated further studies inside the covariant 
framework \cite{fuchs,lutz,goity}. 

In recent works \cite{nos1,nos2} we brought the issue of HBChPT and RBChPT 
to the two-pion exchange nucleon-nucleon potential (TPEP), showing 
that its long distance properties are governed by the same low energy 
region where the HBChPT has convergence problems. In \cite{nos1} we also 
performed an expansion of our results in powers of $1/m$ ($m$ being the 
nucleon mass) and compared with the HB expressions from Entem and 
Machleidt \cite{EM}, where we found a small number of differences. These 
works triggered the importance 
to assess quantitatively the differences between the HB- and RB-ChPT $NN$ 
potentials in asymptotic phase shifts, and if they are comparable to the 
uncertainties in the parameters of the potential, the so called low energy 
constants (LECs). The discrepancies found between our {\em expanded} 
\cite{nos1} and the HB \cite{EM} expressions also 
needs to be understood as they should, in principle, coincide. It is 
the main goal of this work to address both issues. 

The determination of the LECs is of much current interest, as they 
describe the dynamics of QCD at low energies. At the order we are 
working on, it requires the knowledge of the dimension two ($c_i$) and 
three ($d_i$) LECs, present at the order $q^2$ (${\cal L}^{(2)}_N$) and 
order $q^3$ (${\cal L}^{(3)}_N$) pion-nucleon chiral Lagrangians, 
respectively. Obviously, it seems more natural trying to obtain their 
values from $\pi N$ processes. In Ref.~\cite{bkm97} it was performed 
through a fit to threshold and subthreshold parameters, and a physical 
interpretation for their values was given, similar to what has been done 
in the pure mesonic sector \cite{ecker89}. For instance, assuming that 
the LEC $c_1$ is entirely saturated by a scalar meson exchange, a 
perfect agreement for their mass and coupling constant with the ones 
used by the Bonn one boson exchange model \cite{bonn} emerges. 
The other dimension two LECs $c_2$, $c_3$, and $c_4$ where also estimated 
to be dominated by the delta resonance, with relevant contributions from 
rho ($c_4$) and sigma ($c_3$) mesons, and marginal contributions from 
the Roper resonance $N^{*}$(1440). However, description of $\pi N$ 
phase shifts, albeit consistent, was not satisfactory, the same happening 
with the values from Moj\v zi\v s \cite{moj98,datpak}. It was improved 
in subsequent papers \cite{fms98,butik}, but the central values for the 
LECs still vary considerably \cite{evg02}. Moreover, the most reasonable 
results were obtained using the experimental analyses from the Karlsruhe 
group \cite{kochpiet,ka84}, which show disagreements with most recent 
$\pi N$ database. As an alternative to $\pi N$ processes, recently the 
Nijmegen group proposed the extraction of the dimension two LECs $c_1$, 
$c_3$, and $c_4$ using $NN$ partial wave analyses \cite{nij99,nij03}, 
where they obtained values with (statistical) uncertainties much smaller 
than obtained from $\pi N$ processes. Their values were disputed by 
Entem and Machleidt \cite{EM,EMach-com}, and in this work we also try to 
investigate this question. 

An important point here is that we do not try to address details of 
the non-perturbative aspect of the $NN$ interaction; instead, we deal 
only with the two-nucleon irreducible diagrams up to order $q^4$, 
which should be identified with the irreducible kernel in a 
Bethe-Salpeter equation, or the potential in a Lippmann-Schwinger 
equation. As suggested by Weinberg \cite{wein90}, the usual power 
counting of ChPT should be applied to it. The reducible diagrams are 
afflicted with infrared divergences (or, in the language of old-fashioned 
time-ordered perturbation theory, small energy denominators) and, 
depending on the power counting adopted, a resumation of certain classes 
of diagrams have to be performed to all orders. The non-perturbative 
problem is dealt in the more generic framework of effective field 
theory (EFT, for reviews see, for instance, 
Refs.~\cite{bbhps00,kubod99,eftrev,lepage}) and, despite of the progress it 
has achieved in the last years 
\cite{bira,egm,egm04-1,egm04-2,bgr99,cgk04,va04} 
there are still controversies about power counting schemes 
\cite{wein90,ksw,fms,bk02,gege,YH04}. This work follows the same spirit as 
the ones from Kaiser {\em et.al.} \cite{KBW,Kaiser2,Kaiser3}, Entem and 
Machleidt \cite{EM}, and Ballot, Rocha, and Robilotta \cite{BRR}, in 
the sense that we look only at peripheral waves ($l\geq 3$) where the 
Born approximation is expected to be valid \cite{BR,egm04-1}. With this 
procedure we try to investigate how far we are from the dream of establish, 
as in the case of the one-pion exchange (OPE) in the mid-60s, a mathematical 
structure for the two-pion exchange (TPE) which can be called ``unique", 
dictaded by chiral symmetry (by unique here we do not consider 
ambiguities arising from off-shell effects \cite{cf86} or choices 
of field variables \cite{egm}, as discussed in Ref.~\cite{friar99}).

We organize this paper as follows. In Sec.~\ref{secII} we present the 
motivation of our previous works with some detail, showing the contrast 
between RB and HB calculations for the TPEP. There we also address the 
differences we found when comparing our {\em expanding} results with 
the HB expressions, and their origin are better (but still not completely) 
understood. An interesting outcome of this work is a quantitative measure 
of how different ranges of the potential contributes to the phase shifts, 
and this was made possible by the use of the phase function (or variable 
phase) method, described in Sec.~\ref{secIII}. 
This study, as well as the dependence of the phase shifts with some of 
the existing values for the LECs, is shown in Sec.~\ref{secIV}, 
where we found some restrictions in order to achieve consistency with the 
experimental data. We close this work with our conclusions and remarks 
in Sec.~\ref{secV}. 


\section{relativistic versus heavy baryon TPEP} \label{secII}

In our recent works \cite{nos1,nos2} we argued that the problem of 
convergence that appears in the HB formulation manifests itself in the 
TPEP at large distances. This was ilustrated by comparing our basic 
loop functions with their $1/m$ expansion, the latter not being able to 
reproduce the correct asymptotic behavior. In the end of this section we 
show that the same 
occurs to the TPEP decomposed in peripheral partial waves ($J\geq 3$). 
Before that, we present a discussion about the differences we 
found between our {\em expanded} results and the ones obtained in the 
HB framework \cite{KBW,Kaiser2,Kaiser3,EM}. It is worth to emphasize 
that they must in principle coincide, therefore, such discrepancies 
should not exist. To investigate them we revised some of our 
calculations, and updated results are presented in Secs.~\ref{secII2l} 
and \ref{secIIhbexp}. In order to fix the notation and help understanding 
the origin of these discrepancies we outline the main steps on the 
derivation of our TPEP. For further details, we direct the reader to 
our previous works \cite{nos1,nos2}. For numerical evaluations, in this 
section we use the same values for the LECs used by Entem and Machleidt 
\cite{EM} (last column of table \ref{tab3}). 

\subsection{formalism}\label{secIIform}

\begin{figure}[!ht]
  \epsfig{figure=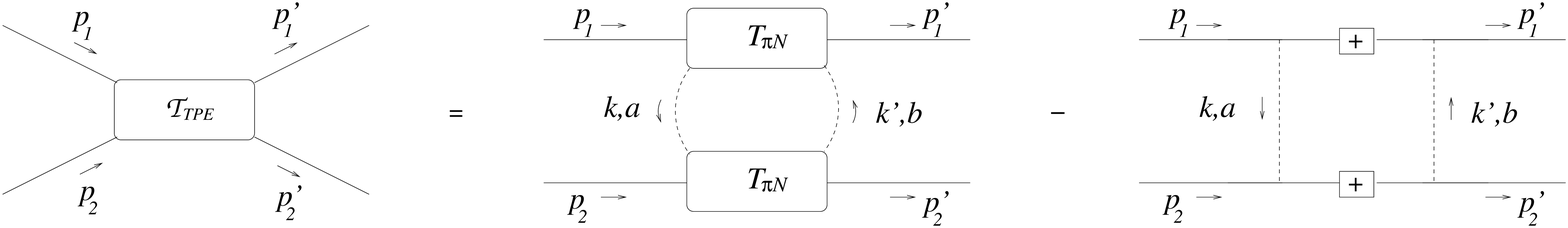, height=0.9in}
  \caption{The irreducible two-pion exchange amplitude.}
\label{figI-1}
\end{figure}

The TPE contribution to the $NN$ interaction can be parametrized in 
terms of $\pi N$ scattering sub-amplitudes, as ilustrated 
by the first graph on the right hand side of Fig.~\ref{figI-1}. 
One can describes the $\pi N$ scattering in terms of two 
invariant amplitudes $D^{\pm}(\nu,t)$ and $B^{\pm}(\nu,t)$, 

\begin{eqnarray}
T_{\pi N}^{ab}&=&\delta_{ab}\,T_{\pi N}^+ 
+ i\epsilon_{bac}\,\tau_c\,T_{\pi N}^-\,,
\nonumber\\[2mm]
T_{\pi N}^{\pm}&=&\bar u({\mbox{\boldmath $p'$}})
\,\Big[D^{\pm}-\frac{i}{2m}\,\sigma_{\mu\nu}(p'-p)^{\mu}\,
\frac{(k+k')^{\nu}}{2}\,B^{\pm}\Big]\,u({\mbox{\boldmath $p'$}})\,,
\end{eqnarray}

\noindent where $m$ is the nucleon mass, with initial and final momenta 
${\mbox{\boldmath $p$}}$ and ${\mbox{\boldmath $p'$}}$, respectively, while 
$\nu=[(p+k)^2-(p-k')^2]/4m$ and $t=(k'-k)^2$ are the usual Mandelstam 
variables. This allows the TPE amplitude to be written as 

\begin{eqnarray}
{\cal T}_{TPE}&=&-\frac{i}{2!}\int[\cdots]
\left[3\,T_{\pi N}^{(1)+}\,T_{\pi N}^{(2)+}
+2\,{\mbox{\boldmath $\tau$}}^{(1)}\cdot{\mbox{\boldmath $\tau$}}^{(2)}\,
T_{\pi N}^{(1)-}\,T_{\pi N}^{(2)-}\right]
\\[3mm]
&=&
\left[\bar uu\right]^{(1)}\left[\bar uu\right]^{(2)}{\cal I}_{DD}^{\pm}
\nonumber\\[2mm]&&
-\left[\bar uu\right]^{(1)}
\left[\bar u\,i\,
\sigma_{\mu\lambda}\,\frac{(p^{\prime}-p)^{\mu}}{2m}\,u\right]^{(2)}
{\cal I}_{DB}^{\lambda\pm}
-\left[\bar u\,i\,
\sigma_{\mu\lambda}\,\frac{(p^{\prime}-p)^{\mu}}{2m}\,u\right]^{(1)}
\left[\bar uu\right]^{(2)}{\cal I}_{DB}^{\lambda\pm}
\nonumber\\[2mm]&&
+\left[\bar u\,i\,\sigma_{\mu\lambda}\,
\frac{(p^{\prime}-p)^{\mu}}{2m}\,u\right]^{(1)}
\left[\bar u\,i\,\sigma_{\nu\rho}
\,\frac{(p^{\prime}-p)^{\nu}}{2m}\,u\right]^{(2)}
{\cal I}_{BB}^{\lambda\rho\pm}\,,
\label{eqI-TPEstruct}
\end{eqnarray}

\noindent where the symbol $\int[\cdots]$ represents the four-dimensional 
integration with two pion propagators, 

\begin{equation}
\int[\cdots]=\int\frac{d^4Q}{(2\pi)^4}\,\frac{1}{(k^2-\mu^2)}
\,\frac{1}{(k'^2-\mu^2)}\,,
\end{equation}

\noindent $\mu$ is the pion mass, $Q=(k+k')/2$, and the profile 
functions ${\cal I}$'s are loop 
integrals written in terms of the amplitudes $D^{\pm}$ and $B^{\pm}$, 

\begin{equation}
\begin{array}{rclcrcl}
{\cal I}_{DD}^{\pm}&=&-\frac{i}{2}\,\int[\cdots]\,D^{(1)\pm}\,D^{(2)\pm}\,,
&{\mbox{\ \ \ \ }}&
{\cal I}_{DB}^{\lambda\pm}&=&-\frac{i}{2}\,\int[\cdots]\,Q^{\lambda}\;
D^{(1)\pm}\,B^{(2)\pm}\,,\\[2mm]
{\cal I}_{BD}^{\lambda\pm}&=&-\frac{i}{2}\,\int[\cdots]\,Q^{\lambda}\;
B^{(1)\pm}\,D^{(2)\pm}\,,&&
{\cal I}_{BB}^{\lambda\rho\pm}&=&-\frac{i}{2}\,\int[\cdots]
\,Q^{\lambda}Q^{\rho}\;B^{(1)\pm}\,B^{(2)\pm}\,.
\end{array}\label{eqI-prf1}
\end{equation}

Using the variables

\begin{equation}
W=p_1+p_2=p'_1+p'_2\,,
\qquad
z=\frac{1}{2}\,\big[(p_1+p'_1)-(p_2+p'_2)\big]\,,
\qquad
q=k'-k=p'_1-p_1=p_2-p'_2\,,
\end{equation}

\noindent we decompose the Lorentz structure of these profile functions 
as 

\begin{eqnarray}
{\cal I}_{DB}^{\lambda\pm}&=&
\frac{W^{\lambda}}{2m}\,{\cal I}_{DB}^{(w)\pm}
+\frac{z^{\lambda}}{2m}\,{\cal I}_{DB}^{(z)\pm}\,,
\\[2mm]
{\cal I}_{BD}^{\lambda\pm}&=&
\frac{W^{\lambda}}{2m}\,{\cal I}_{DB}^{(w)\pm}
-\frac{z^{\lambda}}{2m}\,{\cal I}_{DB}^{(z)\pm}\,,
\\[2mm]
{\cal I}_{BB}^{\lambda\rho\pm}&=&g^{\lambda\rho}\,{\cal I}_{BB}^{(g)\pm}
+\frac{W^{\lambda}W^{\rho}}{4m^2}\,{\cal I}_{BB}^{(w)\pm}
+\frac{z^{\lambda}z^{\rho}}{4m^2}\,{\cal I}_{BB}^{(z)\pm}\,.
\end{eqnarray}

\begin{figure}[!ht]
  \epsfig{figure=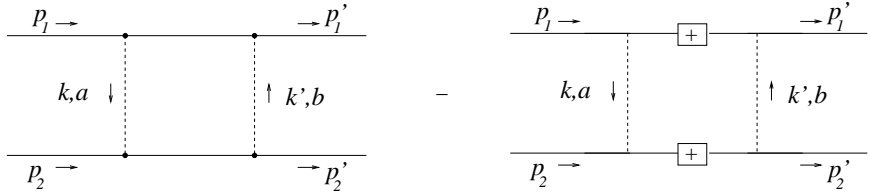, height=1.0in}
  \caption{The plannar box diagram (left) and the positive-energy 
projection of the iterated OPE (right), which has to be subtracted 
to obtain the irreducible TPE amplitude. }
\label{figI-2}
\end{figure}

The evaluation of Eq.(\ref{eqI-prf1}) to leading order gives rise to 
bubble, triangle, 
crossed box, and plannar box diagrams, the latter one containing 
an iteration of the one pion exchange (OPE). 
In order to obtain the irreducible TPE amplitude, this contribution 
has to be subtracted. Schematically it is represented by 
Fig.~\ref{figI-2}, where the symbol ``+" on the second graph means that 
the two-nucleon propagator contains only its 
positive-energy projection. This projection is not uniquely defined 
\cite{RAP99,friar99} and several prescriptions, like the 
Blankenbecler-Sugar \cite{BbS}, Equal-Time \cite{EqT}, or the 
Gross \cite{Gross} equations, could be used. Therefore, when comparing 
our expanded results with the HB ones, we should keep in mind that 
part of the differences might come from the particular prescription 
adopted. 

The iterated OPE can be recasted in the same structure as 
Eq.(\ref{eqI-TPEstruct}), allowing us to define the functions 
$\hat {\cal I}$'s as the ones in Eq.(\ref{eqI-prf1}) subtracted by 
the iterated OPE amplitude. The components of the potential, up 
to $O(q^4)$, can therefore be written in terms of these functions as 

\begin{eqnarray}
V_{C}^{\pm}&=&
\tau^{\pm}\,\frac{m}{E}\Biggl\{\hat{\cal I}_{DD}^{\pm}+\frac{q^2}{4m^2}
\left[2\,\hat{\cal I}^{(w)\pm}_{DB}-\hat{\cal I}_{DD}^{\pm}\right]
\nonumber\\[1mm]&&
\!+\frac{q^2}{4m^2}\biggl[\frac{q^2}{4m^2}\left(\hat{\cal I}^{(w)\pm}_{BB}
-2\,\hat{\cal I}^{(w)\pm}_{DB}+\hat{\cal I}^{(g)\pm}_{BB}\right)
-\frac{z^2}{4m^2}\left(\frac{1}{4}\,\hat{\cal I}_{DD}^{\pm}
+\hat{\cal I}^{(w)\pm}_{DB}+\hat{\cal I}^{(z)\pm}_{DB}\right)\biggr]\Biggr\}\,,
\label{eq:ch4-potSap01}
\\[5mm]
V_{LS}^{\pm}&=&
\tau^{\pm}\,\frac{m}{E}\Biggl\{
-\frac{1}{2}\,\hat{\cal I}_{DD}^{\pm}
+\hat{\cal I}^{(z)\pm}_{DB}+\hat{\cal I}^{(w)\pm}_{DB}
+\frac{q^2}{4m^2}\left[-\frac{1}{2}\,\hat{\cal I}^{(z)\pm}_{DB}
+\frac{3}{2}\,\hat{\cal I}^{(g)\pm}_{BB}
-\frac{3}{2}\,\hat{\cal I}^{(w)\pm}_{DB}
+\frac{1}{8}\,\hat{\cal I}_{DD}^{\pm}+\hat{\cal I}^{(w)\pm}_{BB}\right]
\nonumber\\[1mm]&&
-\frac{z^2}{4m^2}\left[\frac{1}{2}\,\hat{\cal I}^{(z)\pm}_{DB}
+\frac{1}{2}\,\hat{\cal I}^{(w)\pm}_{DB}
+\frac{1}{8}\,\hat{\cal I}_{DD}^{\pm}\right]\Biggr\}\,,
\label{eq:ch4-potSap02}
\\[5mm]
V_{T}^{\pm}&=&\frac{1}{2}\,V_{SS}^{\pm}=\tau^{\pm}\,\frac{m}{E}\Biggl[
-\frac{1}{12}\,\hat{\cal I}^{(g)\pm}_{BB}\Biggr]\,,
\label{eq:ch4-potSap03}
\\[5mm]
V_{Q}^{\pm}&=&
\!\tau^{\pm}\,\frac{m}{E}\Biggl\{-\frac{1}{64}\,\hat{\cal I}_{DD}^{\pm}
+\frac{1}{16}\,\hat{\cal I}^{(z)\pm}_{DB}
-\frac{1}{8}\,\hat{\cal I}^{(g)\pm}_{BB}
+\frac{1}{16}\,\hat{\cal I}^{(w)\pm}_{DB}
+\frac{1}{16}\,\hat{\cal I}^{(z)\pm}_{BB}
-\frac{1}{16}\,\hat{\cal I}^{(w)\pm}_{BB}\Biggr\}\,.
\label{eq:ch4-potSap04}
\end{eqnarray}

\begin{figure}[!ht]
  \epsfig{figure=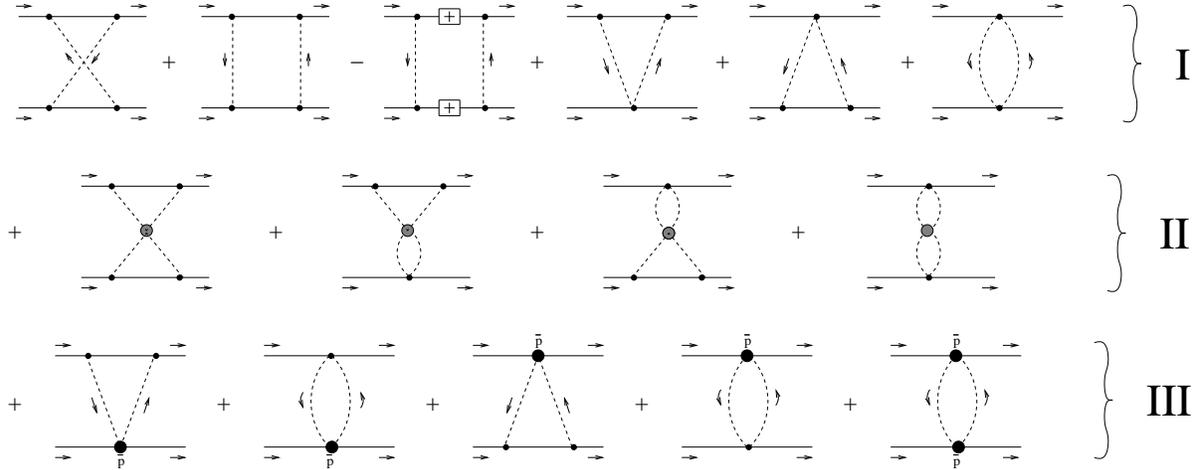, width=6.2in}
  \caption{Dynamics of the relativistic TPEP. The small black dots 
represent vertices from ${\cal L}^{(1)}_{\pi N}$, the big shaded dots, 
the $\pi\pi$ scattering amplitude, and the big black dots, the $\pi N$ 
subthreshold coefficients. The latter contains other two loop 
contributions, as well as vertices from ${\cal L}^{(2)}_{\pi N}$ and 
${\cal L}^{(3)}_{\pi N}$. }
\label{figI-3}
\end{figure}

The dynamics of our relativistic TPEP is classified in terms of three 
families of diagrams, according to their topology. The first line of 
diagrams in Fig.~\ref{figI-3} corresponds to the irreducible one loop 
graphs with vertices from the $O(q^1)$ $\pi N$ chiral Lagrangian, 
${\cal L}^{(1)}_{\pi N}$, with coupling constants at their physical 
values (family I). The second line (family II) 
contains two loop diagrams with an intermediate $\pi \pi$ scattering, 
while the third line (family III) comprises one loop graphs with 
vertices from ${\cal L}^{(2)}_{\pi N}$ and ${\cal L}^{(3)}_{\pi N}$, 
as well as other two loop graphs, parametrized in terms of $\pi N$ 
subthreshold coefficients \cite{hoeler}. 

\subsection{revision of our two loop diagrams}\label{secII2l}

The expressions for our previous profile functions are given in 
appendices D--F of Ref.~\cite{nos1}. While investigating the origin of 
the differences between our expanded and the HB results, we realize a 
small mistake in our two loop contributions (family II). We checked that 
the revised calculation is independent on the parametrization 
of the pion field and leads to 

\begin{eqnarray}
{\cal I}^{+}_{DD}&=&-\frac{1}{(4\pi)^4}\,\frac{\mu^4g_A^4}{16f_{\pi}^6}\,
(1-2t/\mu^2)\,\Big[(1-t/2\mu^2)\,\Pi_t-2\pi\Big]^2\,,
\\[3mm]
{\cal I}^{-}_{DD}&=&
-\frac{1}{(4\pi)^4}\,\frac{\mu^4}{4f_{\pi}^6}\;
\frac{W^2}{4m^2}\,\Bigg\{\frac{1}{3}\,(g_A^2-1)\bigg[(1-t/4\mu^2)\,\Pi_\ell
+2-\frac{t}{4\mu^2}\bigg]-g_A^2\bigg[(1-t/2\mu^2)\,\Pi_\ell+1
-\frac{t}{3\mu^2}\bigg]\Bigg\}^2\,,
\\[3mm]
{\cal I}^{(g)-}_{BB}&=&-\frac{1}{(4\pi)^4}\,\frac{m^2\mu^2}{4f_{\pi}^6}\,
\Big[g_A^2\,(1-t/4\mu^2)\,\Pi_t-\pi\Big]^2\,,
\\[3mm]
{\cal I}^{(g)+}_{BB}&=&{\cal I}^{(z)\pm}_{DB}={\cal I}^{(w)\pm}_{BB}
={\cal I}^{(w)\pm}_{BB}={\cal I}^{(w)\pm}_{DB}=0\,,
\end{eqnarray}

\noindent where $g_A$ and $f_{\pi}$ are respectively the nucleon axial 
coupling and the pion decay constant, and $\Pi_\ell$, $\Pi_t$, 
$\Pi_\times$, $\Pi_b$, $\tilde\Pi_b$, and $\Pi_a$ are our basic 
loop integrals defined in Ref.~\cite{nos1}. These modifications 
alter the expressions of just two components of our {\em non-expanded, 
relativistic} potential: in $V_C^-$, the non-local (nonstatic) term 

\begin{equation}
\left[\frac{\mu}{m}\right]^2\frac{m^2}{256\pi^2f_{\pi}^2}\,\frac{z^2}{\mu^2}\,
g_A^4\left[\left(1-\frac{t}{4\mu^2}\right)\Pi_t-\pi\right]^2
\end{equation}

\noindent does not exist, while in $V_{LS}^-$ there are no contributions 
from family II at all. 

\subsection{$1/m$ expansion and analysis of discrepancies}\label{secIIhbexp}

The second revision we made concerns the HB expansion of our loop 
functions. We found that the plannar and crossed box integrals 
($\Pi_b$ and $\Pi_\times$, respectively) which, in Ref.~\cite{nos1}, 
were given by 

\begin{eqnarray}
\Pi_\times &\rightarrow& -\Pi'_\ell -\frac{\mu}{2m}\,
\frac{\pi}{(1-t/4\mu^2)}-\frac{\mu^2}{12m^2}\,\Big[3\,(1-t/2\mu^2)^2\,
(2\Pi'_\ell-\Pi''_\ell)+2\,(z^2/\mu^2)\,\Pi'_\ell\Big]+O(q^3)\,,
\\[2mm]
\Pi_b &\rightarrow& -\Pi'_\ell -\frac{\mu}{4m}\,
\frac{\pi}{(1-t/4\mu^2)}-\frac{\mu^2}{12m^2}\,\Big[(1-t/2\mu^2)^2\,
(2\Pi'_\ell-\Pi''_\ell)+2\,(z^2/\mu^2)\,\Pi'_\ell\Big]+O(q^3)\;,
\end{eqnarray}

\noindent have to be corrected by 

\begin{eqnarray}
\Pi_\times &\rightarrow& -\Pi'_\ell -\frac{\mu}{2m}\,
\frac{\pi}{(1-t/4\mu^2)}-\frac{\mu^2}{12m^2}\,
\Big[3\,(2\Pi'_\ell-\Pi''_\ell)+2\,(z^2/\mu^2)\,\Pi'_\ell\Big]+O(q^3)\,,
\\[2mm]
\Pi_b &\rightarrow& -\Pi'_\ell -\frac{\mu}{4m}\,
\frac{\pi}{(1-t/4\mu^2)}-\frac{\mu^2}{12m^2}\,
\Big[(2\Pi'_\ell-\Pi''_\ell)+2\,(z^2/\mu^2+t/\mu^2)\,\Pi'_\ell\Big]+O(q^3)\;.
\end{eqnarray}

Therefore, the comparison of our {\em expanded} results with the HB 
expressions (Sec.~X of Ref.~\cite{nos1}) must be updated by the following: 

\begin{eqnarray}
&&V_C=V_C^+=\frac{3 g_A^2}{16\pi f_\pi^4}
\left\{-\frac{g_A^2\;\mu^5}{16m(4\mu^2\!+\!{\mbox{\boldmath $q$}}^2)}
+\left[2\mu^2(2 c_1\!-\!c_3)-{\mbox{\boldmath $q$}}^2c_3\right](2\mu^2\!
+\!{\mbox{\boldmath $q$}}^2)\;A(q)\right.
\nonumber\\[2mm]
&&\left.+\frac{g_A^2(2\mu^2\!
+\!{\mbox{\boldmath $q$}}^2)\;A(q)}{16m}\left[-3{\mbox{\boldmath $q$}}^2
\;{\color{red} \,+(4\mu^2\!+\!{\mbox{\boldmath $q$}}^2)^{\dagger} }\;\right]
\right\}
\nonumber\\[2mm]
&&+\frac{g_A^2\;L(q)}{32\pi^2f_\pi^4m}
\left\{ \frac{24\mu^6}{4\mu^2\!+\!{\mbox{\boldmath $q$}}^2}\;(2c_1\!+\!c_3)
+6\mu^4(c_2\!-\!2c_3)+4\mu^2{\mbox{\boldmath $q$}}^2(6c_1\!+\!c_2\!-\!3c_3)
+{\mbox{\boldmath $q$}}^4(c_2\!-\!6c_3)\right\}
\nonumber\\[2mm]
&&-\frac{3 L(q)}{16\pi^2f_\pi^4}\left\{\left[-4\mu^2 c_1\!+\!
c_3(2\mu^2\!+\!{\mbox{\boldmath $q$}}^2)+c_2(4\mu^2\!
+\!{\mbox{\boldmath $q$}}^2)/6\right]^2
+\frac{1}{45}\,(c_2)^2 (4\mu^2\!+\!{\mbox{\boldmath $q$}}^2)^2\right\}
\nonumber\\[2mm]
&&+\frac{g_A^4}{32\pi^2f_\pi^4m^2}\left\{L(q)\left[
\frac{2\mu^8}{(4\mu^2\!+\!{\mbox{\boldmath $q$}}^2)^2}
+\frac{8\mu^6}{(4\mu^2\!+\!{\mbox{\boldmath $q$}}^2)}-2\mu^4
-{\mbox{\boldmath $q$}}^4\right]+\frac{\mu^6 /2}{(4\mu^2\!
+\!{\mbox{\boldmath $q$}}^2)}\right\}
\nonumber\\[2mm]
&&-\frac{3g_A^4[A(q)]^2}{1024\pi^2f_\pi^6}\;(\mu^2\!
+\!2{\mbox{\boldmath $q$}}^2)\;(2\mu^2\!+\!{\mbox{\boldmath $q$}}^2)^2
\nonumber\\[2mm]
&&-\frac{3g_A^4(2\mu^2\!+\!{\mbox{\boldmath $q$}}^2)\;A(q)}
{1024\pi^2f_\pi^6}\left\{4\mu\;g_A^2\;
(2\mu^2\!+\!{\mbox{\boldmath $q$}}^2)+2\mu\;(\mu^2\!
+\!2{\mbox{\boldmath $q$}}^2)]\right\}\;,
\label{Vc}\\[5mm]
&& V_T=-\frac{3\;V_T^+}{m^2}=\frac{3 g_A^4 \;L(q)}{64 \pi^2 f_\pi^4}
-\frac{g_A^4\;A(q)}{512\pi f_\pi^4m}\left[9(2\mu^2+{\mbox{\boldmath $q$}}^2)
\;{\color{red} \,+3(4\mu^2+{\mbox{\boldmath $q$}}^2)^{\dagger} }\;\right]
\nonumber\\[2mm]
&&-\frac{g_A^4\;L(q)}{32\pi^2f_\pi^4m^2}\left[{\mbox{\boldmath $z$}}^2/4
+5{\mbox{\boldmath $q$}}^2/8
+\frac{\mu^4}{4\mu^2\!+\!{\mbox{\boldmath $q$}}^2}\right]
\nonumber\\[2mm]
&&+\frac{g_A^2\;(4\mu^2+{\mbox{\boldmath $q$}}^2)\;L(q)}{32\pi^2f_\pi^4}
\left[(\tilde{d}_{14}-\tilde{d}_{15})
\;{\color{blue} \,-\left( g_A^4/32\;\pi^2f_{\pi}^2\right)^{\!*} }\;\right]\;,
\label{Vt}\\[5mm]
&& V_{LS}=-\frac{V_{LS}^+}{m^2}=-\frac{3g_A^4\;A(q)}{32\pi f_\pi^4m}
\;\left[(2\mu^2+{\mbox{\boldmath $q$}}^2)
\;{\color{red} \,+(\mu^2+3{\mbox{\boldmath $q$}}^2/8)^{\dagger} }\;\right]
\nonumber\\[2mm]
&&-\frac{g_A^4\;L(q)}{4\pi^2f_\pi^4m^2}
\left[\frac{\mu^4}{4\mu^2\!+\!{\mbox{\boldmath $q$}}^2}
+\frac{11}{32}\,{\mbox{\boldmath $q$}}^2 \right]
-\frac{g_A^2\;c_2\; L(q)}{8 \pi^2 f_\pi^4 m}
\;(4\mu^2+{\mbox{\boldmath $q$}}^2)  \;,
\label{Vls}\\[5mm]
&&V_{\sigma L}=\frac{4\;V_Q^+}{m^4}=
-\frac{g_A^4\;L(q)}{32\pi^2f_\pi^4m^2}\;,
\label{Vq}
\end{eqnarray}

\begin{eqnarray}
&& W_C = V_C^- =\frac{L(q)}{384\pi^2f_\pi^4}\Bigg[4\mu^2\left(
5g_A^4-4g_A^2-1\right)+{\mbox{\boldmath $q$}}^2\left(23g_A^4-10g_A^2-1\right)
+\frac{48 g_A^4 \mu^4}{4\mu^2\!+\!{\mbox{\boldmath $q$}}^2}\Bigg]
\nonumber\\[2mm]
&&-\frac{g_A^2}{128\pi f_\pi^4m}\Bigg\{\frac{3g_A^2\mu^5}{4\mu^2\!
+\!{\mbox{\boldmath $q$}}^2}+A(q)\,(2\mu^2\!+\!{\mbox{\boldmath $q$}}^2)
\bigg[g_A^2\;(4\mu^2\!+\!3{\mbox{\boldmath $q$}}^2)
-2(2\mu^2\!+\!{\mbox{\boldmath $q$}}^2)
\;{\color{red} \,+g_A^2(4\mu^2\!+\!{\mbox{\boldmath $q$}}^2)^{\dagger} }\;
\bigg]\Bigg\}
\nonumber\\[2mm]&&
+\frac{{\mbox{\boldmath $q$}}^2\;c_4
\;L(q)}{192\pi^2f_\pi^4m}\Big[g_A^2(8\mu^2+5{\mbox{\boldmath $q$}}^2)
+(4\mu^2+{\mbox{\boldmath $q$}}^2)\Big]
+\frac{16g_A^4\mu^6}{768\pi^2f_\pi^4m^2}\;\frac{1}{4\mu^2\!
+\!{\mbox{\boldmath $q$}}^2}
\nonumber\\[2mm]
&&
-\frac{L(q)}{768\pi^2f_\pi^4m^2}\Bigg\{(4\mu^2\!
+\!{\mbox{\boldmath $q$}}^2){\mbox{\boldmath $z$}}^2
+g_A^2\Bigg[\frac{48\mu^6}{4\mu^2\!+\!{\mbox{\boldmath $q$}}^2}-24\mu^4-
12\,(2\mu^2\!+\!{\mbox{\boldmath $q$}}^2)\,{\mbox{\boldmath $q$}}^2
+(16\mu^2\!+\!10{\mbox{\boldmath $q$}}^2){\mbox{\boldmath $z$}}^2\Bigg]
\nonumber\\[2mm]&&
+g_A^4\Bigg[{\mbox{\boldmath $z$}}^2\Bigg(\frac{16\mu^4}{4\mu^2\!
+{\mbox{\boldmath $q$}}^2}-7{\mbox{\boldmath $q$}}^2-20\mu^2\Bigg)
-\frac{64\mu^8}{(4\mu^2\!+{\mbox{\boldmath $q$}}^2)^2}-
\frac{48\mu^6}{4\mu^2\!+{\mbox{\boldmath $q$}}^2}
+\frac{16\mu^4{\mbox{\boldmath $q$}}^2}
{4\mu^2\!+\!{\mbox{\boldmath $q$}}^2}
+20{\mbox{\boldmath $q$}}^4
+24\mu^2{\mbox{\boldmath $q$}}^2+24\mu^4\Bigg]\Bigg\}
\nonumber\\[2mm]
&&-\frac{L(q)}{18432\pi^4f_\pi^6}\Bigg\{\Bigg[192\pi^2f_\pi^2\tilde{d}_3
\;{\color{blue} \,-\frac{(15\!+\!7g_A^4)}{5}^{\!*} }\;\Bigg]
(4\mu^2\!+\!{\mbox{\boldmath $q$}}^2)
\Big[2g_A^2(2\mu^2\!+\!{\mbox{\boldmath $q$}}^2)-3/5(g_A^2-1)(4\mu^2\!
+\!{\mbox{\boldmath $q$}}^2)\Big]
\nonumber\\[2mm]&&
+\Big[6g_A^2(2\mu^2\!+\!{\mbox{\boldmath $q$}}^2)-(g_A^2-1)\;(4\mu^2\!
+\!{\mbox{\boldmath $q$}}^2)\Big]
\nonumber\\[2mm]&&
\times\Bigg[384\pi^2f_\pi^2\left((2\mu^2\!+\!{\mbox{\boldmath $q$}}^2)\;
(\tilde{d}_1+\tilde{d}_2)+4\mu^2\;\tilde{d}_5\right)
+L(q)\left(4\mu^2(1+2g_A^2)+{\mbox{\boldmath $q$}}^2(1+5g_A^2)\right)
\nonumber\\[2mm]&&
-\left(\frac{{\mbox{\boldmath $q$}}^2}{3}\,(5+13g_A^2)+8\mu^2(1+2g_A^2)\right)
\;{\color{blue} \,+\left(2g_A^4(2\mu^2+{\mbox{\boldmath $q$}}^2)+\frac{2}{3}\,
{\mbox{\boldmath $q$}}^2(1+2g_A^2)\right)^{\!*} }\;\Bigg]\Bigg\}\;,
\label{Wc}\\[5mm]
&&W_T=-\frac{3}{m^2}\;V_T^-=\frac{g_A^2 A(q)}{32\pi f_\pi^4}
\Bigg\{\left(c_4+\frac{1}{4m}\right)(4\mu^2\!+\!{\mbox{\boldmath $q$}}^2)
-\frac{g_A^2}{8m}\;\Big[10\mu^2+3{\mbox{\boldmath $q$}}^2
\;{\color{red} \,-(4\mu^2\!+\!{\mbox{\boldmath $q$}}^2)^{\dagger} }\;
\Big]\Bigg\}
\nonumber\\[2mm]
&&-\frac{c_4^2\;L(q)}{96\pi^2f_\pi^4}(4\mu^2\!+\!{\mbox{\boldmath $q$}}^2)+
\frac{c_4\;L(q)}{192\pi^2f_\pi^4m}\left[g_A^2(16\mu^2
+7{\mbox{\boldmath $q$}}^2)-(4\mu^2\!+\!{\mbox{\boldmath $q$}}^2)\right]
\nonumber\\[2mm]
&&-\frac{L(q)}{1536\pi^2f_\pi^4m^2}\left[g_A^4\left(28\mu^2
+17{\mbox{\boldmath $q$}}^2+ \frac{16\; \mu^4}{4\mu^2\!
+\!{\mbox{\boldmath $q$}}^2}\right)-g_A^2(32\mu^2+14{\mbox{\boldmath $q$}}^2)+
(4\mu^2\!+\!{\mbox{\boldmath $q$}}^2)\right]
\nonumber\\[2mm]
&&-\frac{[A(q)]^2g_A^4\;(4\mu^2\!+\!{\mbox{\boldmath $q$}}^2)^2}
{2048\pi^2f_\pi^6}-\frac{A(q)g_A^4\;(4\mu^2\!
+\!{\mbox{\boldmath $q$}}^2)}{1024 \pi^2 f_\pi^6}\;\mu(1+2g_A^2)\;,
\label{Wt}\\[5mm]
&&W_{LS}=-\frac{1}{m^2}\;V_{LS}^-=
\frac{A(q)}{32\pi f_\pi^4m}\Big[g_A^2\,(g_A^2-1)\;(4\mu^2
+{\mbox{\boldmath $q$}}^2)
\;{\color{red} \,+g_A^4\;(2\mu^2+3{\mbox{\boldmath $q$}}^2/4)^{\dagger} }\;\Big]
\nonumber\\[2mm]&&
+\frac{c_4\;L(q)}{48\pi^2mf_\pi^4}\Big[g_A^2(8\mu^2+5{\mbox{\boldmath $q$}}^2)
+(4\mu^2\!+\!{\mbox{\boldmath $q$}}^2)\Big]
\nonumber\\[2mm]
&&+\frac{L(q)}{256\pi^2m^2f_\pi^4}\Bigg[(4\mu^2\!
+\!{\mbox{\boldmath $q$}}^2)-16 g_A^2\;(\mu^2+3{\mbox{\boldmath $q$}}^2/8)
+\frac{4 g_A^4}{3}\Bigg(9\mu^2+11{\mbox{\boldmath $q$}}^2/4
-\frac{4\mu^4}{4\mu^2+{\mbox{\boldmath $q$}}^2}\Bigg)\Bigg]\;,
\label{Wls}\\[5mm]
&&W_{\sigma L}\simeq 0\;,
\label{Wq}
\end{eqnarray}

\noindent where $L(q)$ is the usual logaritmic function, 

\begin{equation}
L(q)=-\frac{1}{2}\,\Pi_\ell
=\frac{\sqrt{4\mu^2+q^2}}{q}\,\ln\frac{\sqrt{4\mu^2+q^2}+q}{2\mu}
\end{equation}

\noindent and, from now on, $q=|{\mbox{\boldmath $q$}}|$. 

We can notice that the number of different terms dropped from 14 in 
Ref.~\cite{nos1} down to 9. Now we have a better understanding 
about their origins: those marked with 
${\color{red} [\cdots]^{\dagger} }$ (six terms) are related to our 
choice of the Blankenbecler-Sugar \cite{BbS} prescription to deal with the 
iterated OPEP, which is the same adopted in Refs.~\cite{nos1,nos2,RR,BRR,BR}. 
Different prescriptions lead to different expressions only for these terms, 
as mentioned previously. In the HB calculations of 
Refs.~\cite{KBW,Kaiser3}, the plannar box diagram is expanded in powers 
of $1/m$ and then the iterated OPEP (given by Eq.(20) of Ref.~\cite{KBW}) 
is identified and subtracted. A more detailed study of these aspects will 
be given elsewhere \cite{future}. 

The remaining three different terms, marked with 
${\color{blue} [\cdots]^{\!*} }$, are 
originated from two loop contributions of \cite{Kaiser2}, part of them 
parametrized in our calculations through the subthreshold coefficients
(family III). 
The one in $V_T$ (and, consequently, $V_S$), Eq.(\ref{Vt}), can roughly 
be identified with the unintegrated term $V_T^{(b)}$ of Ref.~\cite{EM}, 
which was not considered while performing the above comparison. In 
order to show that 
this can be the case, notice first that their structures are similar. 
The HB expressions from Entem and Machleidt read \footnote{Note that, 
due to our convention for the scattering matrix, there is a global minus 
sign of difference in our potential relative to the one in Ref.~\cite{EM}.}

\begin{eqnarray}
V_T^{(a)}(q)&=&\frac{g_A^2}{32\pi^2f_{\pi}^4}\,
{\mbox{\boldmath $q$}}^4\,\int_{2\mu}^{\infty}
\frac{4d\alpha}{\alpha^4(\alpha^2+{\mbox{\boldmath $q$}}^2)}\,
3\kappa^3\int_0^1dx(1-x^2)\,(\bar d_{14}-\bar d_{15})
\nonumber\\[1mm]&=&
\frac{g_A^2L(q)}{32\pi^2f_{\pi}^4}\,
(4\mu^2+{\mbox{\boldmath $q$}}^2)\,(\bar d_{14}-\bar d_{15})\,,
\\[2mm]
V_T^{(b)}(q)&=&-\frac{g_A^2}{32\pi^2f_{\pi}^4}
\left[\frac{g_A^4}{32\pi^2f_{\pi}^2}\right]
{\mbox{\boldmath $q$}}^4\,
\nonumber\\[1mm]&\times&
\int_{2\mu}^{\infty}
\frac{4d\alpha}{\alpha^4(\alpha^2+{\mbox{\boldmath $q$}}^2)}\,
2\kappa^3\int_0^1dx(1-x^2)\,\Bigg[\,\frac{1}{6}-\frac{\mu^2}{\kappa^2x^2}
+\Bigg(1+\frac{\mu^2}{\kappa^2x^2}\Bigg)^{3/2}\ln\,
\frac{\kappa x+\sqrt{\mu^2+\kappa^2x^2}}{\mu}\,\Bigg]\,,
\label{vt-hb}
\end{eqnarray}

\noindent where $\kappa\equiv \sqrt{\alpha^2/4-\mu^2}$. 
Our equivalent terms are 

\begin{eqnarray}
V_T^{(a)}(q)&=&\frac{g_A^2L(q)}{32\pi^2f_{\pi}^4}\,
(4\mu^2+{\mbox{\boldmath $q$}}^2)\,(\bar d_{14}-\bar d_{15})\,,
\\[2mm]
V_T^{(b)}(q)&=&-\frac{g_A^2L(q)}{32\pi^2f_{\pi}^4}\,
(4\mu^2+{\mbox{\boldmath $q$}}^2)
\left[\frac{g_A^4}{32\pi^2f_{\pi}^2}\right]\,.
\label{vt-re}
\end{eqnarray}

In the first and second columns of table~\ref{tab1} we display, 
respectively, the values of Eqs.(\ref{vt-re}) and (\ref{vt-hb}) in 
configuration space, as a function of the internucleon distance $r$. 
We can see that the difference between them decreases with $r$, being 
around 20\% at 1fm, 10\% at 2fm, and remaining in a few \% for 
$r\geq 3$fm. As both expressions are numerically consistent at large 
distances one can say that this particular discrepancy is now understood. 
Looking at the third column, which shows the contribution of the term 
$V_T^{(a)}$, one can notice that such a difference is tiny compared to 
the whole two loop contributions, close to 3\% at 1fm and less than 
1\% for $r\geq 2$fm. To complete this analysis, we present in the fourth 
column the total value of the isoscalar tensor potential, which clearly 
shows that the whole two loop contributions (parametrized in our 
calculations inside family III) becomes less important at large 
distances. This behavior is already known, as ilustrated in Fig.~5(c) of 
Ref.~\cite{nos2}. Therefore the difference of using either Eq.(\ref{vt-hb}) 
or Eq.(\ref{vt-re}) compared to the total $V_T$ is numerically 
insignificant. 

\begin{table}[htb]
\begin{center}
\caption{Values of the isoscalar tensor potential (MeV) in configuration 
space. The first, second, and third columns are defined and explained in 
the text, while the last column is its total value. 
\label{tab1}
}
\begin{tabular} {|c|c|c|c|c|}
\hline
$r$(fm) & $V_T^{(b)}$, Eq.(\ref{vt-re}) & $V_T^{(b)}$, Eq.(\ref{vt-hb}) & 
$V_T^{(a)}$ & $V_T$ (total) 
\\ \hline
1.00 & -4.5479 E+00 & -5.4716 E+00 & -2.4737 E+01 & 
-3.9311 E+01 \\ \hline
2.00 & -2.2078 E-02 & -2.4062 E-02 & -1.2008 E-01 & 
-3.2551 E-01 \\ \hline
3.00 & -6.6798 E-04 & -7.0387 E-04 & -3.6332 E-03 & 
-1.5170 E-02 \\ \hline
4.00 & -4.0679 E-05 & -4.2189 E-05 & -2.2126 E-04 & 
-1.2858 E-03 \\ \hline
5.00 & -3.5582 E-06 & -3.6576 E-06 & -1.9353 E-05 & 
-1.4573 E-04 \\ \hline
6.00 & -3.8725 E-07 & -3.9586 E-07 & -2.1063 E-06 & 
-1.9562 E-05 \\ \hline
7.00 & -4.8725 E-08 & -4.9619 E-08 & -2.6502 E-07 & 
-2.9308 E-06 \\ \hline
8.00 & -6.7941 E-09 & -6.8999 E-09 & -3.6954 E-08 & 
-4.7418 E-07 \\ \hline
9.00 & -1.0225 E-09 & -1.0363 E-09 & -5.5614 E-09 & 
-8.1184 E-08 \\ \hline
10.00 & -1.6319 E-10 & -1.6514 E-10 & -8.8763 E-10 & 
-1.4515 E-08 \\ \hline
\end{tabular}
\end{center}
\end{table}

The two remaining discrepancies, which appear in $W_C$, are more 
difficult to trace. Analytically the identification of terms is not 
as clear as in $V_T$ and, in principle, does not appear to be 
equivalent. Despite of that, we follow the same numerical procedure 
as before, identifying $W_C^{(b)}$ in our calculations as 

\begin{eqnarray}
&& W_C^{(b)} = 
-\frac{L(q)}{18432\pi^4f_\pi^6}\Bigg\{-\frac{(15\!+\!7g_A^4)}{5}\;
(4\mu^2\!+\!{\mbox{\boldmath $q$}}^2)
\Big[2g_A^2(2\mu^2\!+\!{\mbox{\boldmath $q$}}^2)-3/5(g_A^2-1)(4\mu^2\!
+\!{\mbox{\boldmath $q$}}^2)\Big]
\nonumber\\[2mm]&&
+\Big[6g_A^2(2\mu^2\!+\!{\mbox{\boldmath $q$}}^2)-(g_A^2-1)\;(4\mu^2\!
+\!{\mbox{\boldmath $q$}}^2)\Big]
\left[2g_A^4(2\mu^2+{\mbox{\boldmath $q$}}^2)+\frac{2}{3}\,
{\mbox{\boldmath $q$}}^2(1+2g_A^2)\right]\Bigg\}\;,
\label{wc-re}
\end{eqnarray}

\noindent while the HB expressions are given by 

\begin{eqnarray}
W_C^{(b)}&=&-\frac{2q^6}{\pi}\,\int_{2\mu}^{\infty}d\alpha\;
\frac{\mbox{Im}\,W_C^{(b)}(i\alpha)}{\alpha^5(\alpha^2+q^2)}\,,
\\[2mm]
\mbox{Im}\,W_C^{(b)}(i\alpha)&=&-\frac{2\kappa}{3\alpha(8\pi f_{\pi}^2)^3}
\int_0^1dx\left[g_A^2(2\mu^2-\alpha^2)+2(g_A^2-1)\kappa^2x^2\right]
\nonumber\\[1mm]&&
\times\Bigg\{-3\kappa^2x^2+6\kappa x\sqrt{\mu^2+\kappa^2x^2}\,
\ln\,\frac{\kappa x+\sqrt{\mu^2+\kappa^2x^2}}{\mu}
\nonumber\\[1mm]&&
+g_A^4(\alpha^2-2\kappa^2x^2-2\mu^2)\,
\Bigg[\,\frac{5}{6}+\frac{\mu^2}{\kappa^2x^2}
-\Bigg(1+\frac{\mu^2}{\kappa^2x^2}\Bigg)^{3/2}\ln\,
\frac{\kappa x+\sqrt{\mu^2+\kappa^2x^2}}{\mu}\,\Bigg]\Bigg\}\,.
\label{wc-hb}
\end{eqnarray}

In table~\ref{tab2} we show the analogous of table~\ref{tab1} for 
$W_C$, where one cannot observe the same consistency as before: 
at large distances ($r\geq 4$fm), our result for (what we initially 
supposed to be) $W_C^{(b)}$ is almost 
twice larger than the HB one. Fortunately this difference is not 
numerically important for the total value of $W_C$ (as one can check 
by looking at the last column), but is somehow relevant for the whole 
two loop contributions ($W_C^{(a)}+W_C^{(b)}$). This numerical 
comparison indicates that these two mentioned discrepancies still 
persist, and should be considered the last small detail to be solved 
in order to reach an unique structure for the $O(q^4)$, chiral two-pion 
exchange component of the nucleon-nucleon potential. 

\begin{table}[htb]
\begin{center}
\caption{Values of the isovector central potential (MeV) in configuration
space, analogously to what was done in table~\ref{tab1} for the 
isoscalar tensor component.
\label{tab2}
}
\begin{tabular} {|c|c|c|c|c|}
\hline
$r$(fm) & $W_C^{(b)}$, Eq.(\ref{wc-re}) & $W_C^{(b)}$, Eq.(\ref{wc-hb}) &
$W_C^{(a)}$ & $W_C$ (total)
\\ \hline
1.00 & -5.8826 E+00 & -5.5372 E+00 & -7.6350 E+01 & -5.3305 E+01 \\ \hline
2.00 & -3.3949 E-02 & -2.4611 E-02 & -3.2322 E-01 & -4.5217 E-01 \\ \hline
3.00 & -1.2397 E-03 & -8.0784 E-04 & -8.9086 E-03 & -2.7783 E-02 \\ \hline
4.00 & -9.0177 E-05 & -5.5440 E-05 & -5.0169 E-04 & -3.0848 E-03 \\ \hline
5.00 & -9.2668 E-06 & -5.4899 E-06 & -4.0838 E-05 & -4.4010 E-04 \\ \hline
6.00 & -1.1654 E-06 & -6.7289 E-07 & -4.1468 E-06 & -7.1764 E-05 \\ \hline
7.00 & -1.6697 E-07 & -9.4596 E-08 & -4.8701 E-07 & -1.2702 E-05 \\ \hline
8.00 & -2.6176 E-08 & -1.4617 E-08 & -6.3325 E-08 & -2.3757 E-06 \\ \hline
9.00 & -4.3816 E-09 & -2.4189 E-09 & -8.8692 E-09 & -4.6224 E-07 \\ \hline
10.00 & -7.7072 E-10 & -4.2156 E-10 & -1.3135 E-09 & -9.2646 E-08 \\ \hline
\end{tabular}
\end{center}
\end{table}

\subsection{clean comparison between relativistic and HB TPEP}
\label{secIIcomp}

Here we present the consequences of expanding our TPEP in powers of 
$1/m$. In order to isolate only this particular effect, we use the 
fact that our {\em expanded} results have to coincide with the HB 
expressions. As the discrepancies in Eqs.(\ref{Vc})--(\ref{Wq}) are 
not actually related to the problem of the HB expansion, and using 
the fact that they can easily be identified in our {\em non-expanded} 
expressions, we simply ignore them in our relativistic calculations 
from now on. With this procedure we can also compare our phase shifts 
computed in configuration space with the ones from Entem and Machleidt 
\cite{EM}, performed in momentum space. 

\begin{figure}[!ht]
  \epsfig{figure=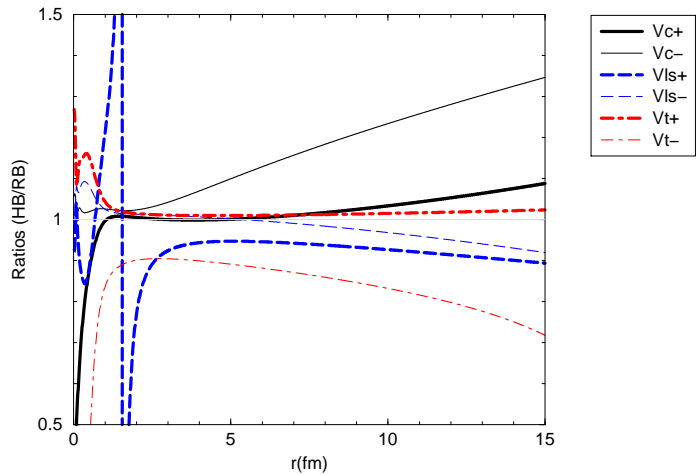, height=2.5in}
  \caption{Ratios between the HB and RB components of the TPEP, in 
configuration space.}
\label{figpots}
\end{figure}

In Fig.~\ref{figpots} we plot the ratios of the HB over the RB 
components of the TPEP. Even though the distance considered is 
quite large for hadronic interactions, we 
will show latter in Sec.~\ref{secIII} that peripheral phase shifts at 
sufficiently small energies are sensitive to distances as large as 
10fm. One can see that the pattern observed in the $1/m$ expansion of 
our loop functions 
\cite{nos2} reflects more on the isovector (--) channels, but 
produces deviations smaller than 10\% on the isoscalar (+) ones. 
Note that in the isoscalar central potential it remains below 5\% 
up to 10fm. We omit the ratios of the spin-spin components, which 
are almost the same as for the corresponding tensor ones. Before 
commenting further on these results, it is interesting to look 
how they manifest in the partial wave decomposition. 

\begin{figure}[hb]
\begin{center}
\begin{tabular}{ccccc}
\epsfig{figure=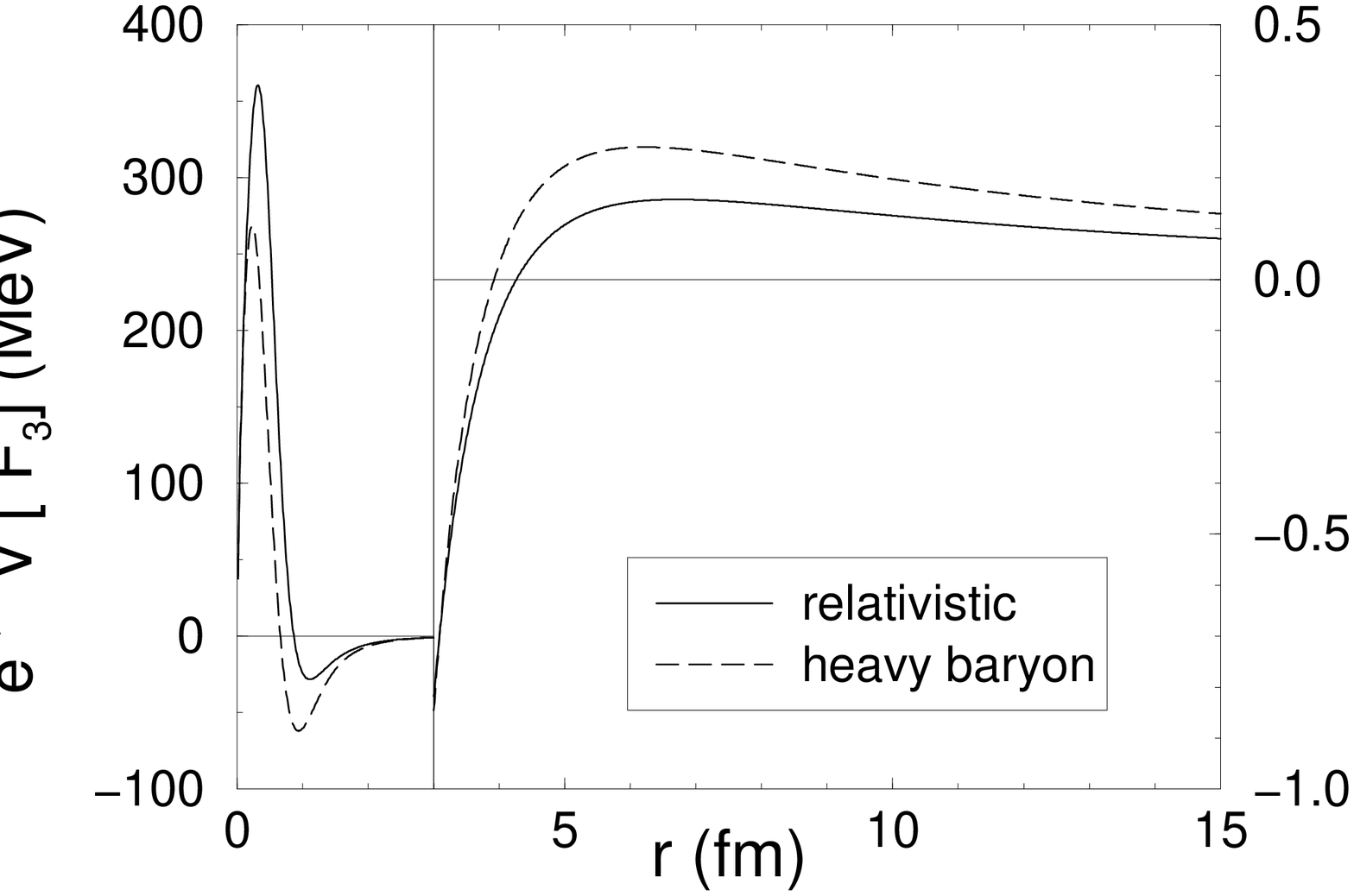, height=1.2in} \hspace{0.1in} &
\epsfig{figure=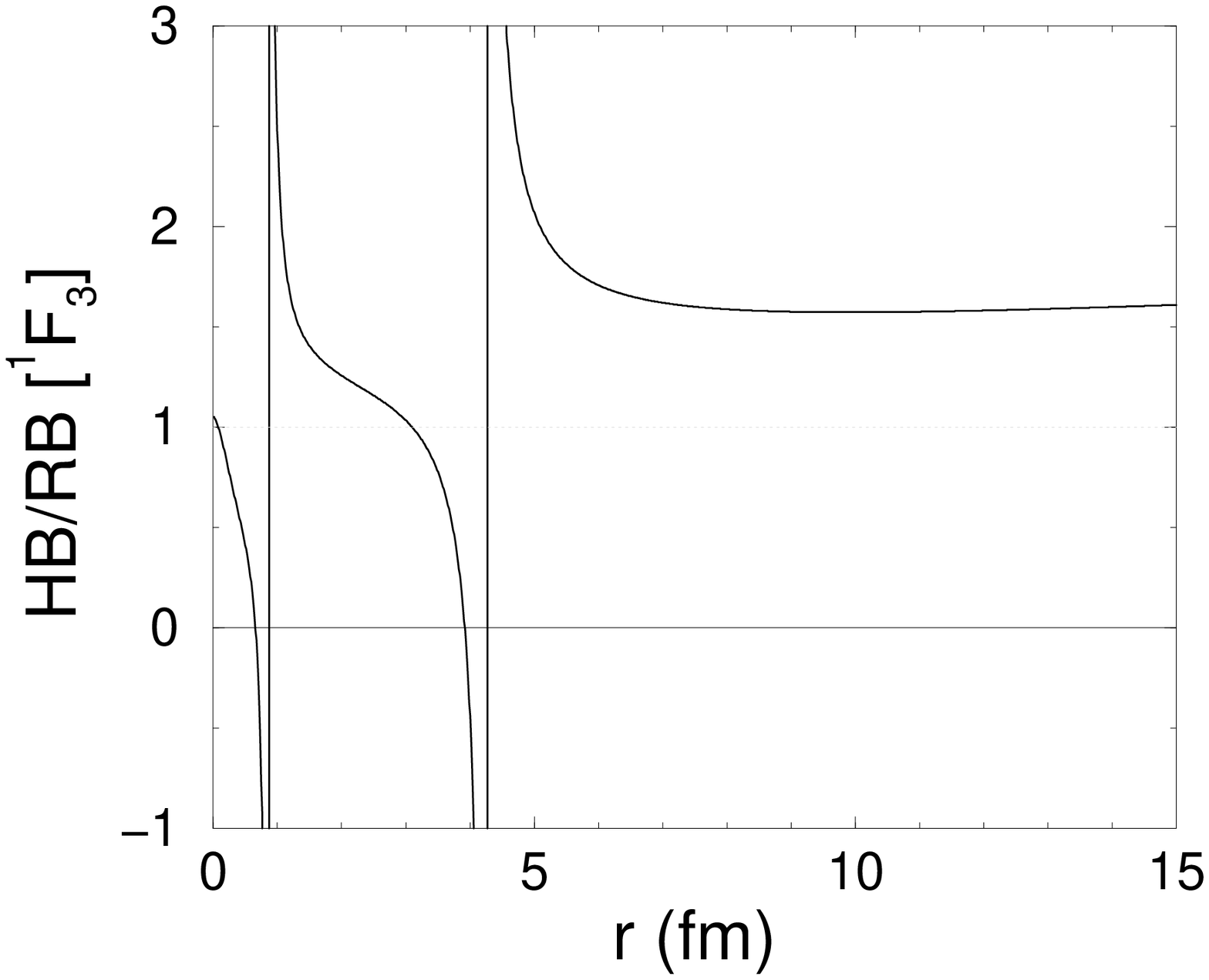, height=1.2in}& \hspace{0.2in} &
\epsfig{figure=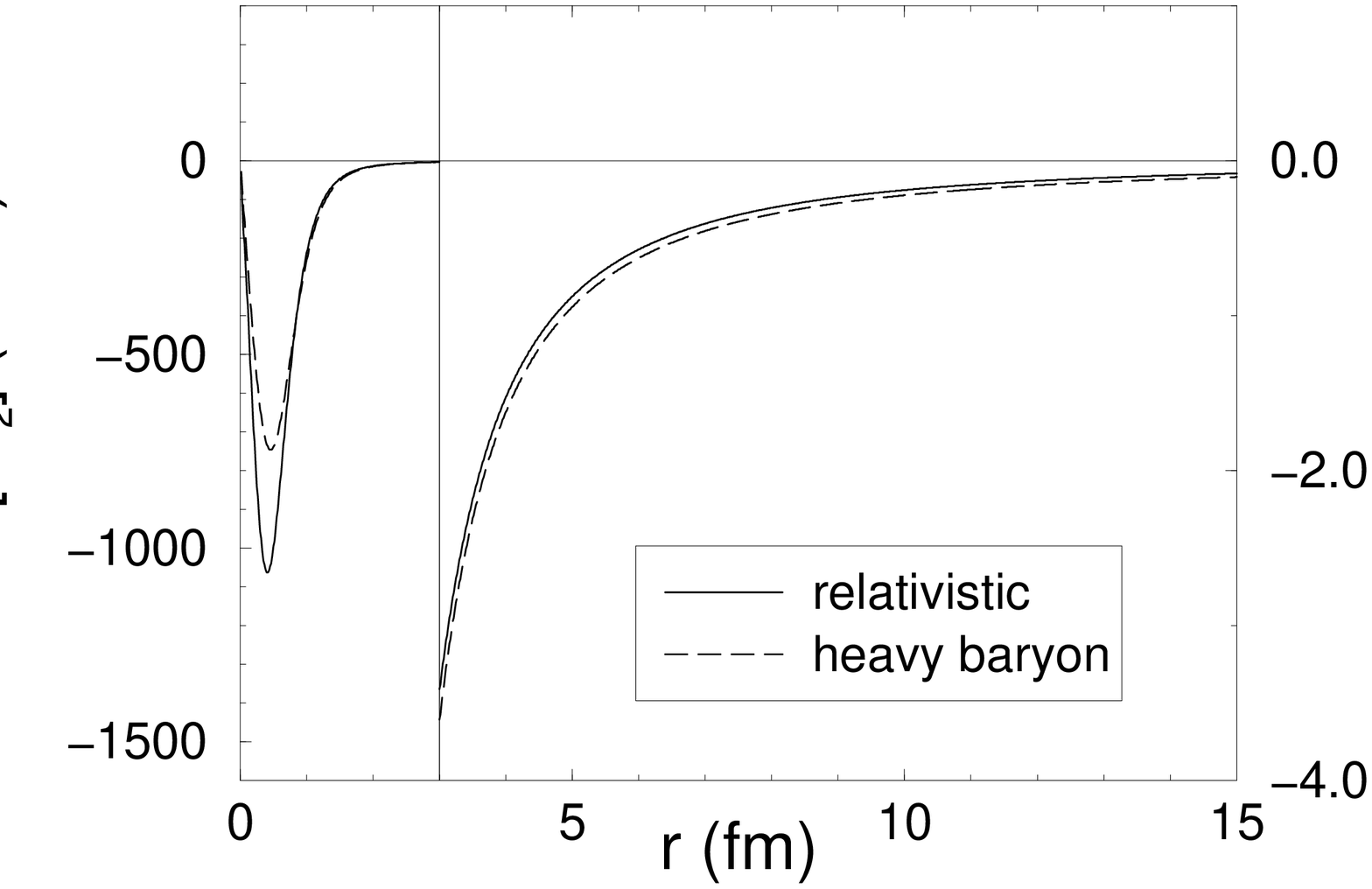, height=1.2in} \hspace{0.1in} &
\epsfig{figure=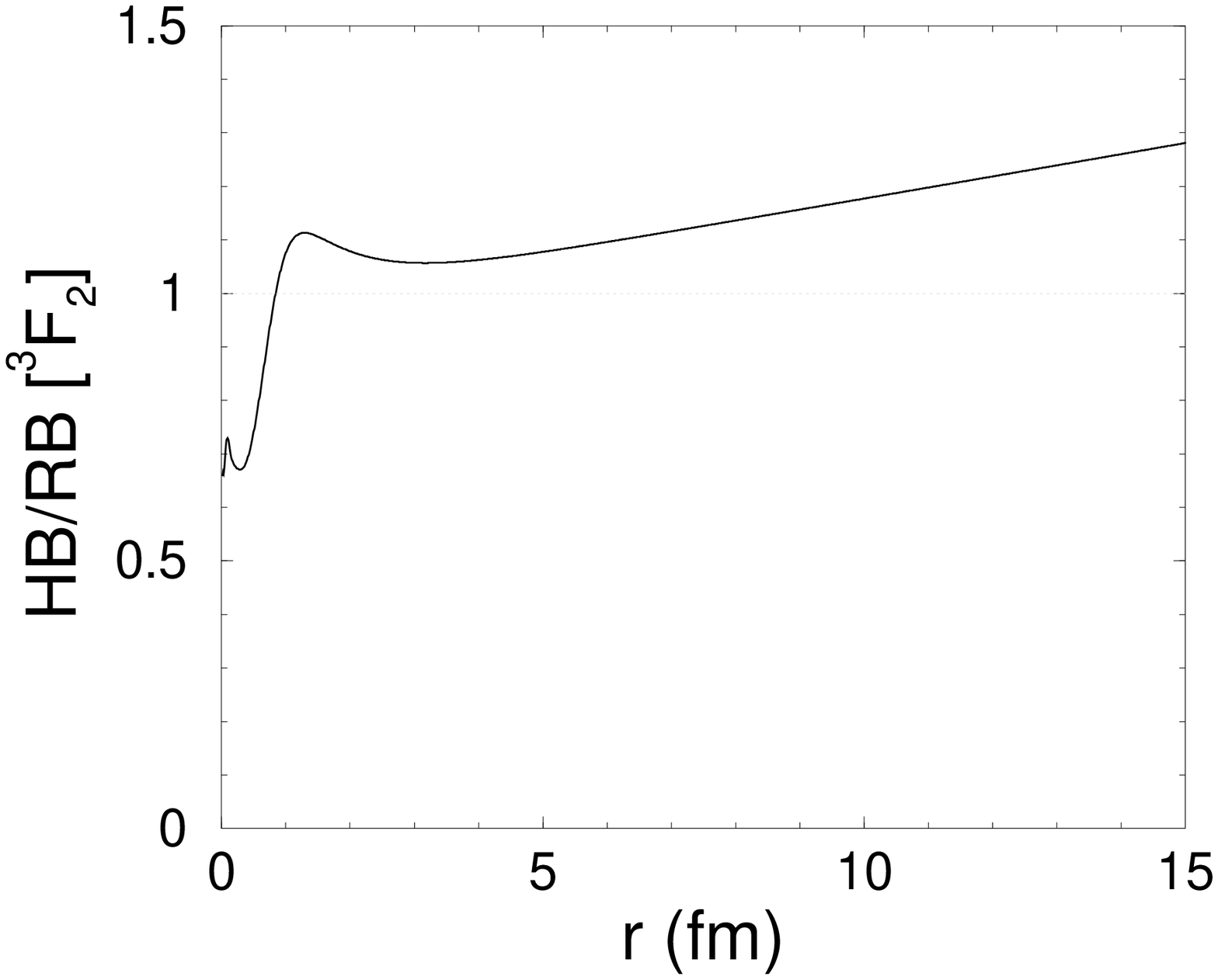, height=1.2in}\\[2mm]
\epsfig{figure=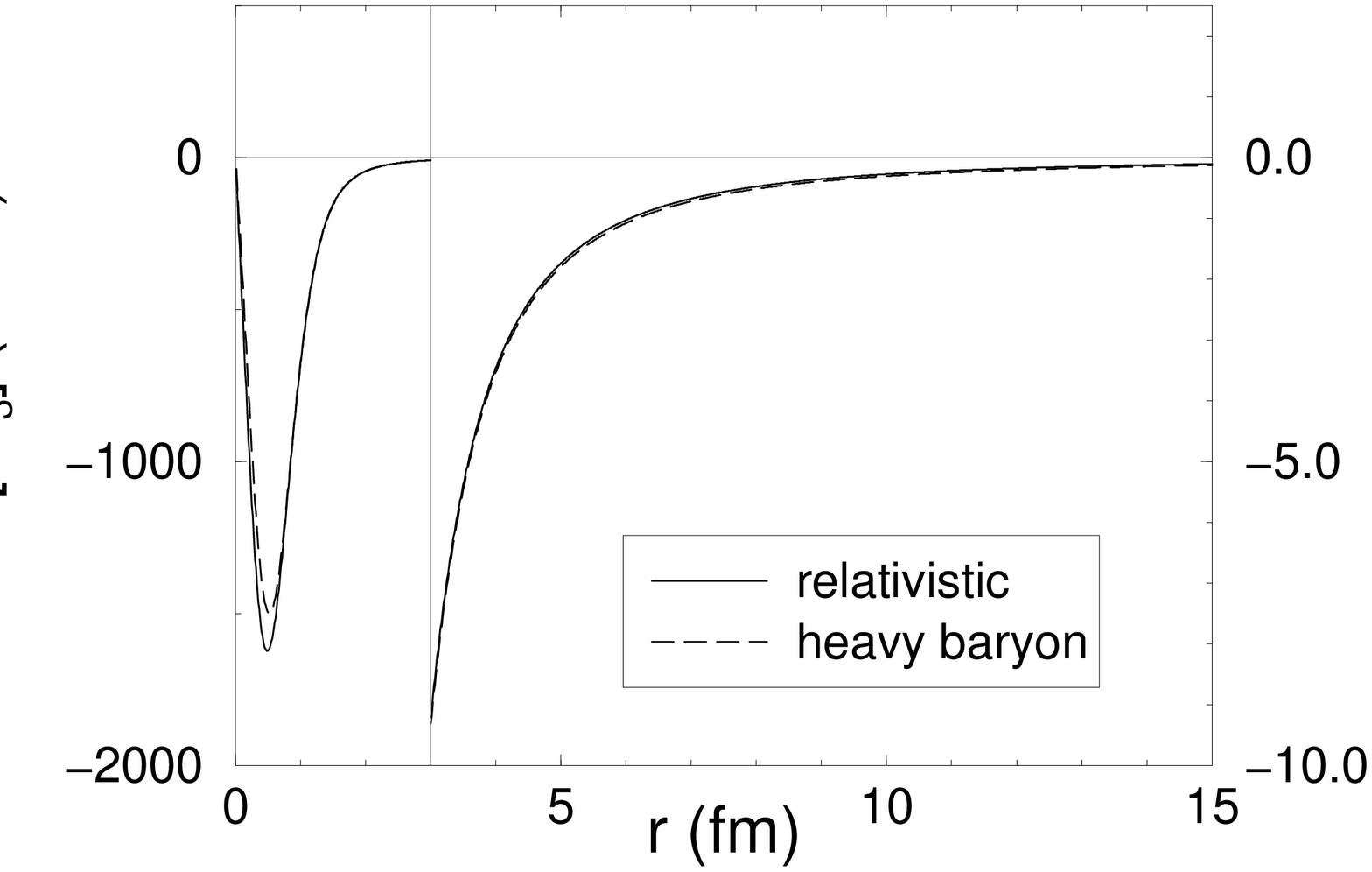, height=1.2in}&
\epsfig{figure=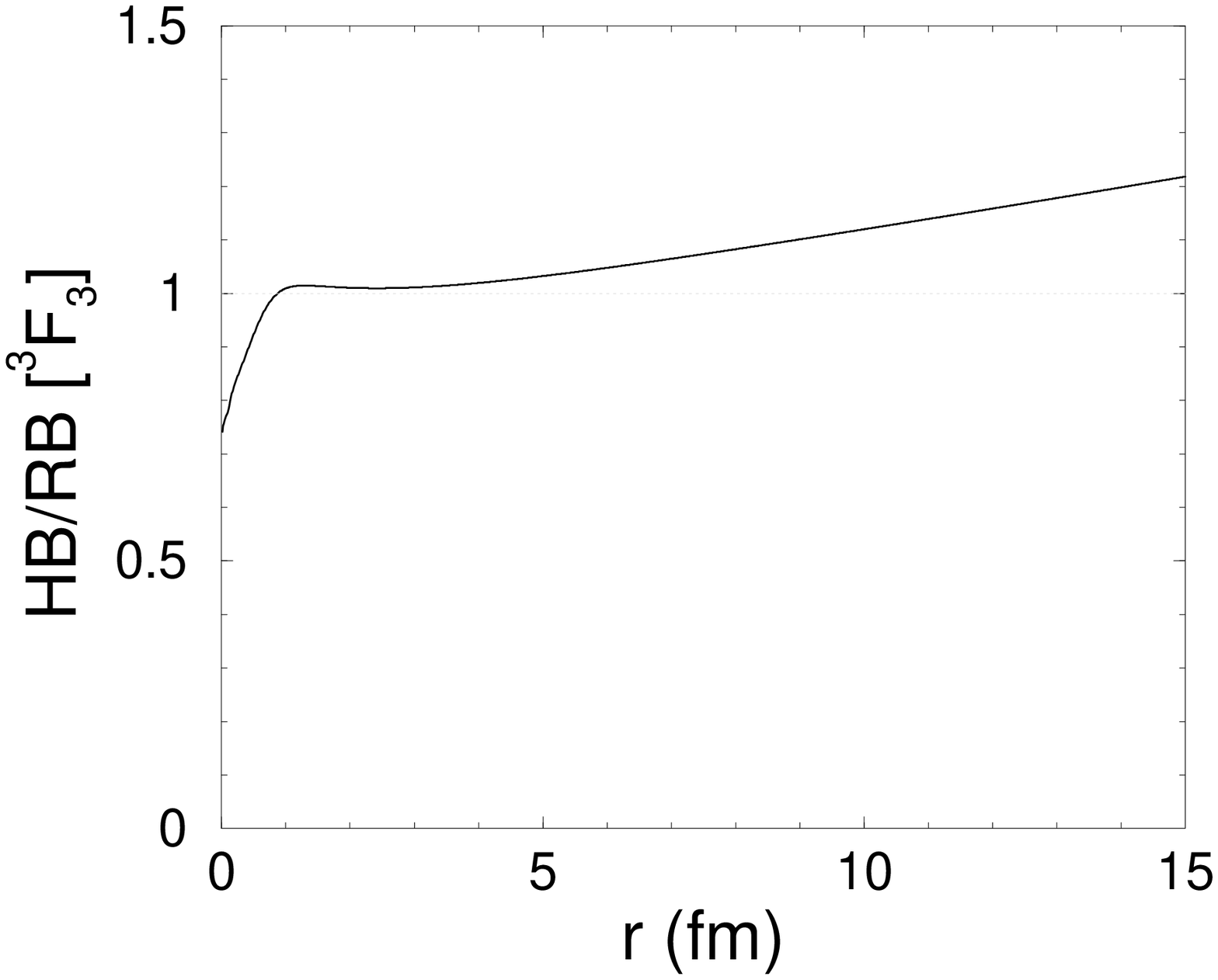, height=1.2in}&&
\epsfig{figure=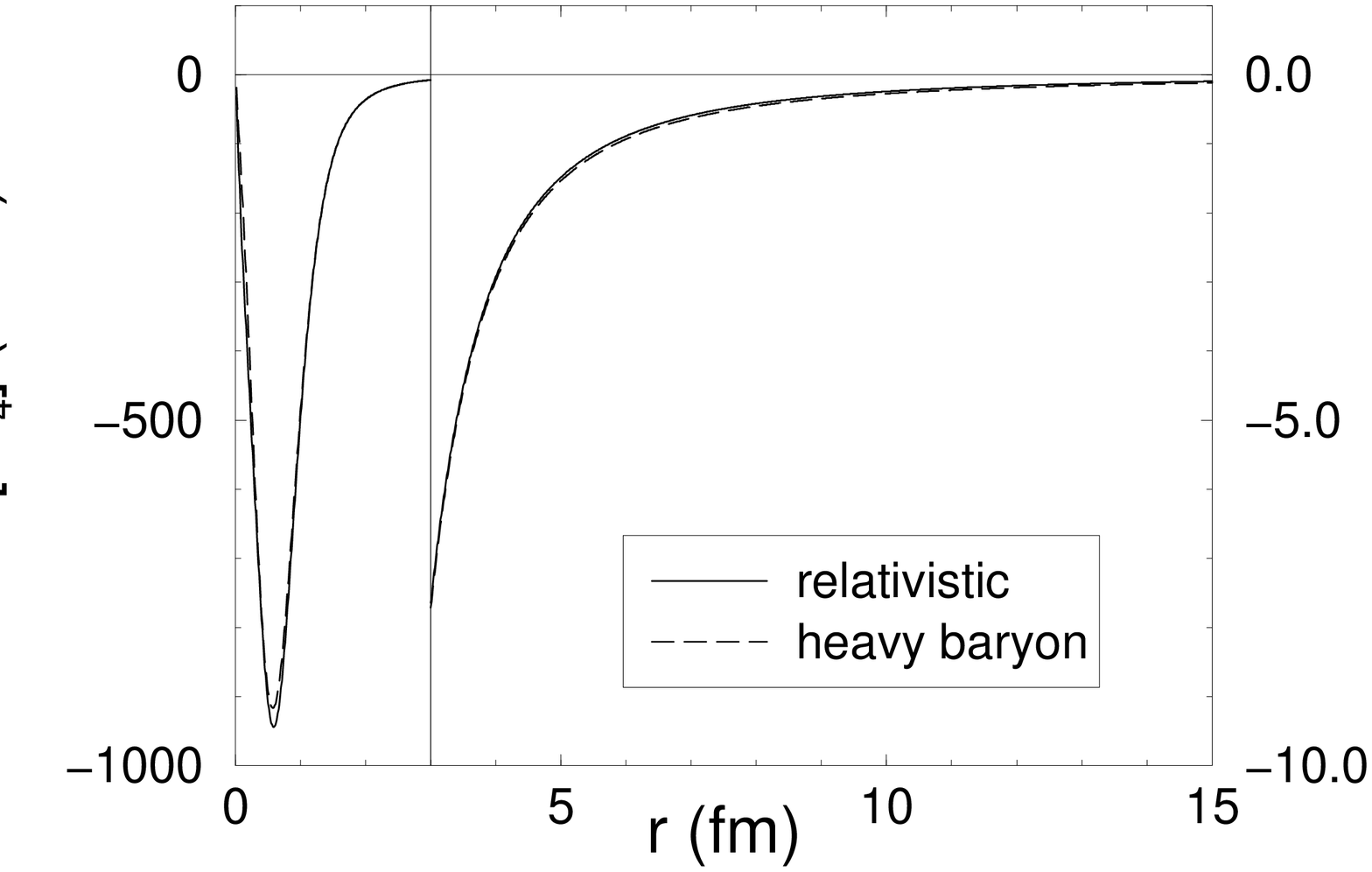, height=1.2in}&
\epsfig{figure=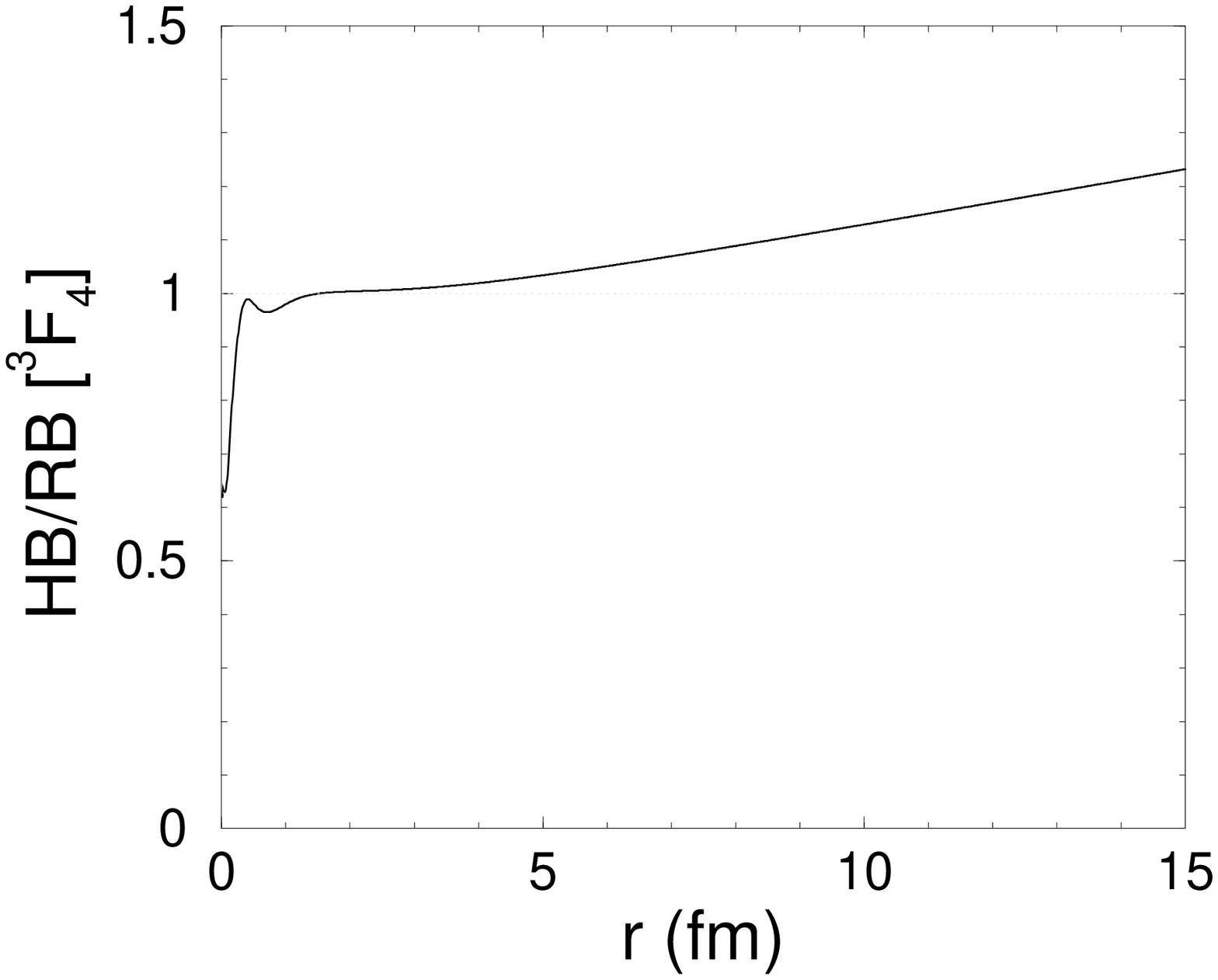, height=1.2in}\\[2mm]
\epsfig{figure=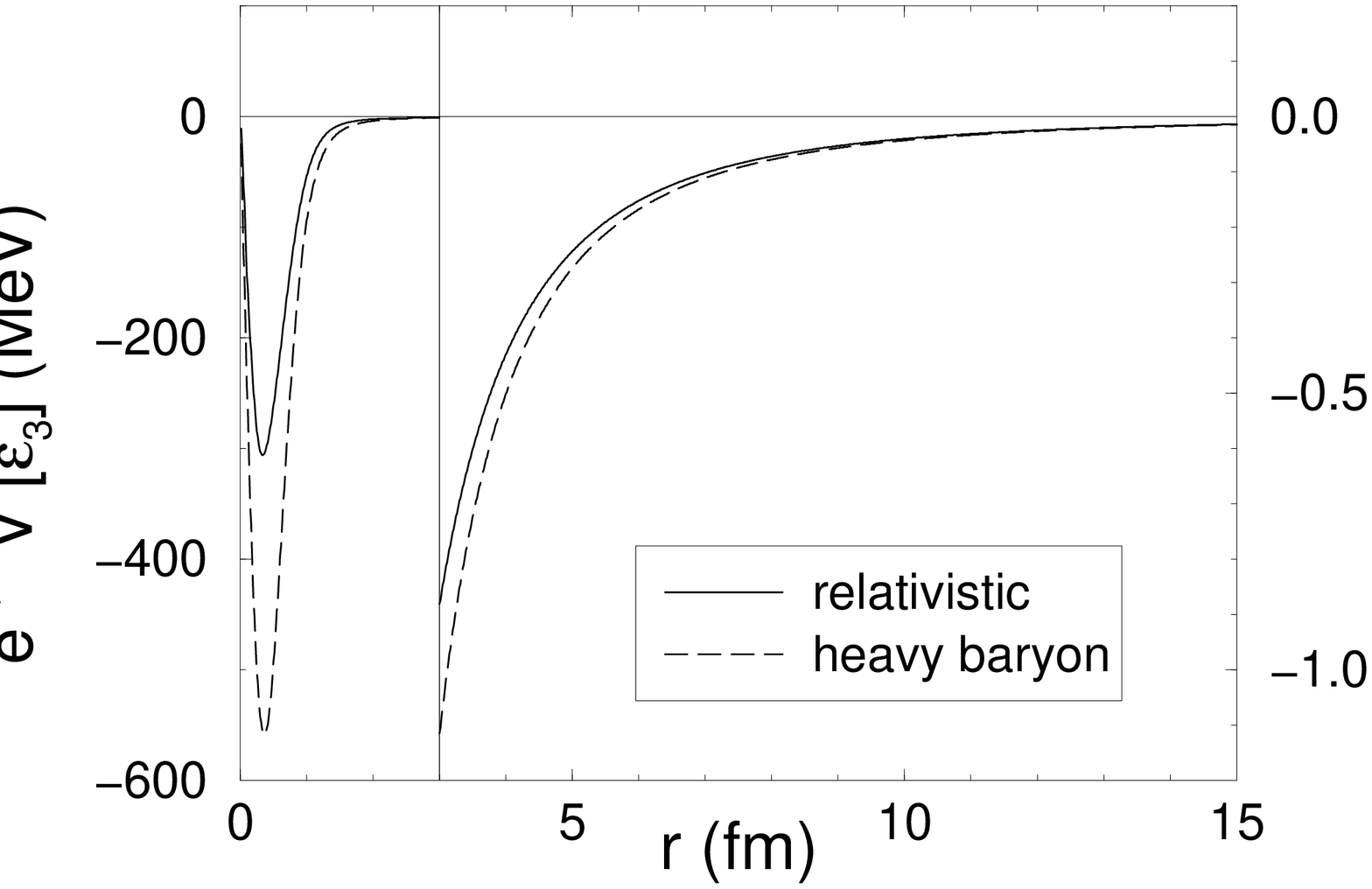, height=1.2in}&
\epsfig{figure=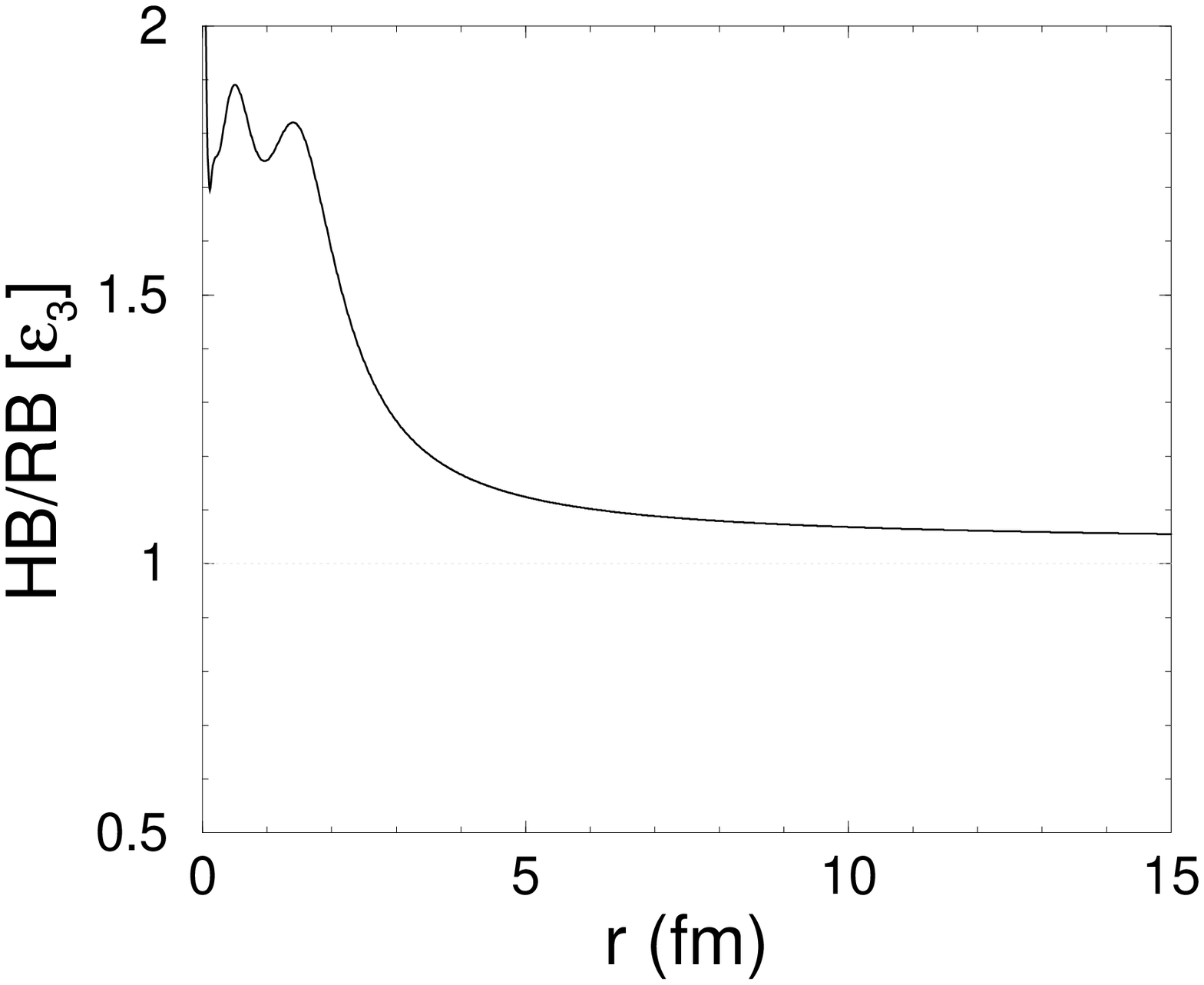, height=1.2in}&&&
\end{tabular}
\end{center}
\caption{Values of the RB and HB potentials [multiplied by 
$\exp(2\mu r)$], as well as their ratios, for $F$ waves.}
\label{figFwaves}
\end{figure}

\begin{figure}[!tbp]
\begin{center}
\begin{tabular}{ccccc}
\epsfig{figure=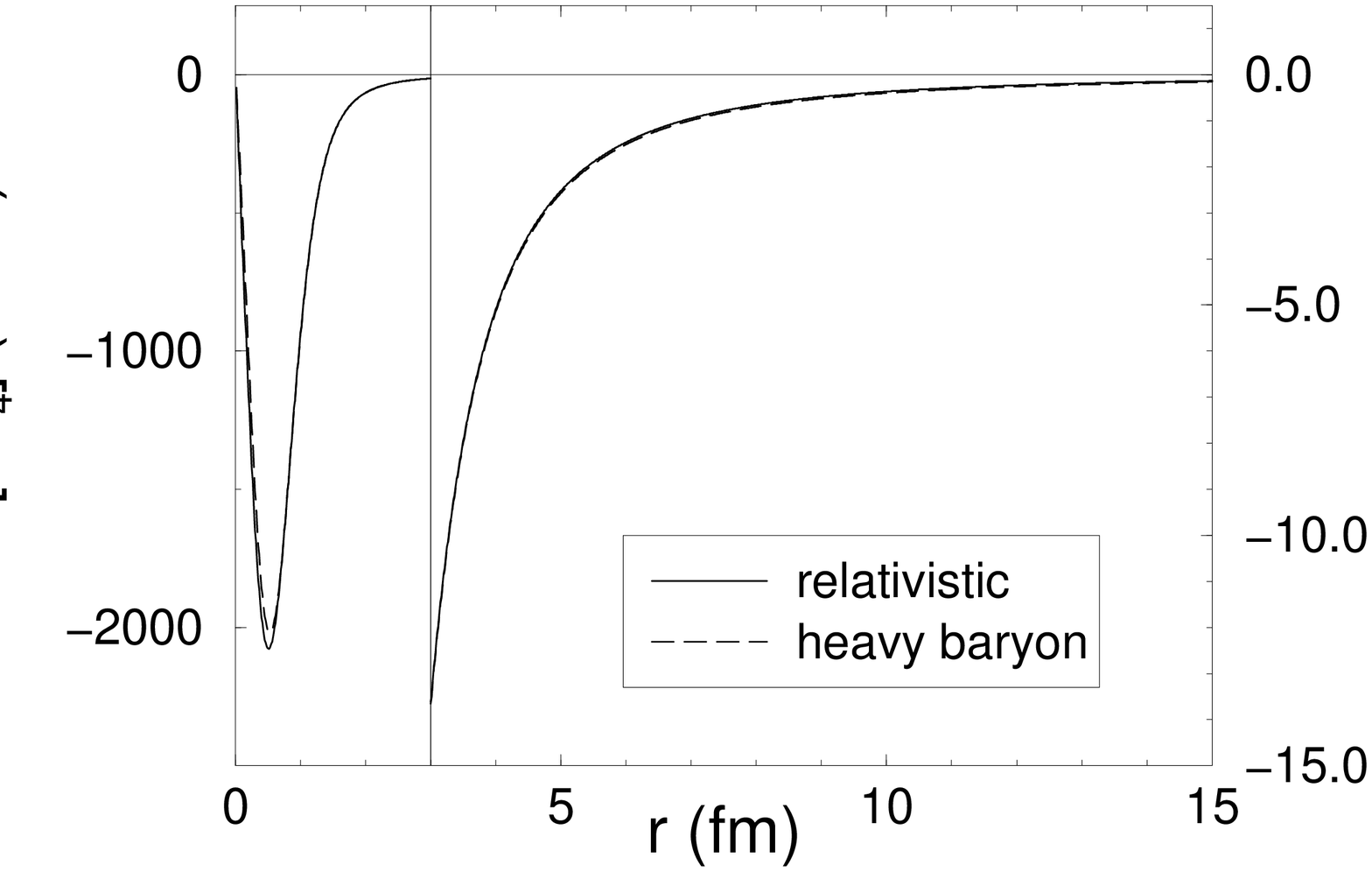, height=1.2in} \hspace{0.1in} &
\epsfig{figure=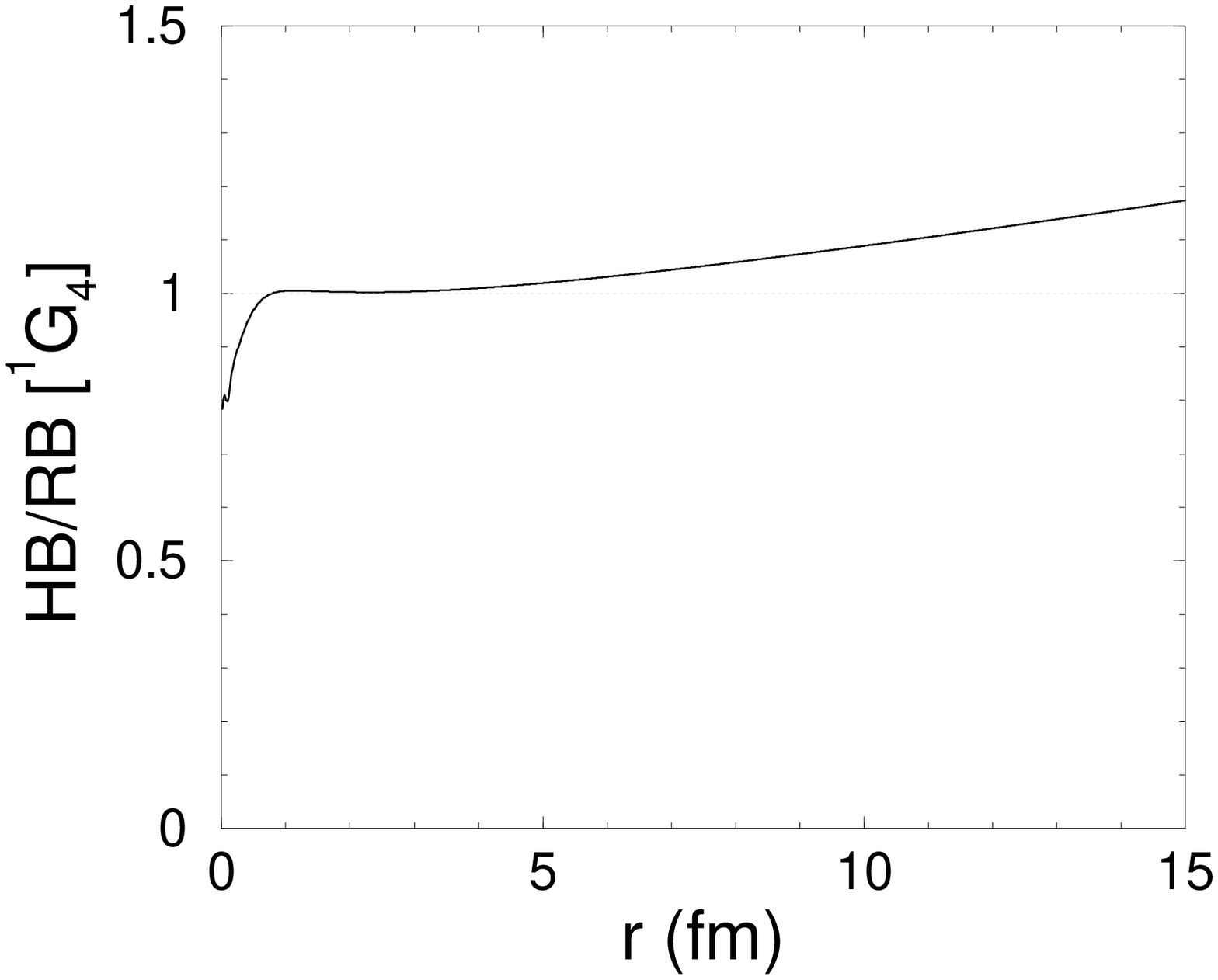, height=1.2in}& \hspace{0.2in} &
\epsfig{figure=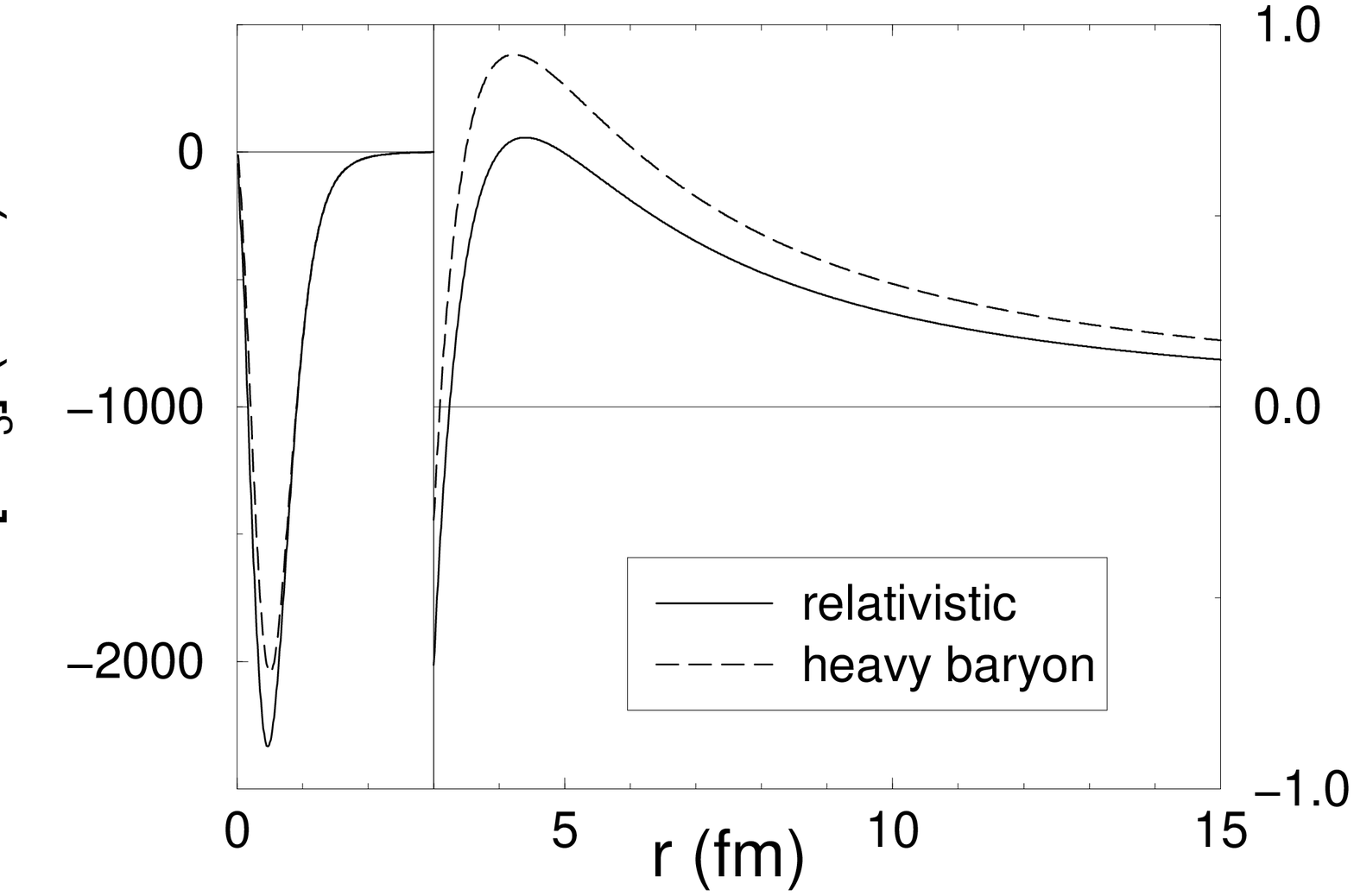, height=1.2in} \hspace{0.1in} &
\epsfig{figure=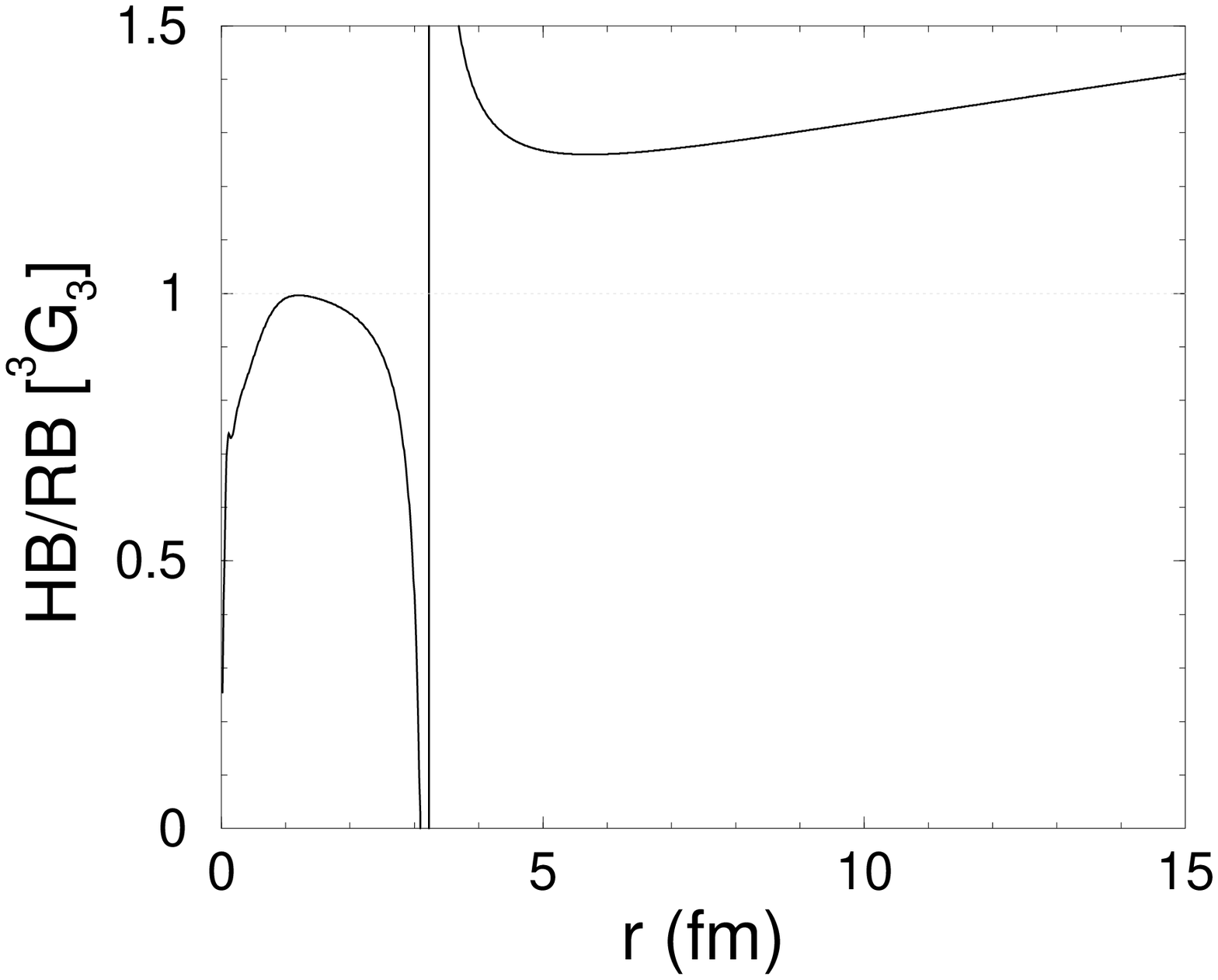, height=1.2in}\\[2mm]
\epsfig{figure=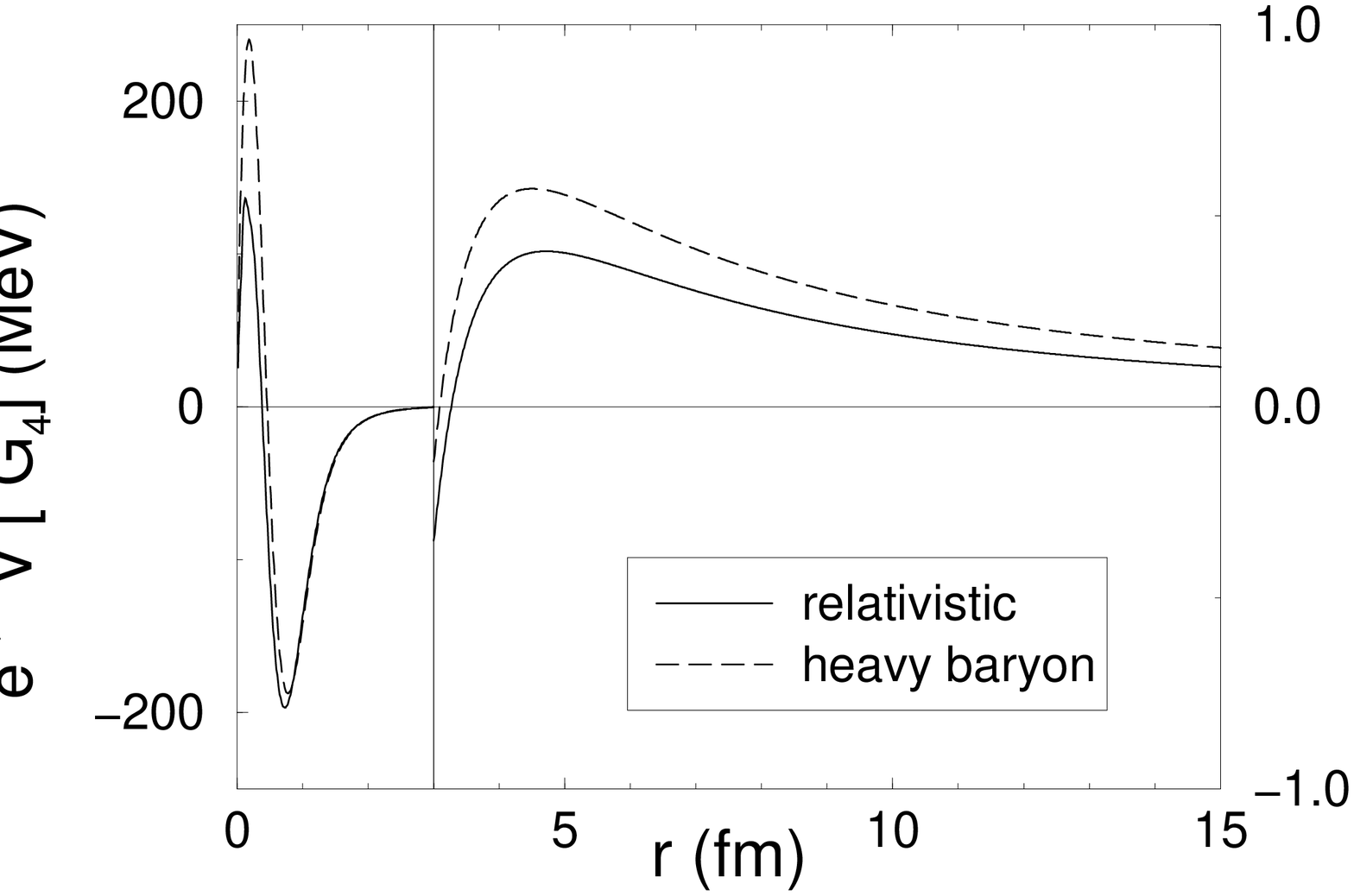, height=1.2in}&
\epsfig{figure=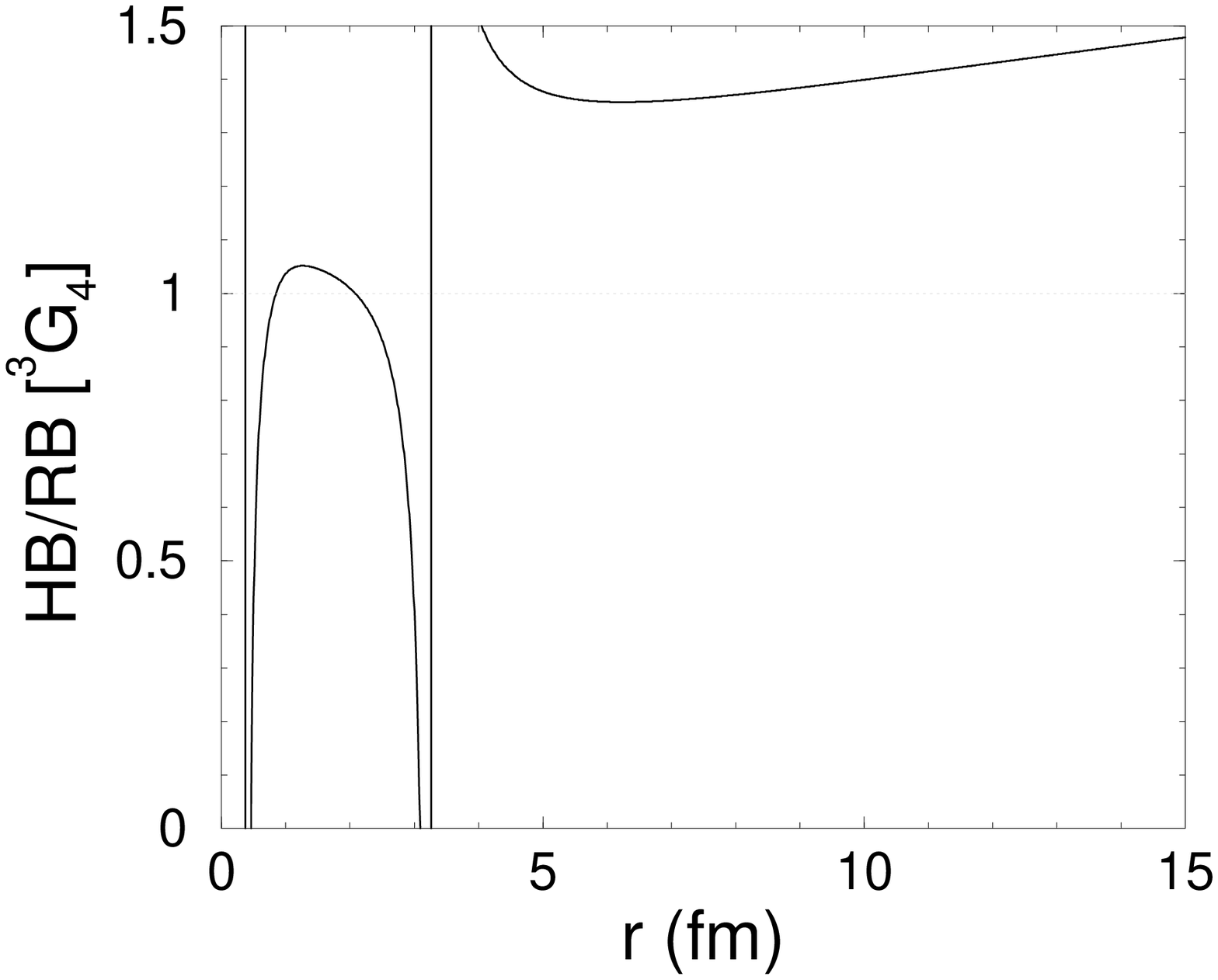, height=1.2in}&&
\epsfig{figure=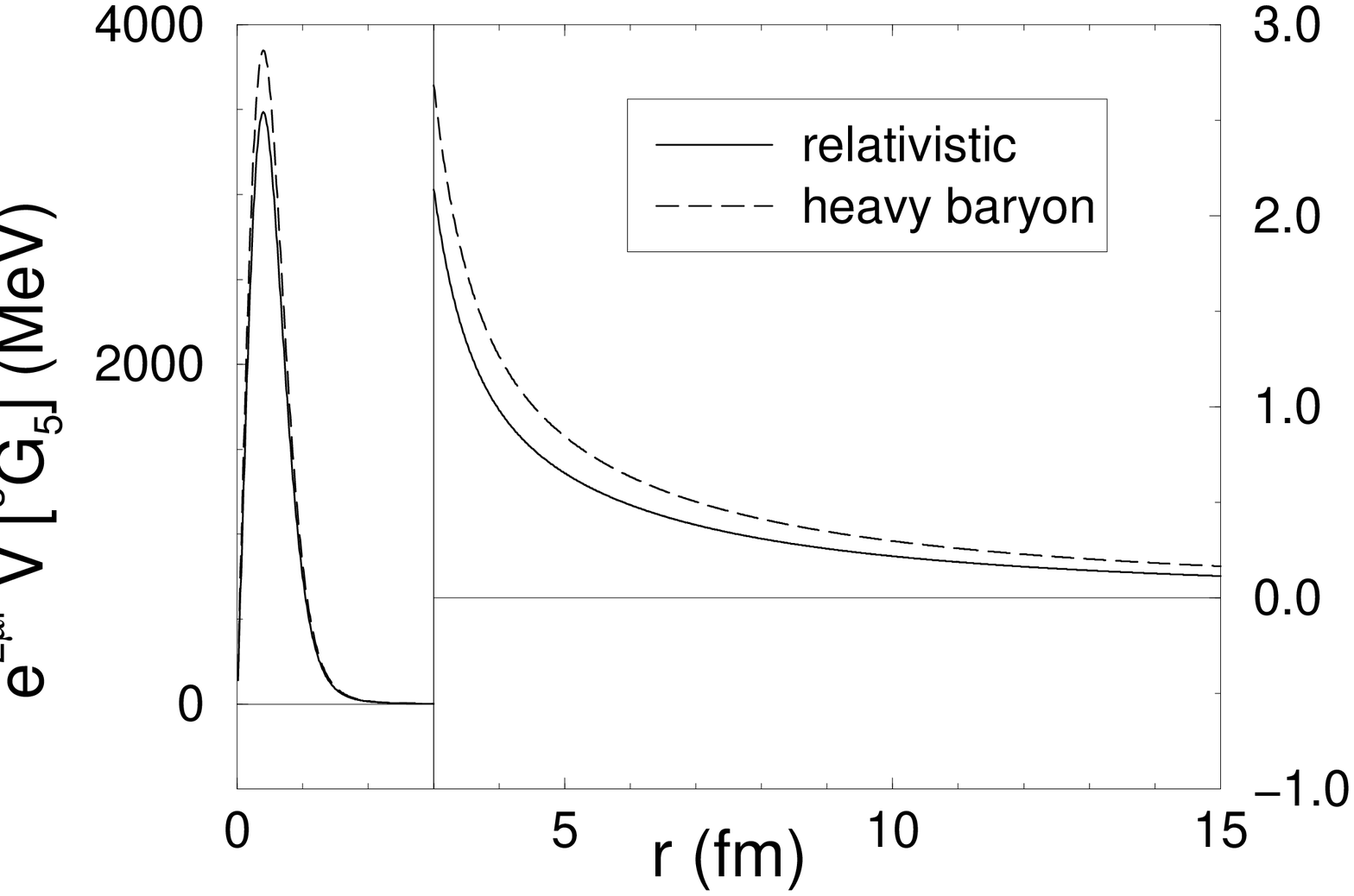, height=1.2in}&
\epsfig{figure=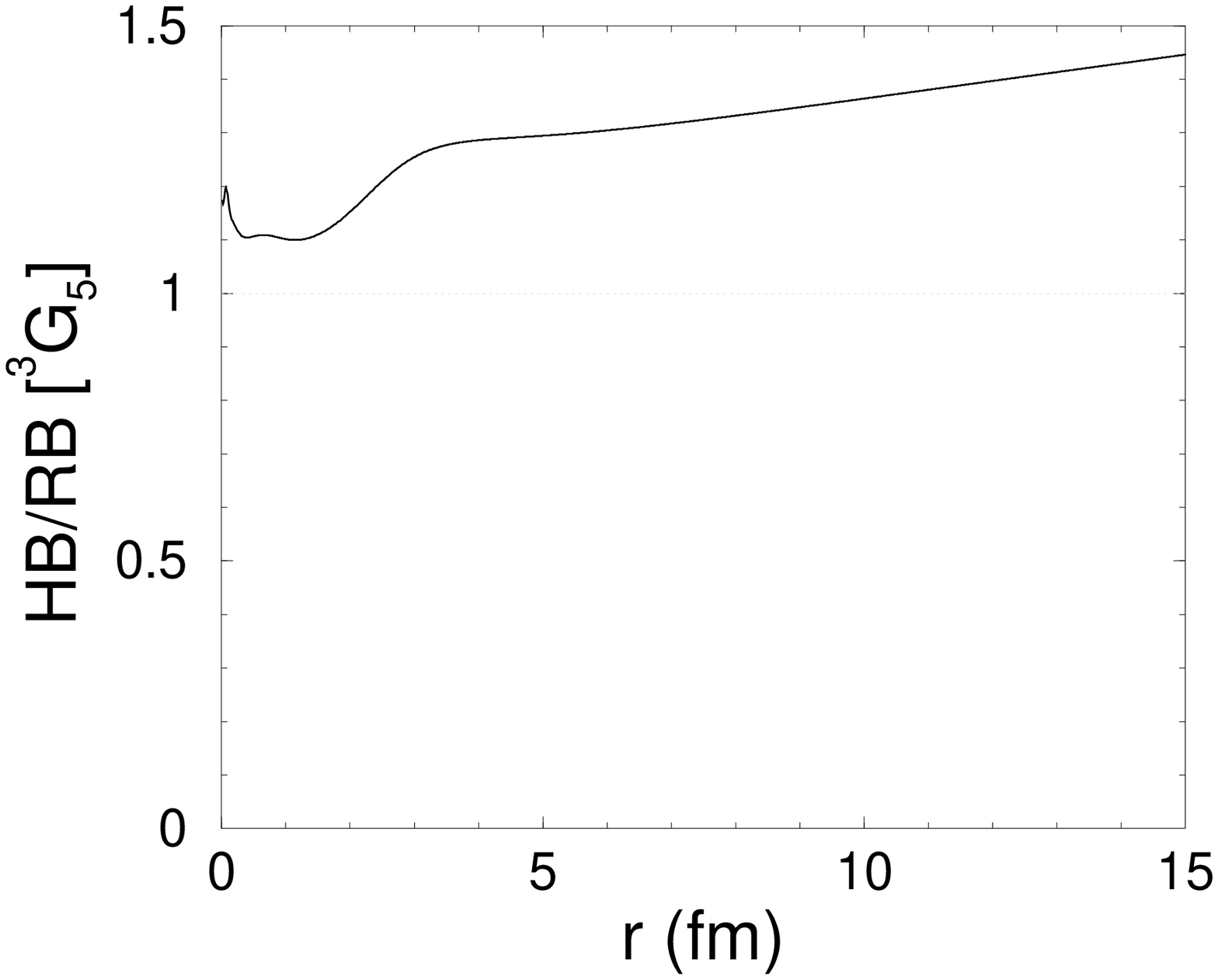, height=1.2in}\\[2mm]
\epsfig{figure=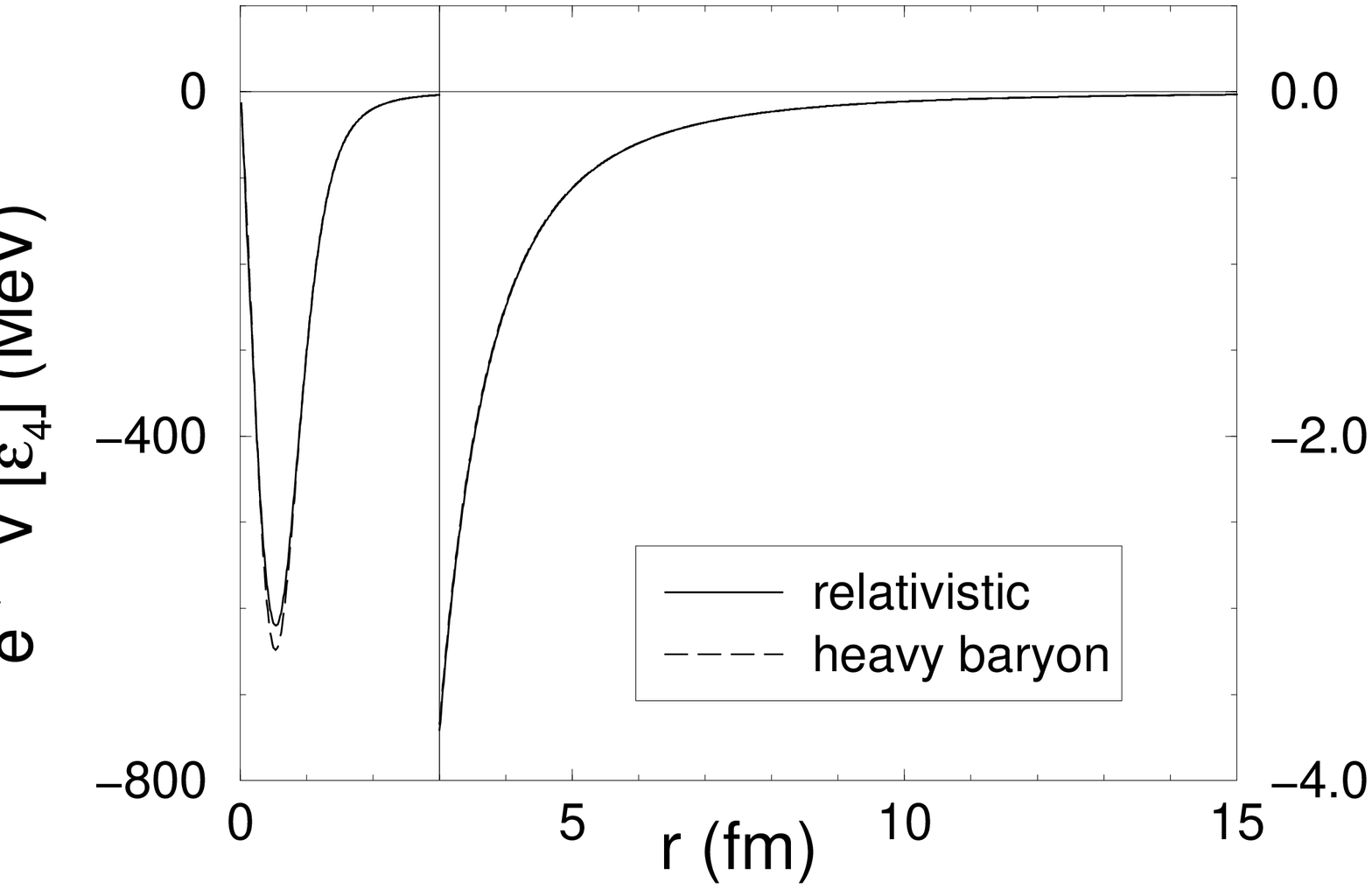, height=1.2in}&
\epsfig{figure=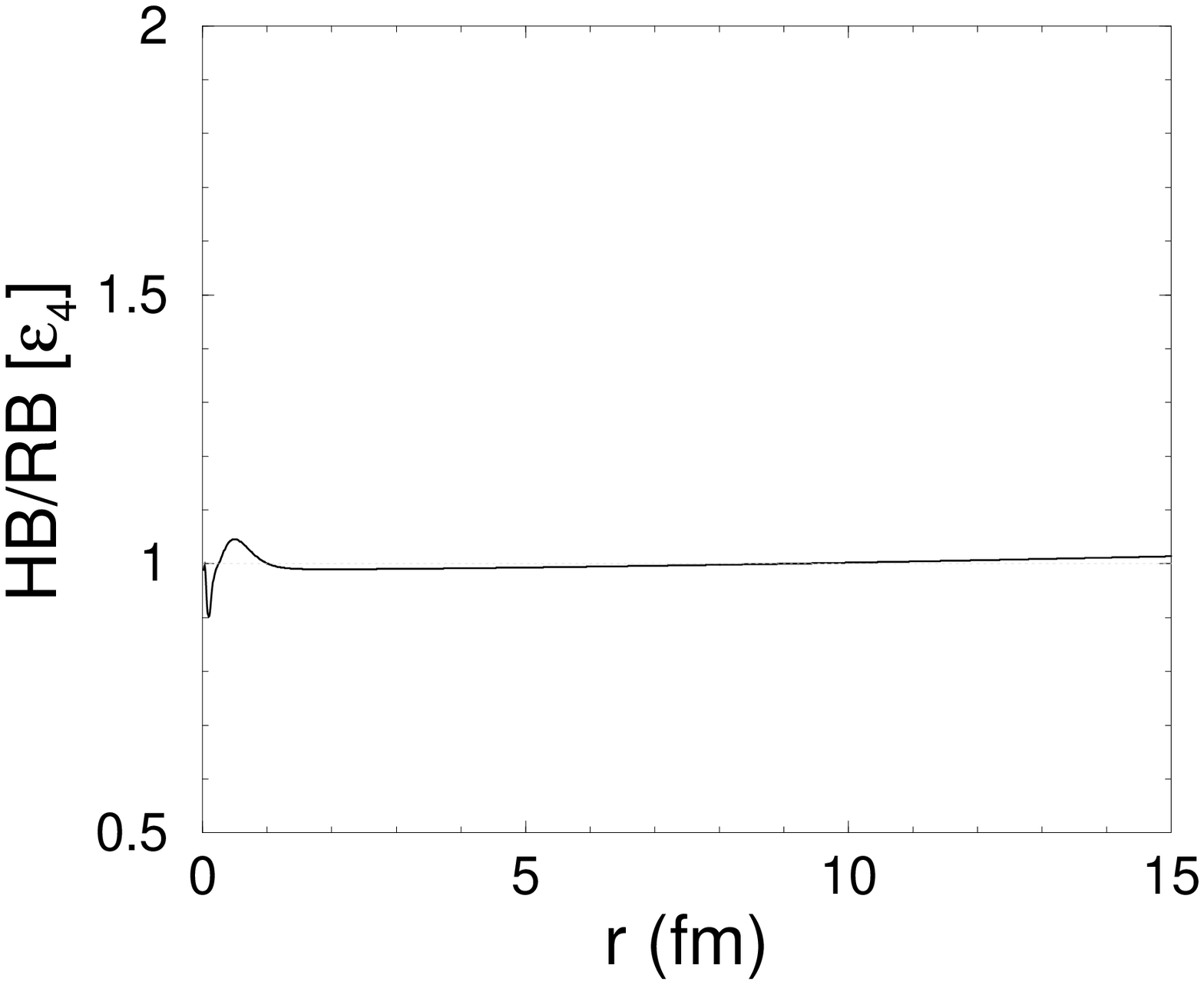, height=1.2in}&&&
\end{tabular}
\end{center}
\caption{Same as in Fig.~\ref{figFwaves}, for $G$ waves.}
\label{figGwaves}
\end{figure}

\begin{figure}[!tbp]
\begin{center}
\begin{tabular}{ccccc}
\epsfig{figure=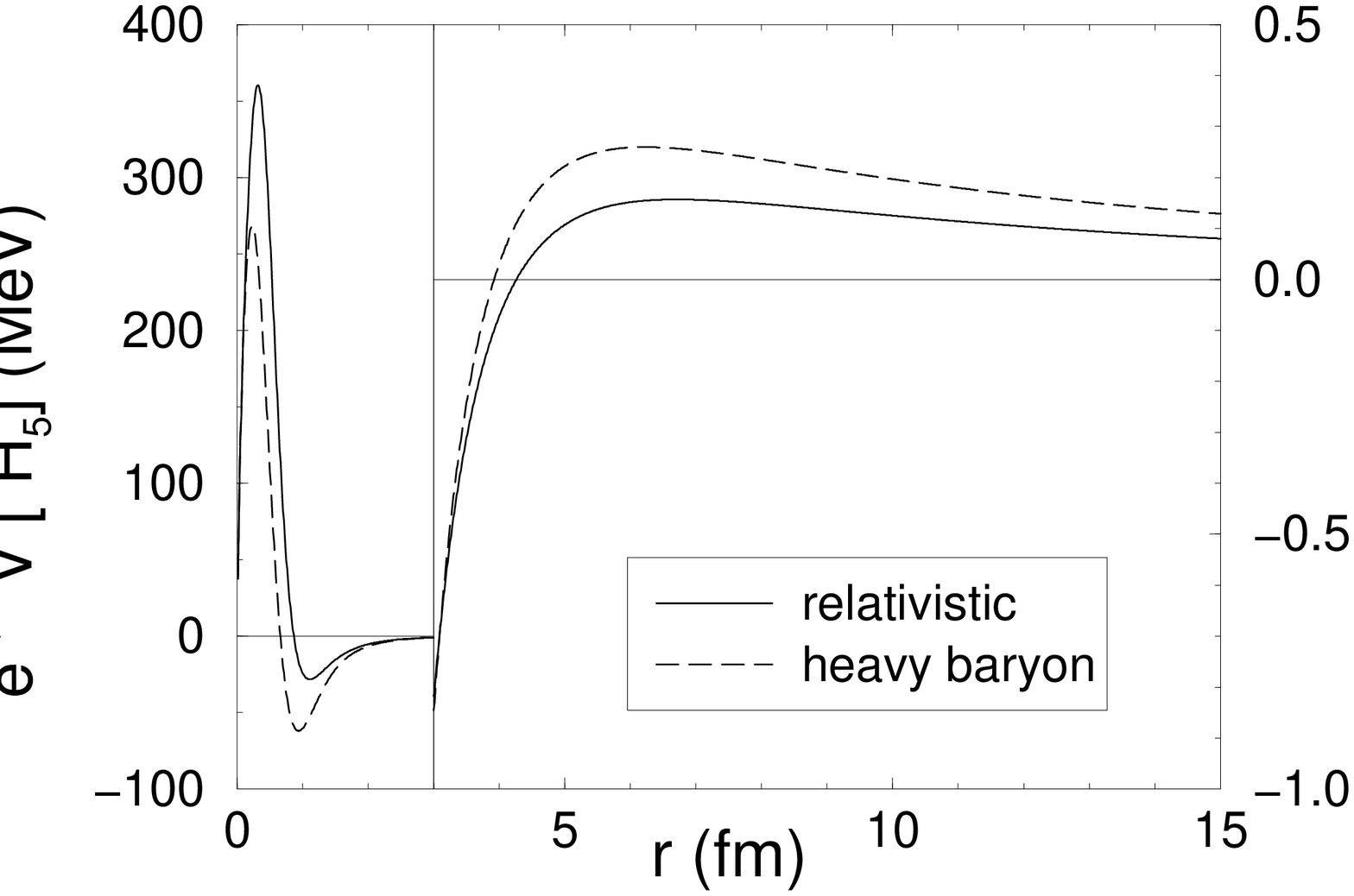, height=1.2in} \hspace{0.1in} &
\epsfig{figure=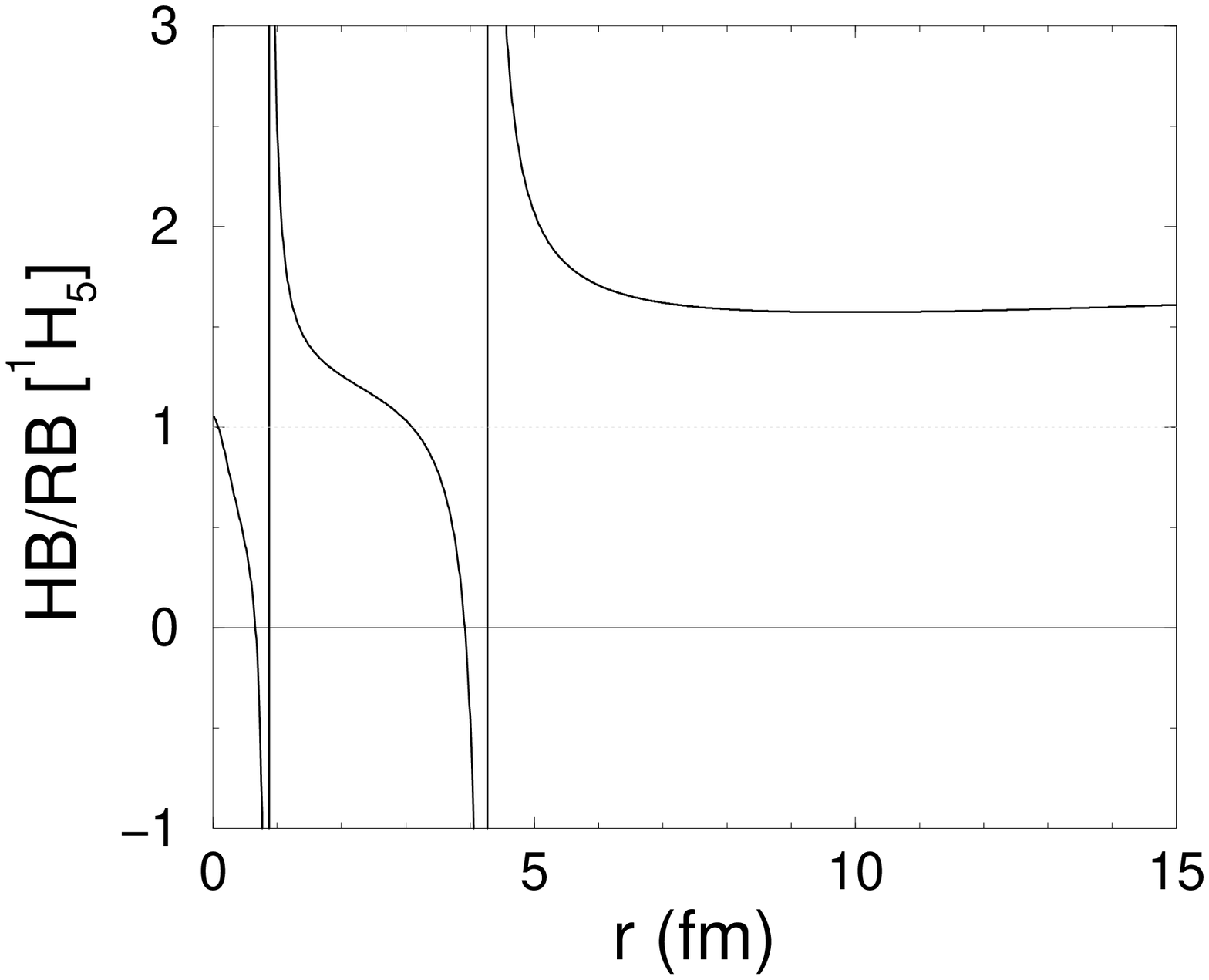, height=1.2in}& \hspace{0.2in} &
\epsfig{figure=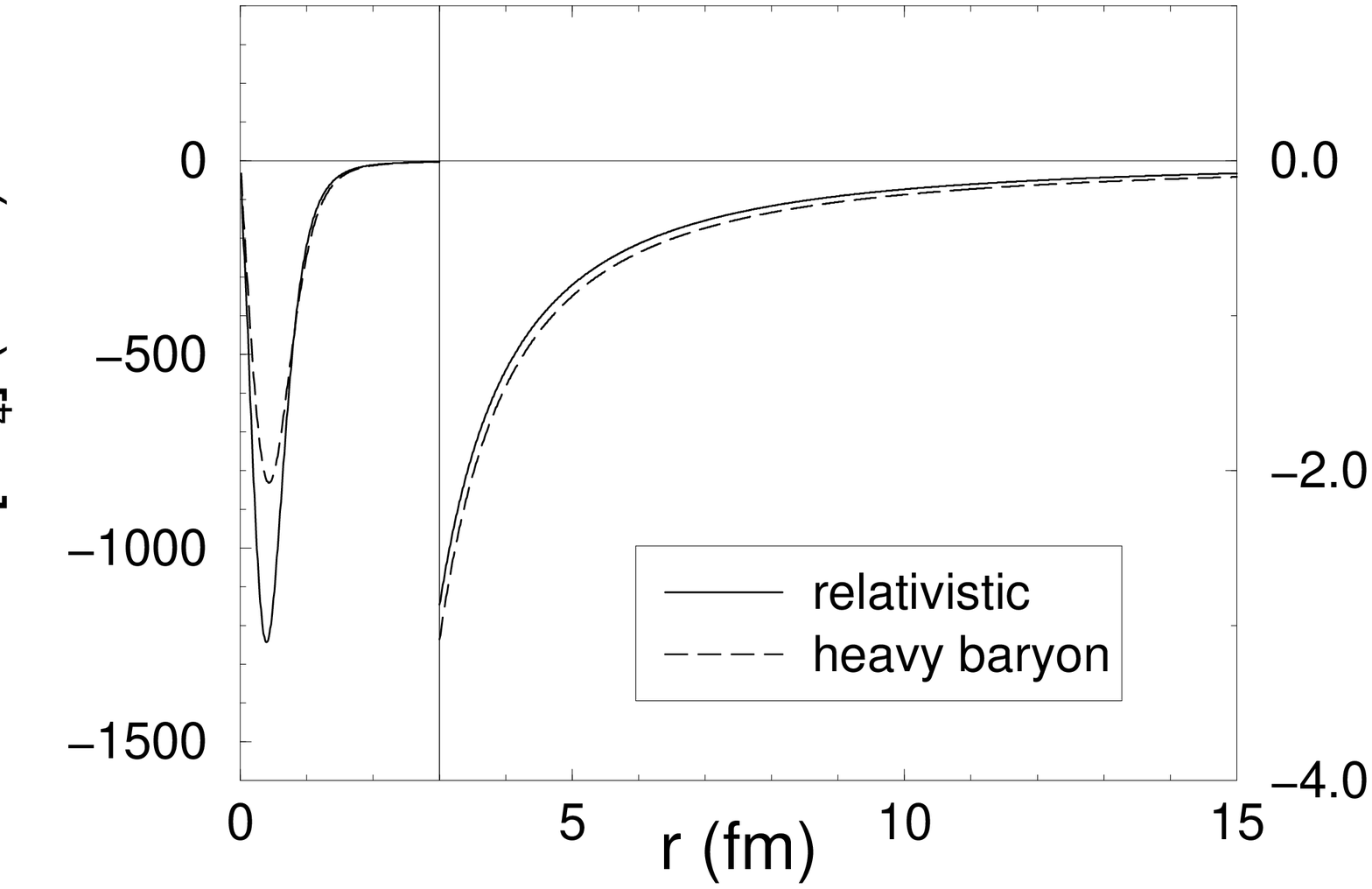, height=1.2in} \hspace{0.1in} &
\epsfig{figure=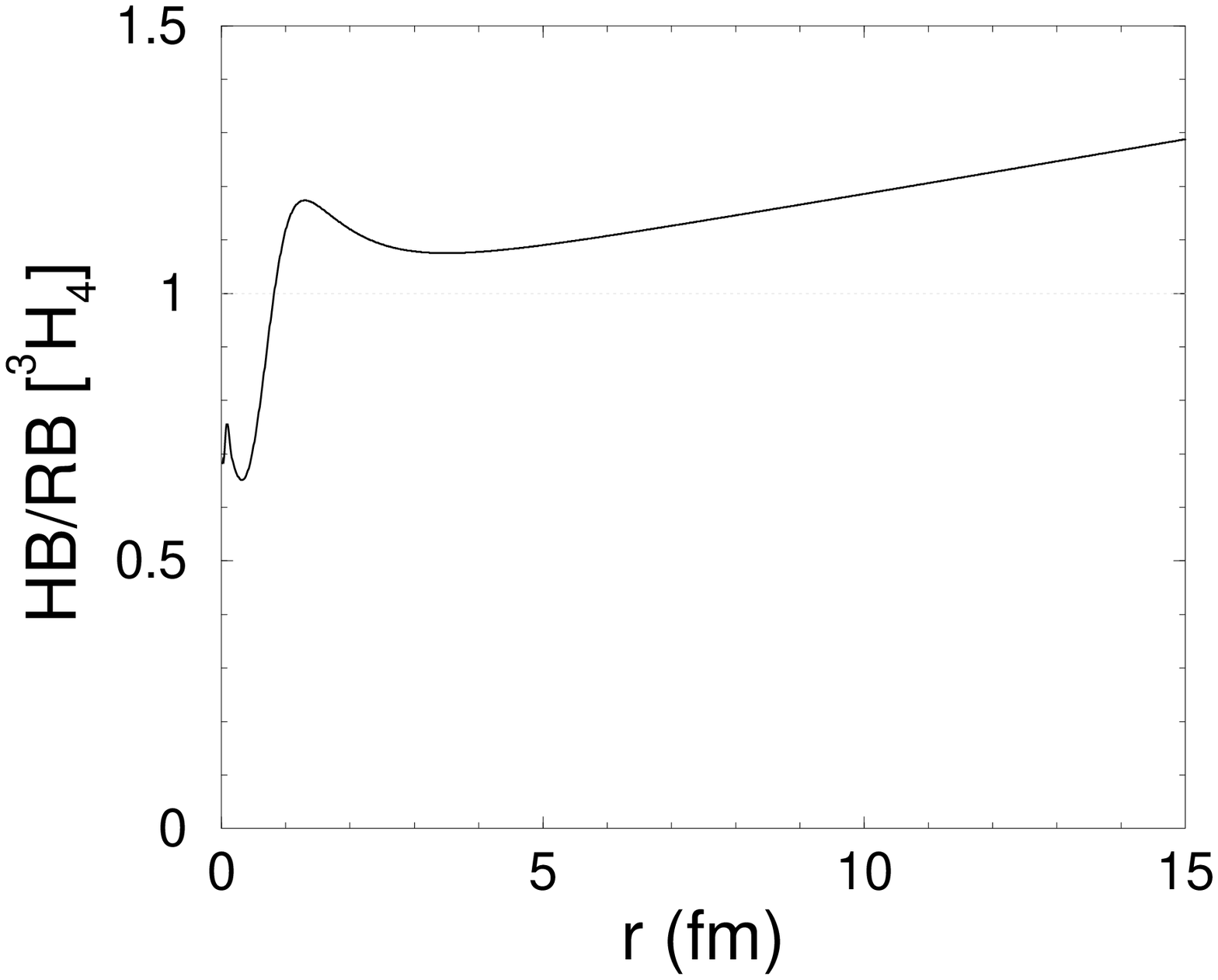, height=1.2in}\\[2mm]
\epsfig{figure=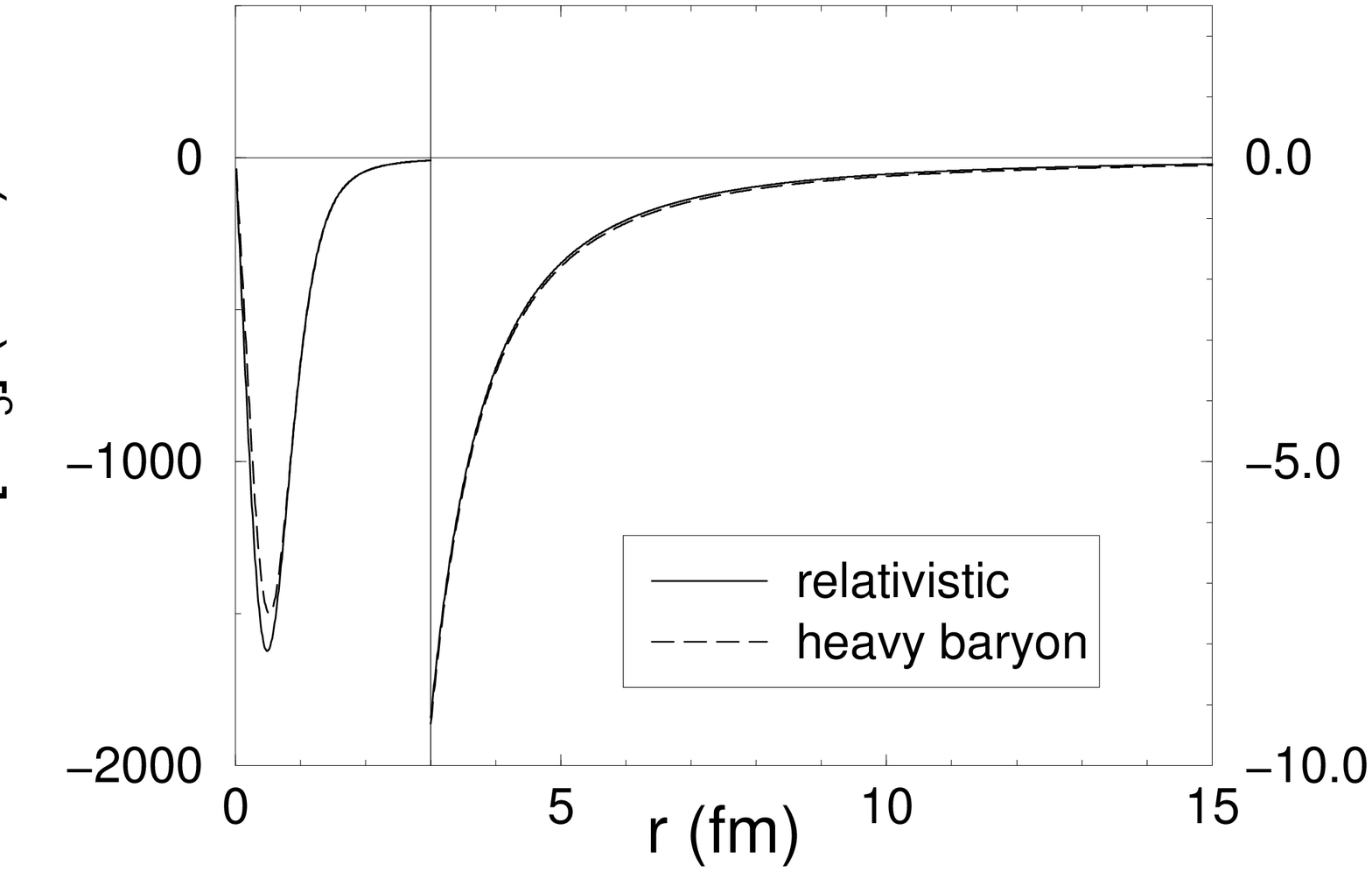, height=1.2in}&
\epsfig{figure=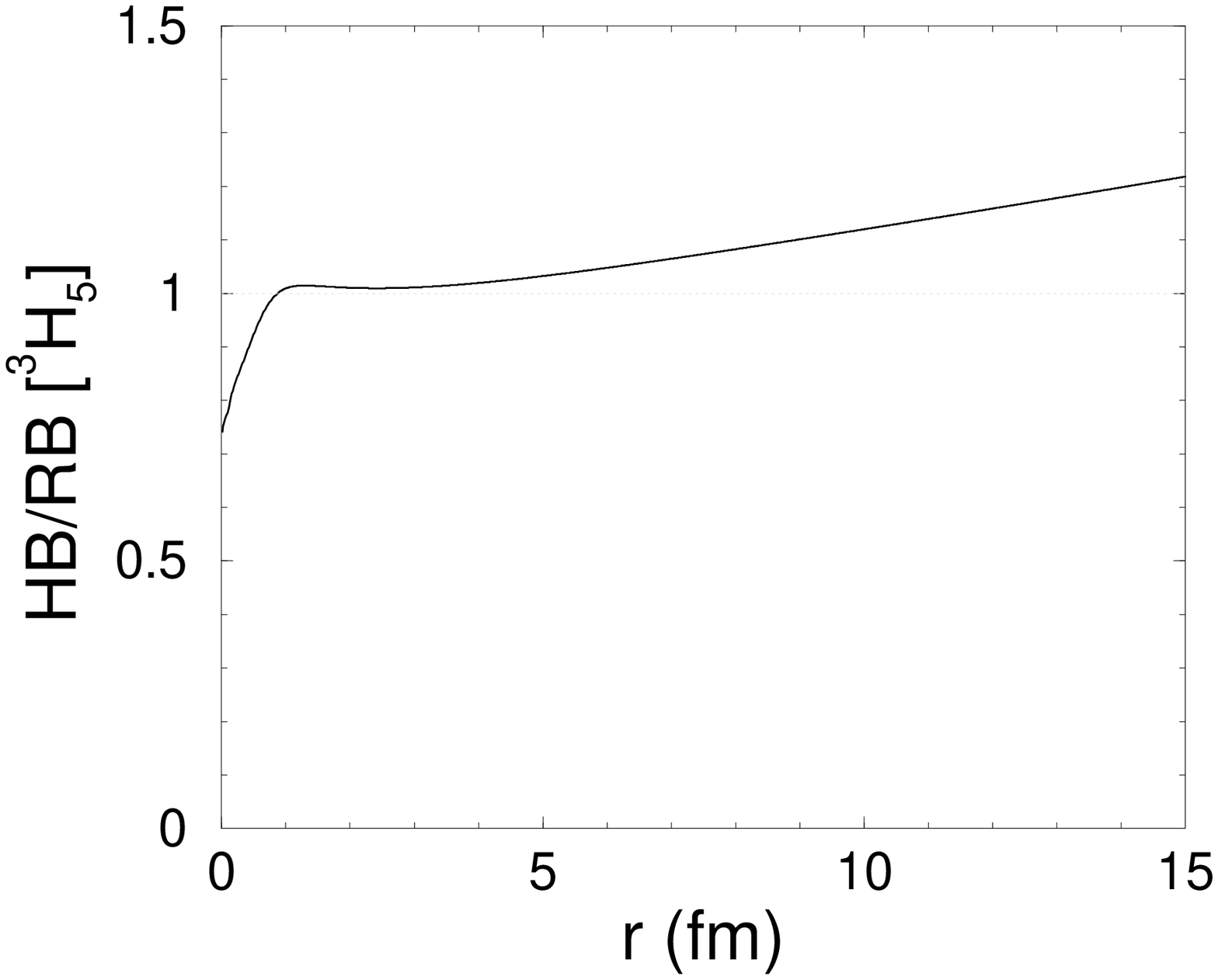, height=1.2in}&&
\epsfig{figure=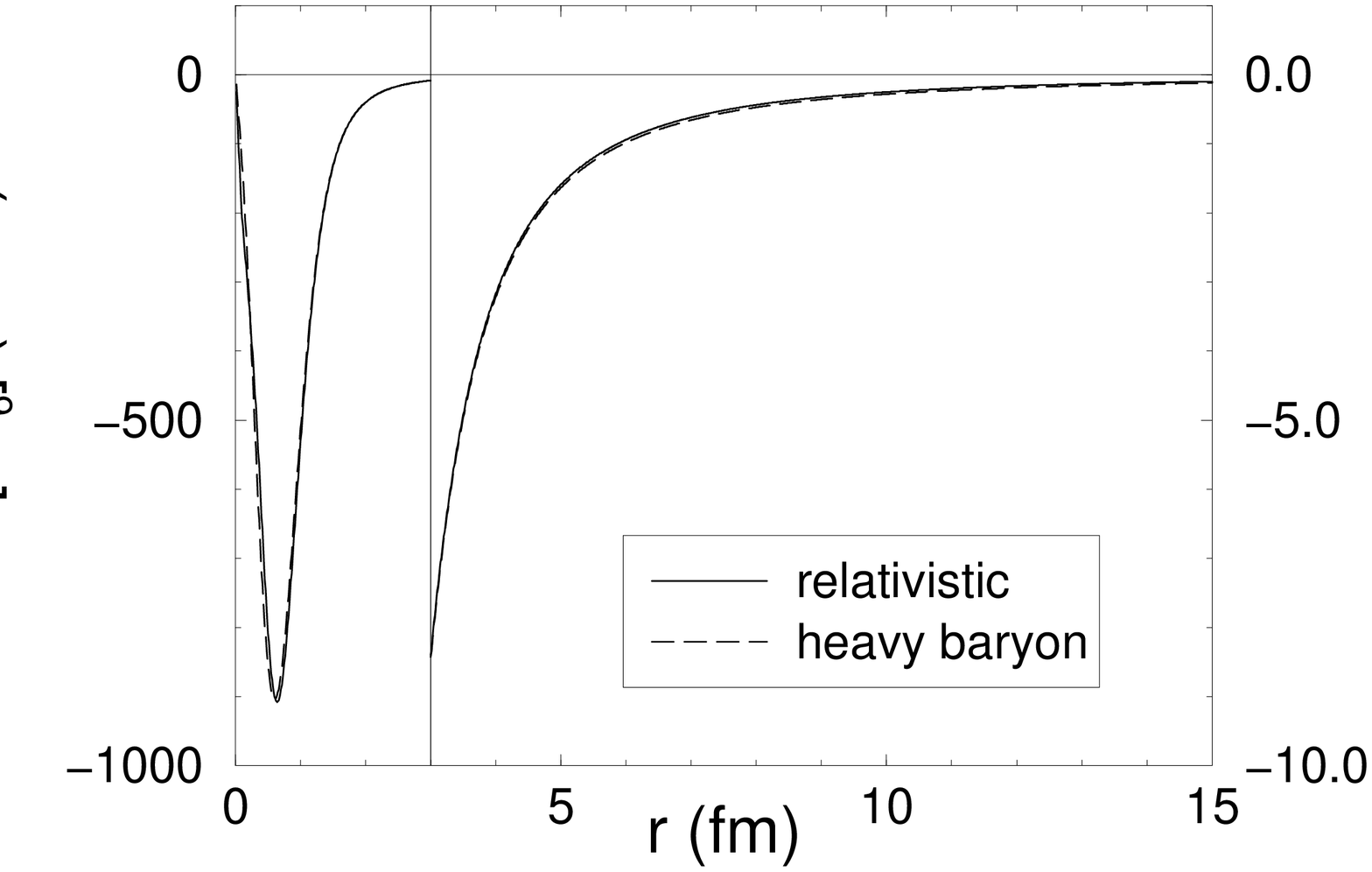, height=1.2in}&
\epsfig{figure=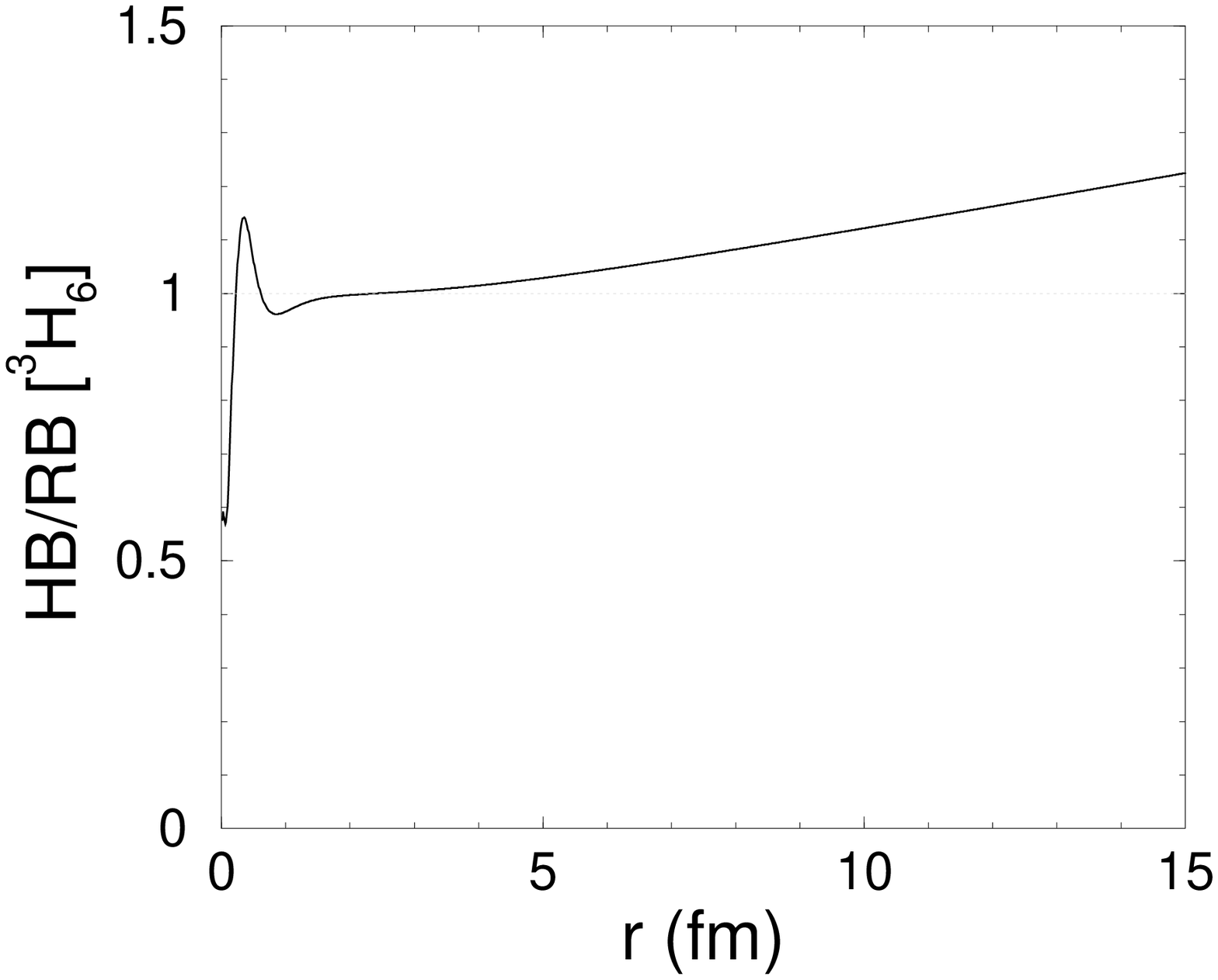, height=1.2in}\\[2mm]
\epsfig{figure=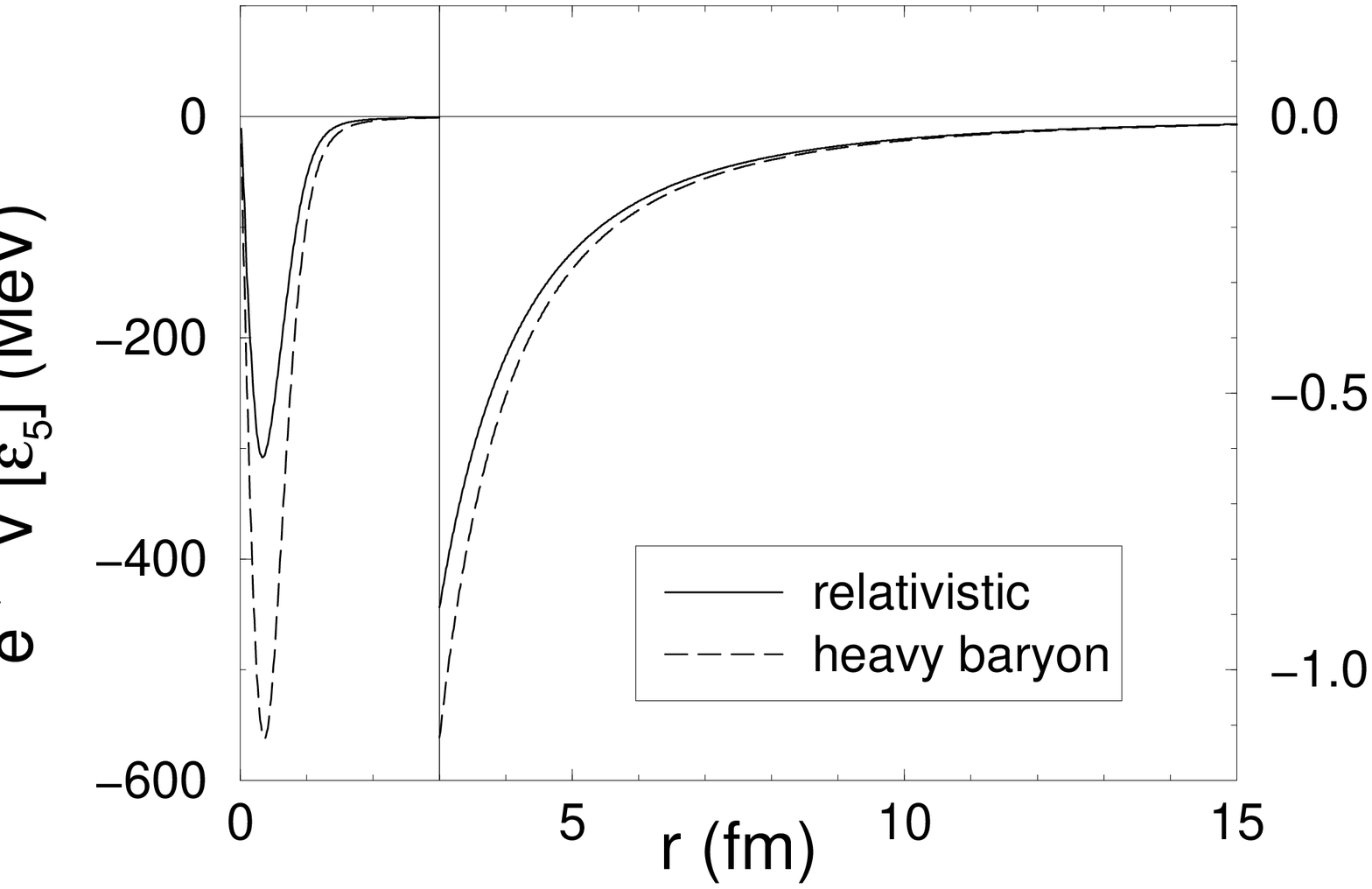, height=1.2in}&
\epsfig{figure=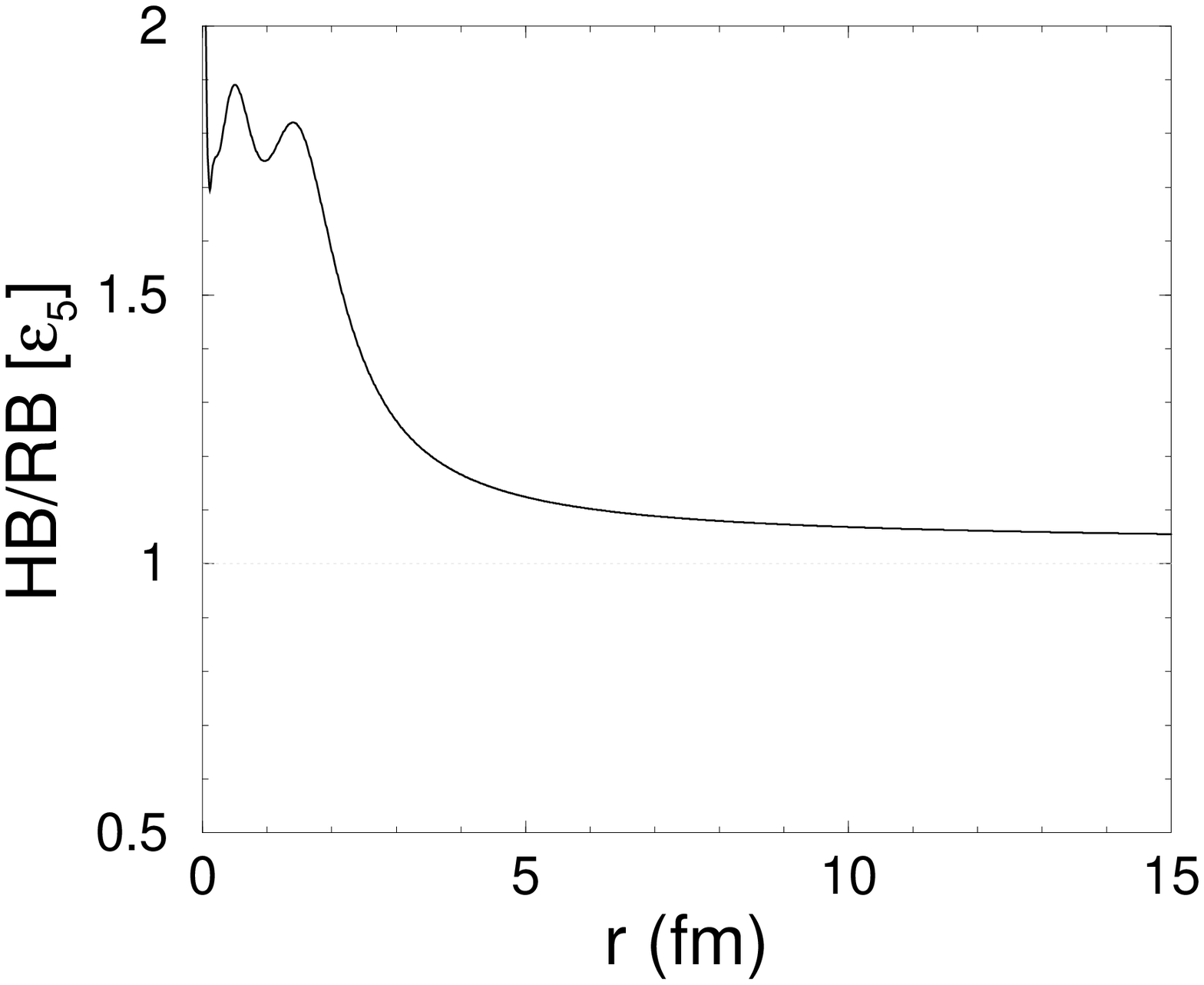, height=1.2in}&&&
\end{tabular}
\end{center}
\caption{Same as in Fig.~\ref{figFwaves}, for $H$ waves.}
\label{figHwaves}
\end{figure}

Figs.~\ref{figFwaves}, \ref{figGwaves}, and \ref{figHwaves} show, for 
$F$, $G$, and $H$ partial wave channels, respectively, the 
values of the RB and HB versions of the TPEP multiplied by a common 
factor $\exp(2\mu r)$, as well as their corresponding ratios. 
The notation is similar to the one used for phase shifts --- for a given 
total and orbital angular momenta $J$ and $L$, the single channel 
potentials are identified by $J=L$ waves, while for coupled channels 
the $2\times 2$ matrix potential has diagonal components 
$\left[{}^3(J-1)_{J},{}^3(J+1)_{J}\right]$ and non-diagonal component 
$\epsilon_{J}$. 
A chiral TPEP in configuration space has the drawback of being divergent 
at the origin and, in order to avoid numerical complications while 
computing phase shifts, we multiply it by a phenomenological regulator, 

\begin{equation}
V_{\rm TPEP,\;reg}=\left[1-e^{-cr^2}\right]^4\,V_{\rm TPEP,\;div}\,,
\label{eq:regul}
\end{equation}

\noindent which is similar to the one used by the Urbana \cite{urb} and 
Argonne \cite{av14,av18} potentials. For the regulator we adopted the 
value from AV14 \cite{av14}, $c=2.0$fm${}^{-2}$. We will address 
this question with more details in Sec.~\ref{secV}. 

Besides the non-diagonal potentials, the ratios clearly show the failure 
of the HB in describing the correct asymptotic behavior of the TPEP. 
Such deviations get more dramatic for ${}^1F_3$, ${}^3G_3$, 
${}^3G_4$, ${}^3G_5$, and ${}^1H_5$ waves (not by accident), being 
significant already at $r=5$fm. From Fig.~\ref{figpots} it seems strange 
that differences of the order of 35\% turn into $\sim 50-100$\% observed 
for these waves. The reason is that they are channels with total isospin 
$T=0$, leading to cancelations between isoscalar and isovector components. 
As an example, for total spin $S=0$ states the tensor and spin-orbit 
components don't contribute and the potential is given by 

\begin{equation}
V\left[{}^1L_J\right]=
V_C^++(4T-3)\,V_C^--3\big[V_{SS}^++(4T-3)\,V_{SS}^-\big]=
V_C^+-3\,V_C^--3\,\big[V_{SS}^+-3\,V_{SS}^-\big]\,.
\end{equation}

\noindent At $r=6$fm one has, for the RB and HB cases, 

\begin{eqnarray}
V\left[{}^1L_J\right]_{RB}=
\big[-150.0-3(-78.7)\big]-3\big[30.1-3(4.28)\big]
\approx86.1-51.8\approx34.3
\label{eq:singrb}
\\[1mm]
V\left[{}^1L_J\right]_{HB}=
\big[-150.7-3(-88.8)\big]-3\big[30.4-3(3.78)\big]
\approx115.7-57.2\approx58.5
\label{eq:singhb}
\end{eqnarray}

\noindent (in units of $10^{-6}$MeV). From Eqs.(\ref{eq:singrb}) and 
(\ref{eq:singhb}) one can understand the mechanism that leads to a 
deviation of roughly 70\% in the final result: it comes first from the 
structure [(isoscalar)-3(isovector)] (which generates more than 30\% 
of deviation from the central component), and is further increased by 
the sum of central and spin-spin contributions. 


\newpage

\newpage

\newpage

\section{the phase function method} \label{secIII}

In the calculation of phase shifts we employ the formalism of the 
variable phase, well studied by a couple of authors in the 
early sixties \cite{calog,calogrev,babikov,babikovrev,kynch,coxperl} 
but, surprisingly, not very popular in the nuclear physics community 
(in the last decade, from the best of our knowledge, only a few works 
took advantage of this method \cite{nalpha,BRR,BR,va04}). Due to its 
simplicity, here we outline the main steps to obtain the phase function 
equation, where its physical interpretation immediately arises. A more 
detailed description, including scattering with tensor forces, as well 
as extensions of these ideas to other processes, observables, and 
kinematics can be found in a review paper by Babikov 
\cite{babikovrev}, as well as in the book from Calogero \cite{calogrev}. 

Here we consider the case of a central potential with a regular behavior, 
{\em i. e.}, the ones which behave as $r^{-n}$ close to the origin, 
with $n<2$. The radial Scr\"odinger equation is written as 

\begin{equation}
u''_L(r)+\big[k^2-L(L+1)/r^2-U(r)\big]\,u_L(r)=0\,, 
\label{eq:radwfeq}
\end{equation}

\noindent with $U(r)=2m_{red}\,V(r)$, 
and $k=\sqrt{2m_{red}{\cal E}}$, $V(r)$ being the central potential, 
${\cal E}$, the energy eigenvalue in the Sch\"odinger equation, 
$m_{red}$, the reduced mass of the system, and $u_L(r)$, the radial 
wave function. 

The basic idea of the phase function method consists in introducing the 
functions $\delta_L(r)$ and $A_L(r)$ in such a way that the radial wave 
function and its derivative are written as 

\begin{eqnarray}
u_L(r)&=&A_L(r)\big[\cos\delta_L(r)\,\hat{\jmath}_L(kr)
-\sin\delta_L(r)\,\hat{n}_L(kr)\big]\,,
\label{eq:radwf1}
\\[1mm]
u'_L(r)&=&A_L(r)\big[\cos\delta_L(r)\,\hat{\jmath}'_L(kr)
-\sin\delta_L(r)\,\hat{n}'_L(kr)\big]\,,
\label{eq:radwf2}
\end{eqnarray}

\noindent where $\hat{\jmath}_L(x)$ and $\hat{n}_L(x)$ are the usual 
spherical Bessel functions, ${j}_L(x)$ and ${n}_L(x)$, multiplied by 
the argument (so called Riccati-Bessel functions). Eq.(\ref{eq:radwf1}) 
is just a general parametrization 
of $u_L(r)$, while Eq.(\ref{eq:radwf2}) imposes the condition 

\begin{eqnarray}
A'_L(r)\big[\cos\delta_L(r)\,\hat{\jmath}_L(kr)
-\sin\delta_L(r)\,\hat{n}_L(kr)\big]-\delta'_L(r)\,A_L(r)
\big[\sin\delta_L(r)\,\hat{\jmath}_L(kr)
+\cos\delta_L(r)\,\hat{n}_L(kr)\big]=0\,.
\label{eq:radwfcond}
\end{eqnarray}

Substituting Eq.(\ref{eq:radwf1}) in Eq.(\ref{eq:radwfeq}), using 
the condition (\ref{eq:radwfcond}), and trigonometric relations we 
end up with a first order diffential equation for $\delta_L(r)$, 
the ``phase function":

\begin{equation}
\delta'_L(r)=-\frac{U(r)}{k}\,\Big[\cos\delta_L(r)\,\hat{\jmath}_L(kr)
-\sin\delta_L(r)\,\hat{n}_L(kr)\Big]^2\,. 
\label{phafunct}
\end{equation}

Notice that beyond the point where it is assumed to be the range of 
the potential this function becomes a constant, and the form of 
Eq.(\ref{eq:radwf1}) tells us that its value is precisely the phase shift. 
Furthermore, the condition (\ref{eq:radwfcond}) 
assures the continuity of the derivative of the wave function at 
this point. 
But the phase function means more than that --- from Eq.(\ref{phafunct}) 
one can easily check that, for a given distance $R$, 
the quantity $\delta_L(R)$ indeed {\em is} the phase shift associated 
to a potential cut at this point, $V(r)\,\theta(R-r)$. 

\begin{figure}[!ht]
  \begin{minipage}{3in}
  \epsfig{figure=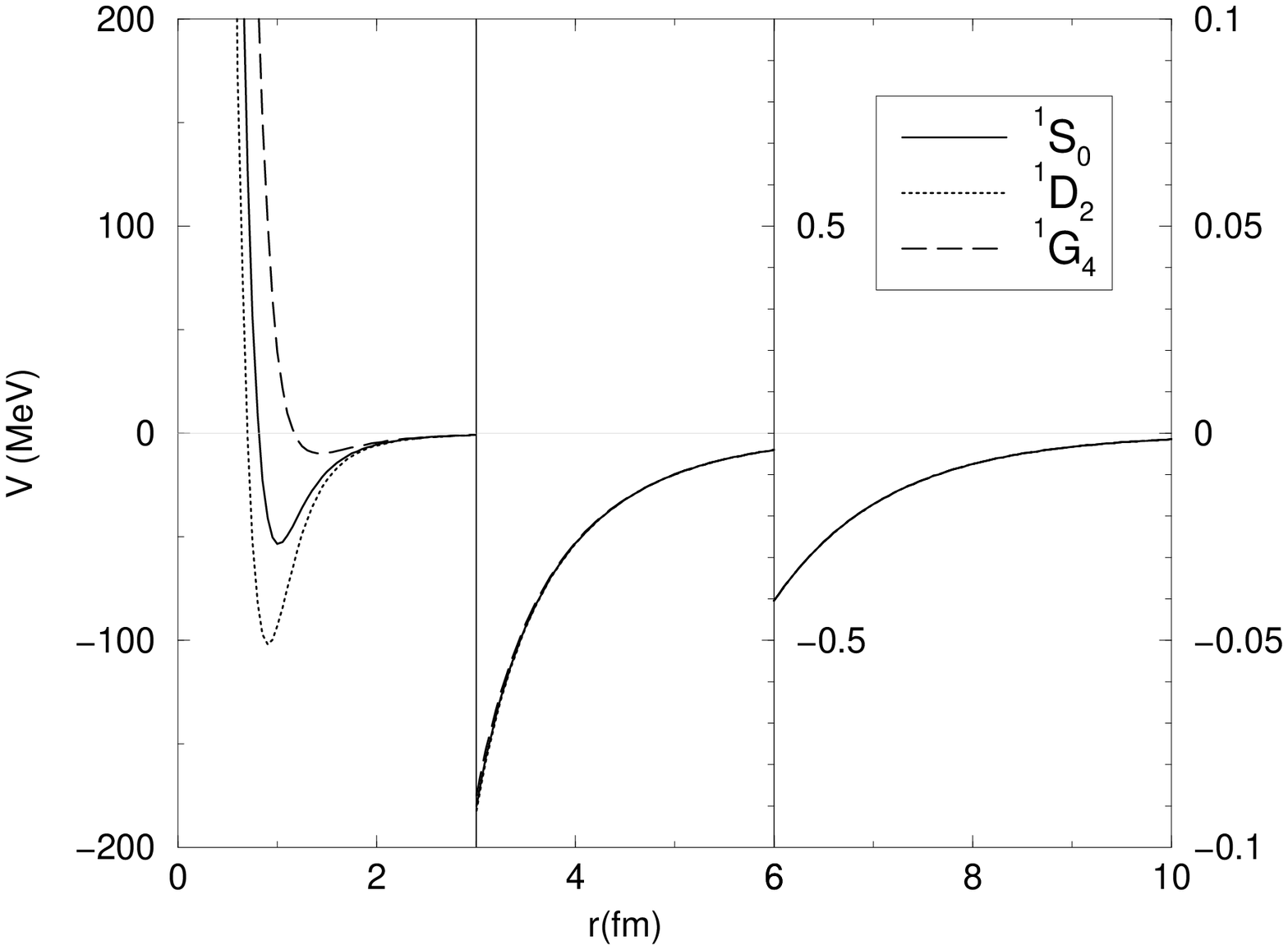, height=50mm}
  \caption{Projections of the AV14 $NN$ potential into ${}^1S_0$, 
${}^1D_2$, and ${}^1G_4$ partial waves.}
  \label{figargpot}
  \end{minipage}\hspace{10mm}
  \begin{minipage}{3in}
  \epsfig{figure=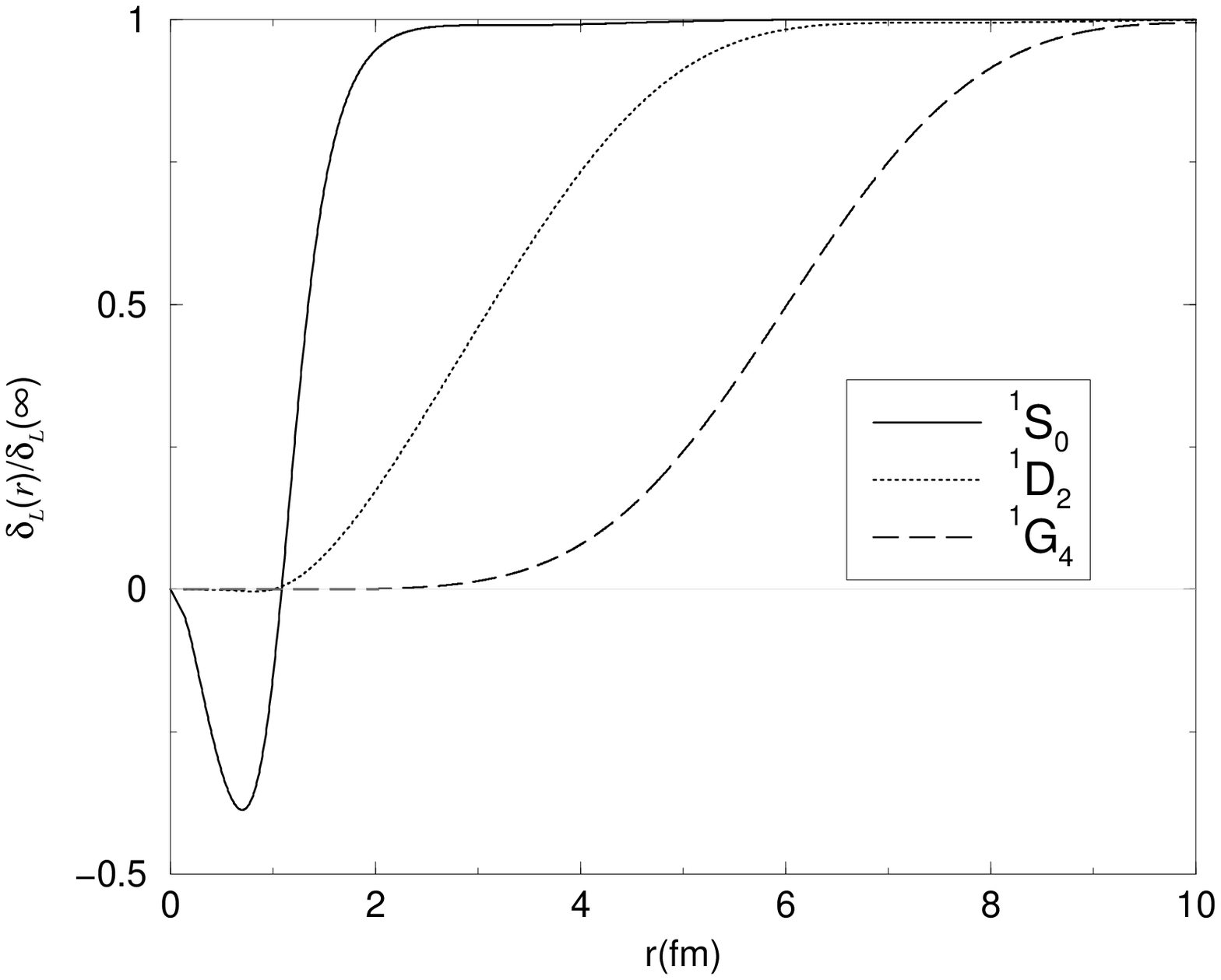, height=50mm}
  \caption{Normalized phase function at $E_{lab}=50$MeV associated to 
the partial waves from Fig.~\ref{figargpot}.}
  \label{figargph}
  \end{minipage}
\end{figure}

This property is particularly interesting to assess the importance of 
different regions of the potential. For example, in Fig.~\ref{figargph} 
we show the normalized phase function, $\delta_L(r)/\delta_L(\infty)$, 
for nucleon-nucleon scattering in 
${}^1S_0$, ${}^1D_2$, and ${}^1G_4$ partial waves, with the laboratory 
energy at 50MeV. The corresponding potentials, displayed in 
Fig.~\ref{figargpot}, are partial wave projections of the Argonne V14 
potential \cite{av14}. 
Note that, for ${}^1S_0$, the repulsive potential in the region 
$r\lesssim~0.8$fm is 
described by a decrease in the value of the phase function up to a 
minimum at this point. Beyond that the potential is attractive, resulting 
in positive increments to $\delta_L(r)$. 
For the other waves, the phase function also shows nicely the effect of 
the centrifugal barrier which, for sufficiently large $L$ and small 
energy, supress the dependence of the phase on the short range structure 
of the potential. In the next section we take advantage of this 
feature to determine the ranges of relevance when extracting the 
phase shifts. 


\section{LECs and phase shifts} \label{secIV}

In this section we present our studies for peripheral phase shifts. 
The calculations are performed in configuration space using the 
phase function method described in the previous section, which is 
fully equivalent to the Schr\"odinger equation. The final $NN$ potential 
is the sum of TPEP and OPEP and we disregard the exchange of three or 
more pions, which are much more weaker and smaller in range 
\cite{egm04-2,Kaiser4,joel}. We use the Fourier transform of the same 
charge-dependent OPE for neutron-proton scattering used by Entem and 
Machleidt \cite{EM}, 

\begin{equation}
V_{OPE}=-V_{\pi}(\mu_{\pi^0})+(-1)^{T+1}\,2\,V_{\pi}(\mu_{\pi^{\pm}})\,,
\end{equation}

\noindent with $\mu_{\pi^0}=134.9766$ MeV, $\mu_{\pi^{\pm}}=139.5702$ MeV, 
and the function $V_{\pi}(\mu)$ defined as 

\begin{equation}
V_{\pi}(\mu)=\frac{\mu^2g_A^2}{4f_{\pi}^2}\,
\frac{e^{-\mu r}}{\mu r}\left\{\left[1+\frac{3}{\mu r}
+\frac{3}{(\mu r)^2}\right]S_{12}+{\mbox{\boldmath $\sigma$}}^{(1)}
\cdot{\mbox{\boldmath $\sigma$}}^{(2)}\right\}\,,
\end{equation}

\noindent where $S_{12}$ is the usual tensor operator in configuration 
space. This is the same charge-dependent OPEP used in 
Refs.~\cite{nij93,av18}, 
provided we set the scaling mass $m_s$ and the couplings as 

\begin{equation}
m_s=\mu_{\pi^{\pm}}\,,\qquad\qquad\quad
f_{pp}=-f_{nn}=\frac{1}{\sqrt{4\pi}}\,\frac{\mu_{\pi^{\pm}}\,g_A}{2f_{\pi}}\,.
\end{equation}

\noindent For our TPEP we use the average nucleon and pion mass from the 
AV18 potential \cite{av18}, $m=938.9190$~MeV and $\mu=138.0363$~MeV. 
The remaining constants besides the LECs are 
$\hbar c=197.3270$~MeV.fm, $f_{\pi}=92.4$~MeV, and 
$g_A=1.3187$, the latter being higher than the experimental 
value in order to accommodate the Goldberger-Treiman discrepancy. 
In this work we did not consider the quadratic spin-orbit term, given by 
Eq.(\ref{Vq}). Its effect on the phase shifts is expected to be 
insignificant for the waves we are considering as we could estimate, for 
instance, by switching on and off the $L^2$ and $(L\cdot S)^2$ components 
of the Argonne V14 potential and comparing the results. 

\begin{table}[htb]
\begin{center}
\caption{Values for the dimension two ($c_i$) and three ($\bar d_i$) 
LECs considered in this work, given respectively in GeV${}^{-1}$ and 
GeV${}^{-2}$.
\label{tab3}
}
\begin{tabular} {|c|c|c|c|c|c|}
\hline
LEC & Ref.~\cite{moj98} & Ref.~\cite{fms98} (fit 1) & Ref.~\cite{butik} & 
Ref.~\cite{nij03} & Ref.~\cite{EM} \\ \hline
$c_1$                     & -0.94 & -1.23 & -0.81  & -0.76 & -0.81 \\ \hline
$c_2$                     & 3.20  & 3.28  & 8.43   & 3.20  & 3.28  \\ \hline
$c_3$                     & -5.40 & -5.94 & -4.70  & -4.78 & -3.40 \\ \hline
$c_4$                     & 3.47  & 3.47  & 3.40   & 3.96  & 3.40  \\ 
\hline\hline
$\bar d_1+\bar d_2$       & 2.40  & 3.06  & 3.06   & 2.40  & 3.06  \\ \hline
$\bar d_3$                & -2.80 & -3.27 & -3.27  & -2.80 & -3.27 \\ \hline
$\bar d_5$                & 1.40  & 0.45  & 0.45   & 1.40  & 0.45  \\ \hline
$\bar d_{14}-\bar d_{15}$ & -6.10 & -5.65 & -5.65  & -6.10 & -5.65 \\ \hline
\end{tabular}
\end{center}
\end{table}

Concerning the LECs, in table~\ref{tab3} we display their values from 
five different studies. The first column was extracted from a work 
from Moj\v zi\v s \cite{moj98} where he presented the first complete 
$O(q^3)$ expressions 
for $\pi N$ scattering amplitude and pinned down the LECs using the 
Karlsruhe-Helsinki \cite{kochpiet} threshold parameters in combination 
with the pion-nucleon $\sigma$ term. The second column was determined 
by Fettes {\em et.al.} \cite{fms98} through a direct fit to $S$- and 
$P$- wave $\pi N$ phase shifts between 40 and 100 MeV pion lab momentum, 
using the Karlsruhe partial wave analysis \cite{ka84} (fit 1). The 
third column shows the results from B\"uttiker and Mei\ss ner \cite{butik}, 
where the $\pi N$ amplitude was reconstructed inside the Mandelstam 
triangle via dispersion relations, then the LECs were obtained through a 
fit to this amplitude, in two different points (because their results 
for the dimension three LECs were not consistent, we adopt here the same 
values from the second column). These three works rely on old Karlsruhe 
dispersion 
analyses, which have presumably some inconsistent data included 
\cite{pinlett} \footnote{Refs.~\cite{fms98,butik} also considered the VPI 
analysis \cite{vpiol}, but B\"uttiker and Mei\ss ner showed that they 
raise some issues, like a 10\% violation in the Adler consistency 
condition, and a large value for the $\pi N$ $\sigma$ term of about 
200 MeV. Besides that, it also leads to a value for 
$c_3\sim 6$ GeV$^{-1}$, much larger than the constraint imposed by $NN$ 
peripheral phase shifts.}. Different from them, the dimension two LECs 
on the fourth column were determined by the Nijmegen group \cite{nij03} 
from their new partial wave analyses of 5109 proton-proton and 4786 
neutron-proton scattering data below 500 MeV. As the dimension three 
LECs were not considered in their work, for this set we adopted the 
same values from Moj\v zi\v s. Finally, on the last column we 
have the values used by Entem and Machleidt in their calculations 
of neutron-proton phase shifts \cite{EM}. 

In the calculation of phase shifts we use the familiar nuclear-bar 
convention \cite{dbar}, where the coupled $S$ matrix with total 
angular momentum $J$ is parametrized as 

\begin{equation}
S_J=\pmatrix{
e^{2i\delta_{J-1}}\,\cos2\epsilon_J & 
i\,e^{2i(\delta_{J-1}+\delta_{J+1})}\,\sin2\epsilon_J\cr 
i\,e^{2i(\delta_{J-1}+\delta_{J+1})}\,\sin2\epsilon_J & 
e^{2i\delta_{J+1}}\,\cos2\epsilon_J}\,.
\end{equation}

\begin{figure}[ht]
\begin{center}
\begin{tabular}{ccc}
\epsfig{figure=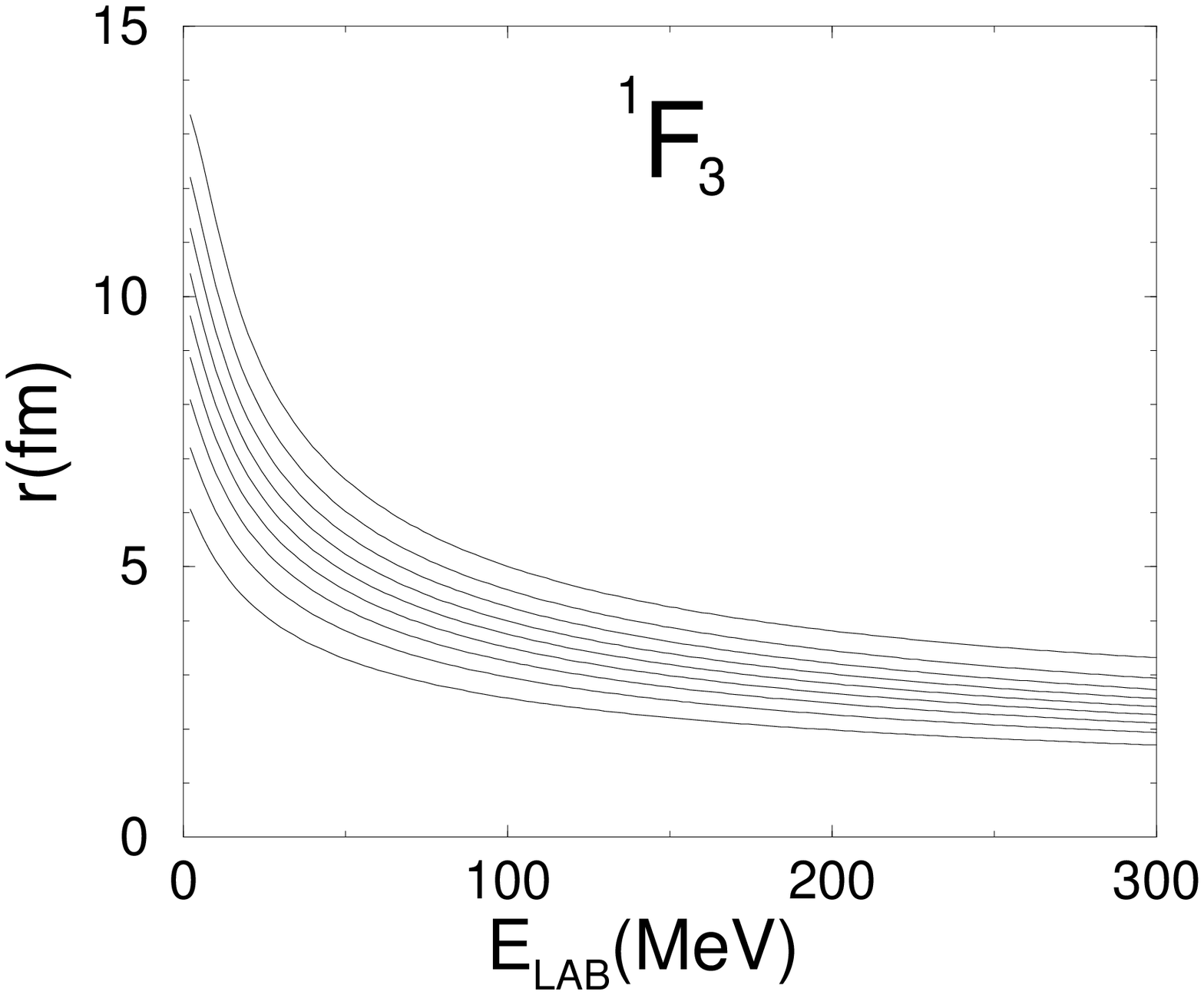, height=1.65in} \hspace{0.1in} &
\epsfig{figure=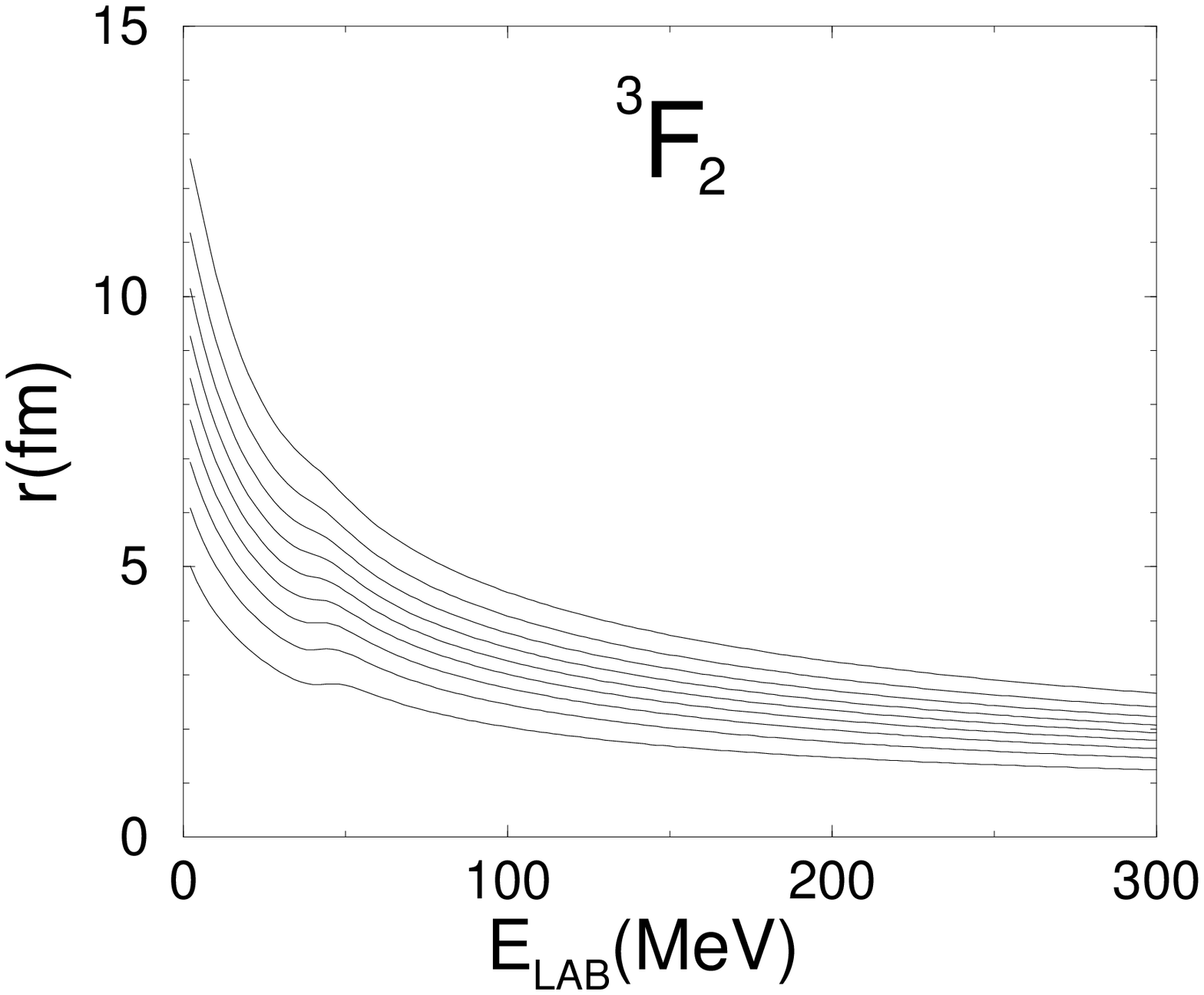, height=1.65in} \hspace{0.3in} &
\epsfig{figure=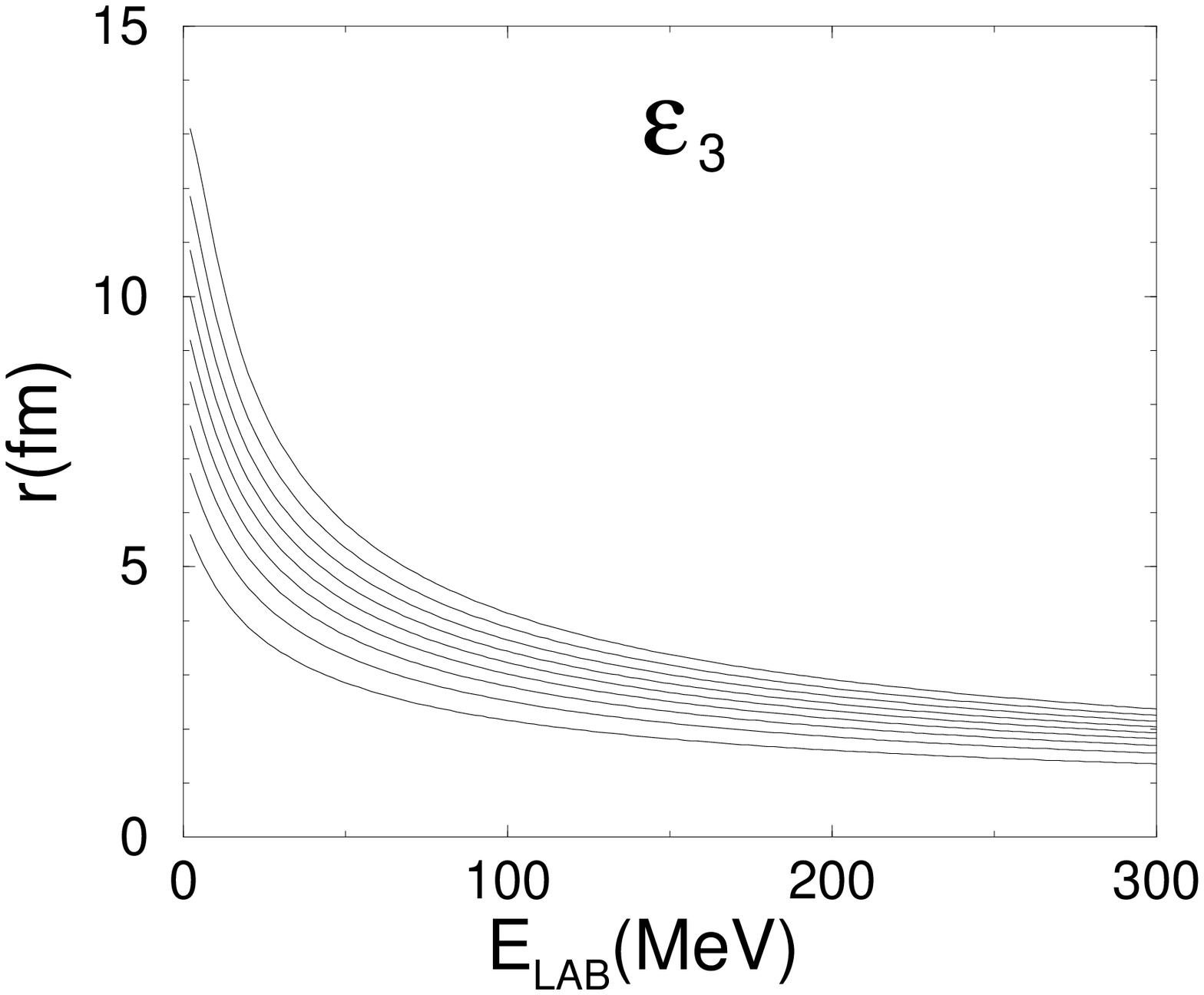, height=1.65in}\\[2mm]
\epsfig{figure=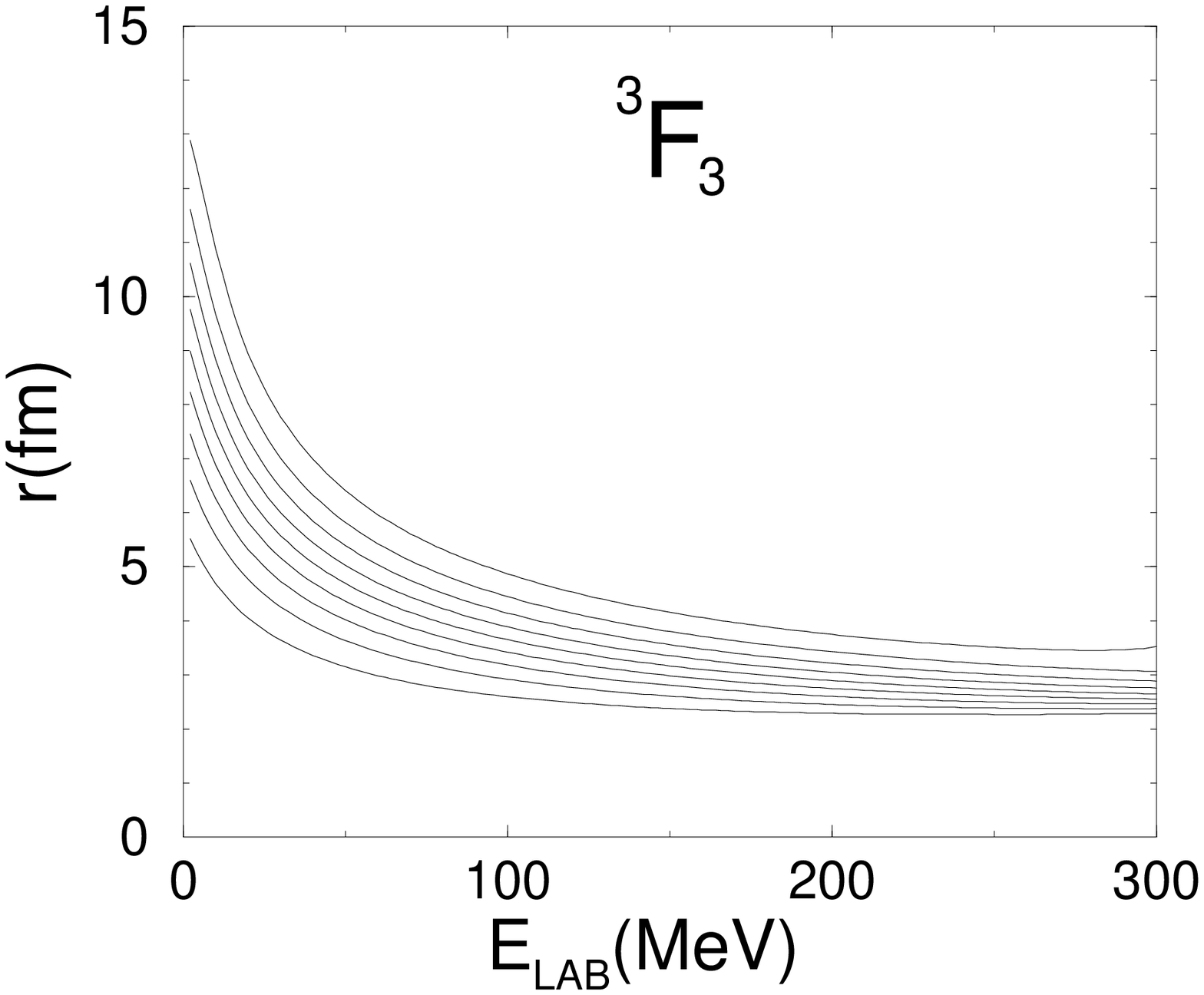, height=1.65in} \hspace{0.1in} &
\epsfig{figure=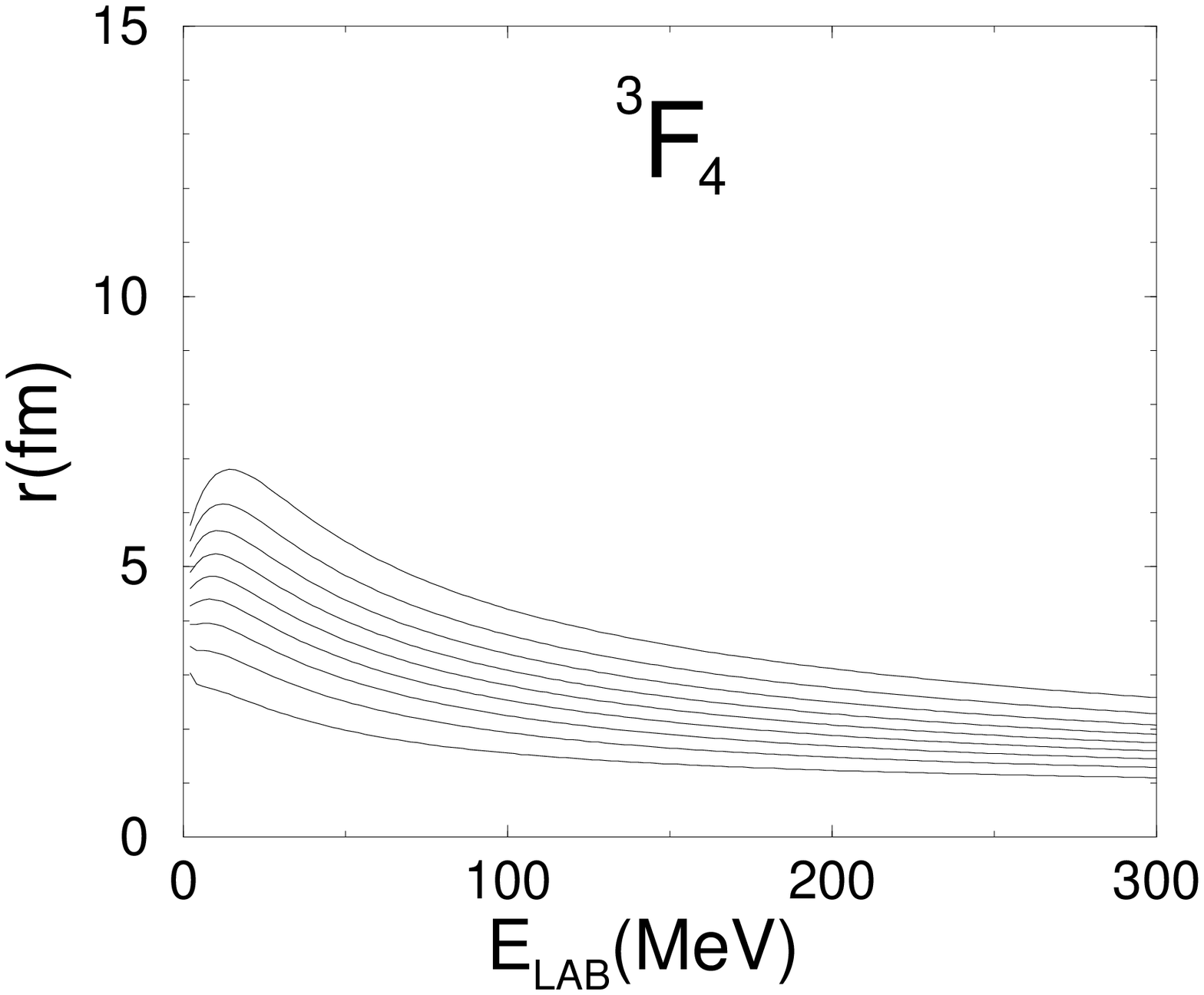, height=1.65in} \hspace{0.3in} &
\end{tabular}
\end{center}
\caption{Distances, for $F$ waves, where the phase shift (or mixing 
parameter) achieves $X$\% of its value, as functions of the laboratory 
energy. The curves correspond, from bottom to top, to 
$X=$10, 20, $\cdots$, 90, respectively.}
\label{figPF-F}
\end{figure}

\begin{figure}[ht]
\begin{center}
\begin{tabular}{ccc}
\epsfig{figure=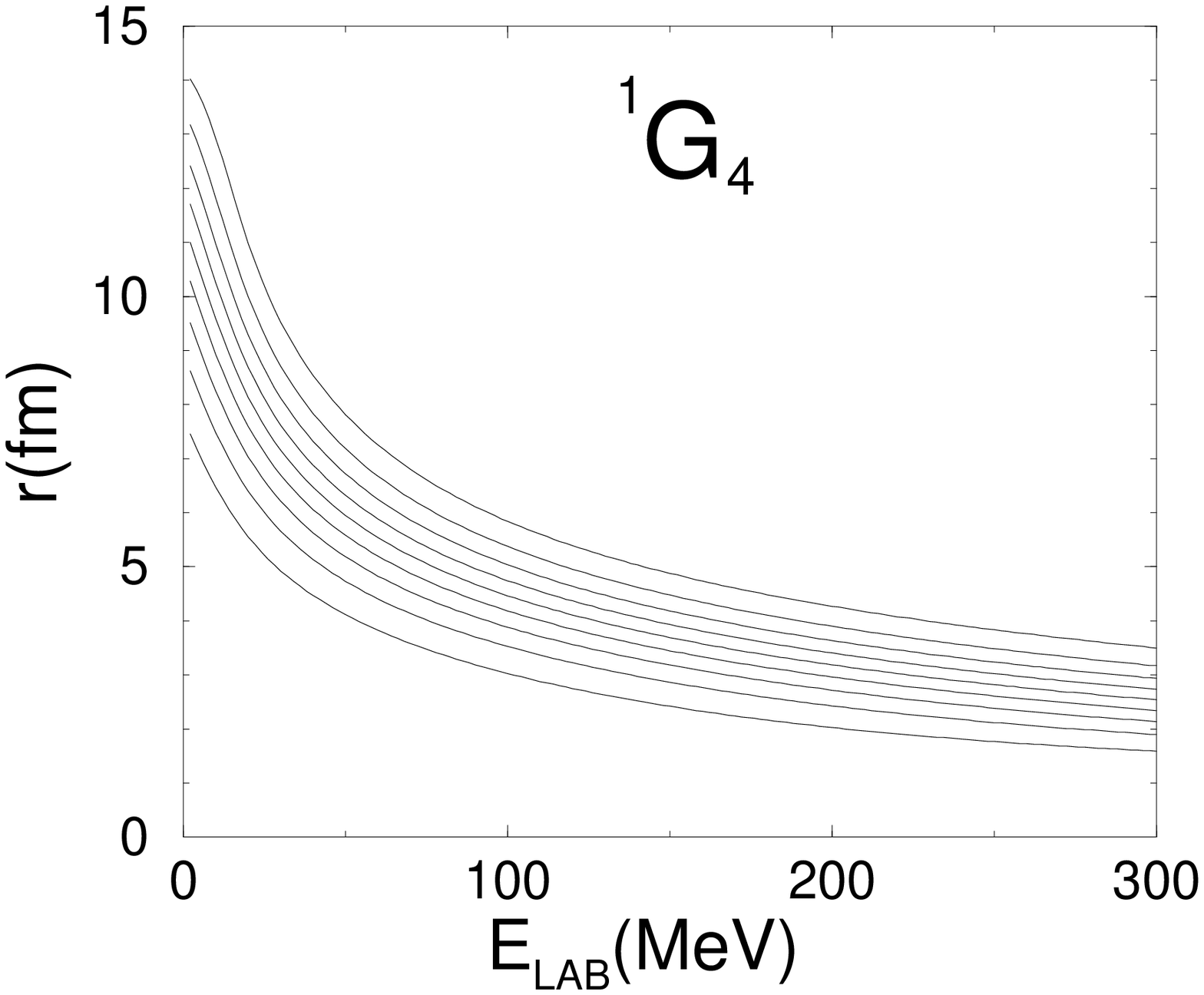, height=1.65in} \hspace{0.1in} &
\epsfig{figure=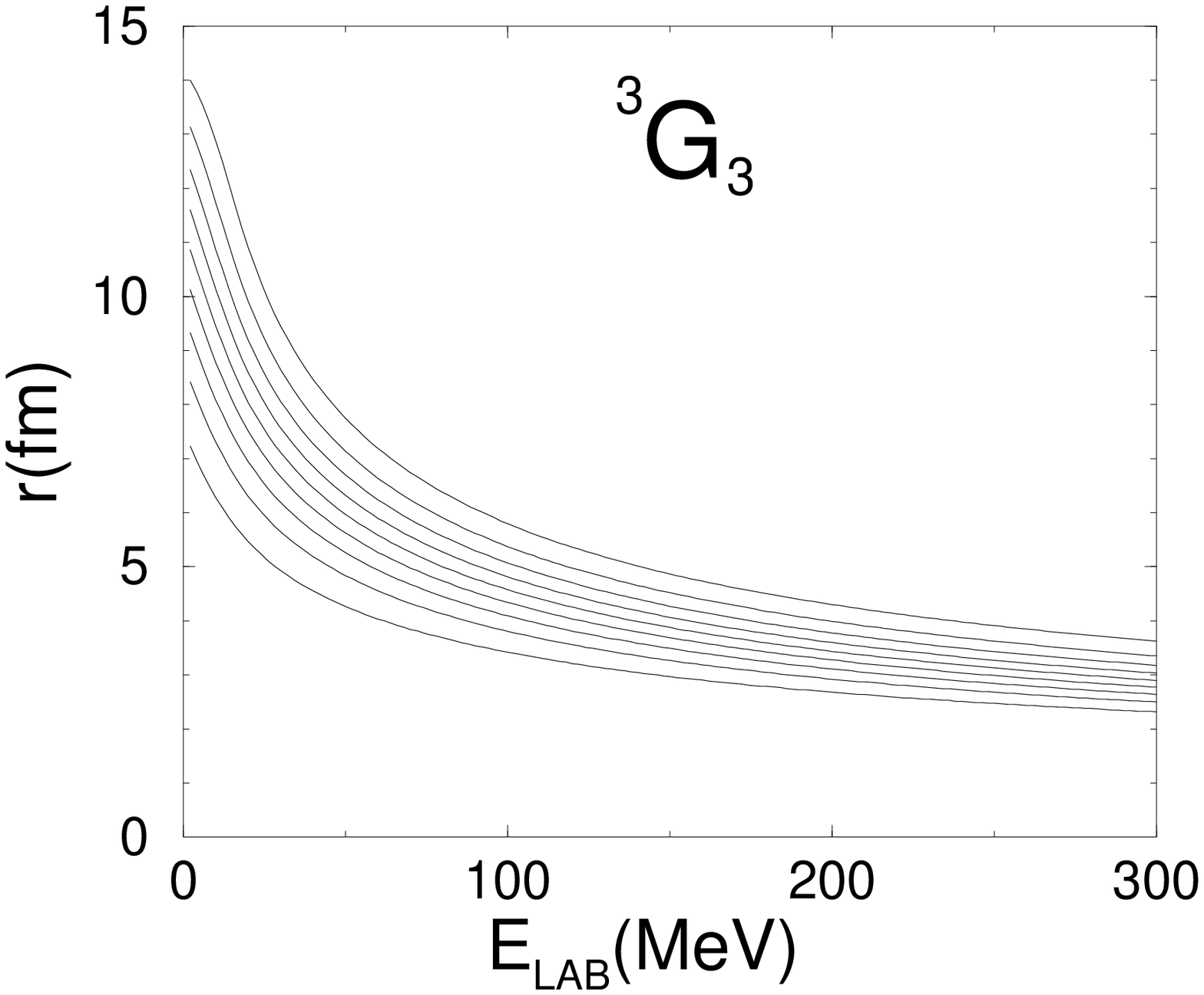, height=1.65in} \hspace{0.3in} &
\epsfig{figure=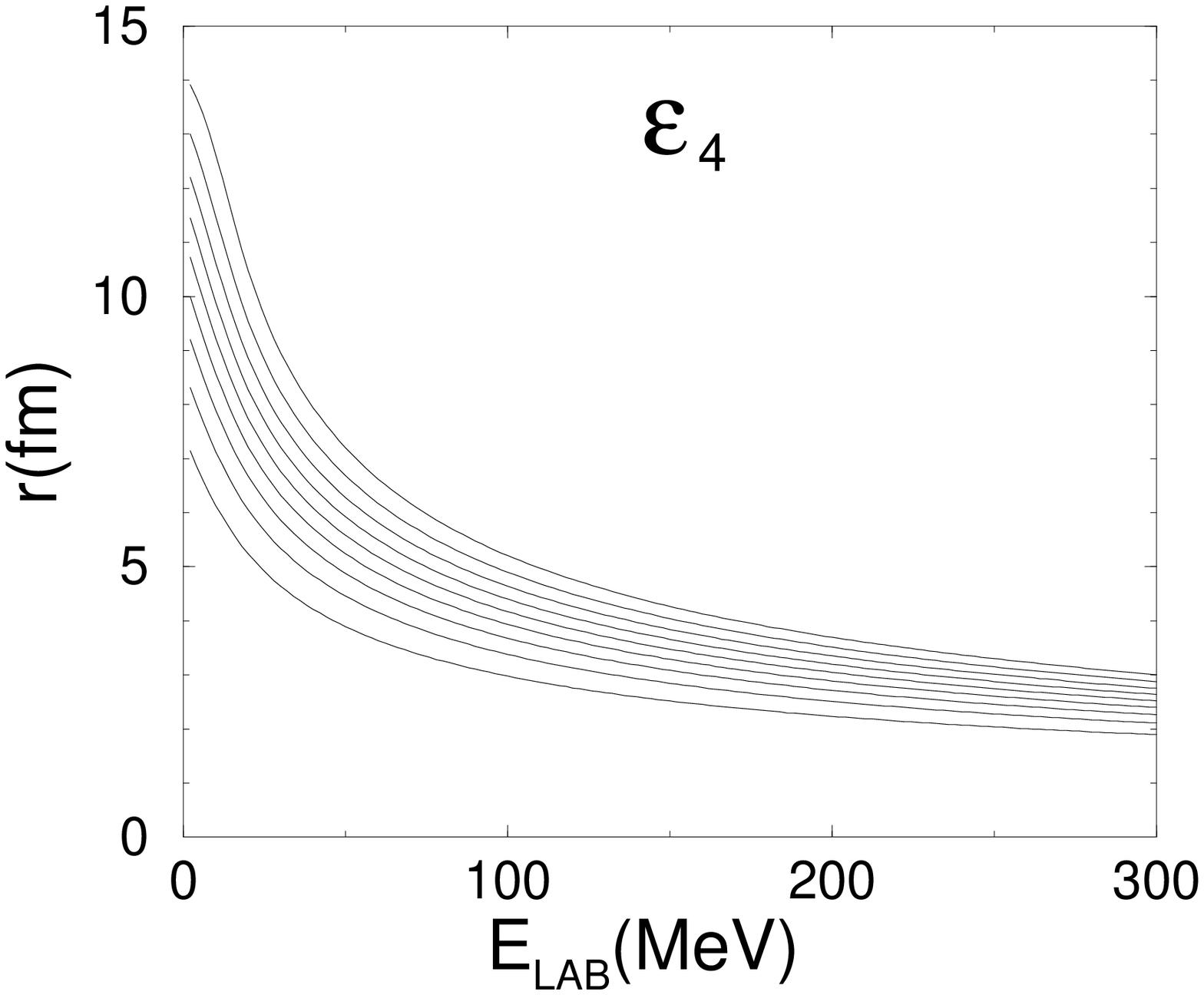, height=1.65in}\\[2mm]
\epsfig{figure=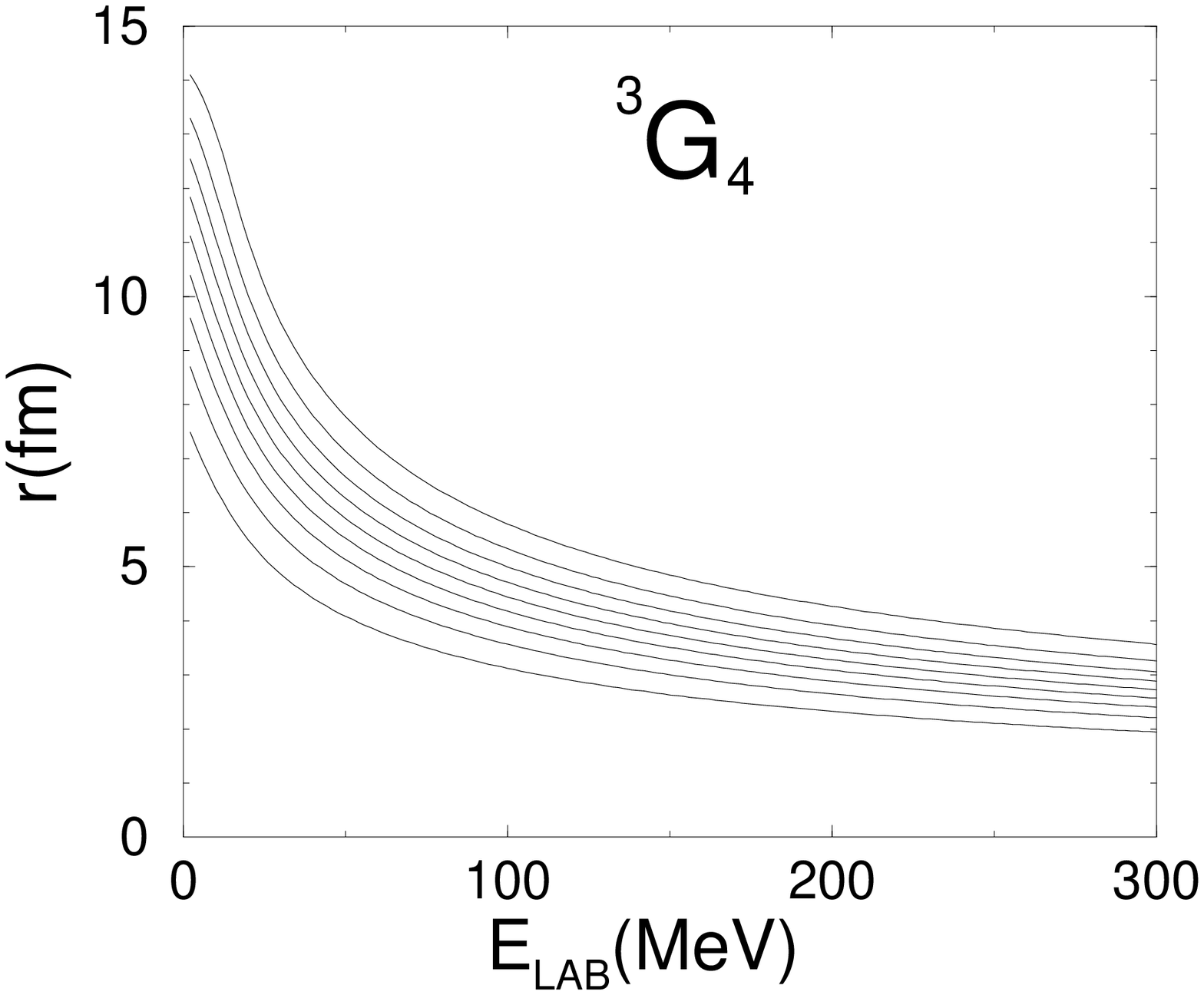, height=1.65in} \hspace{0.1in} &
\epsfig{figure=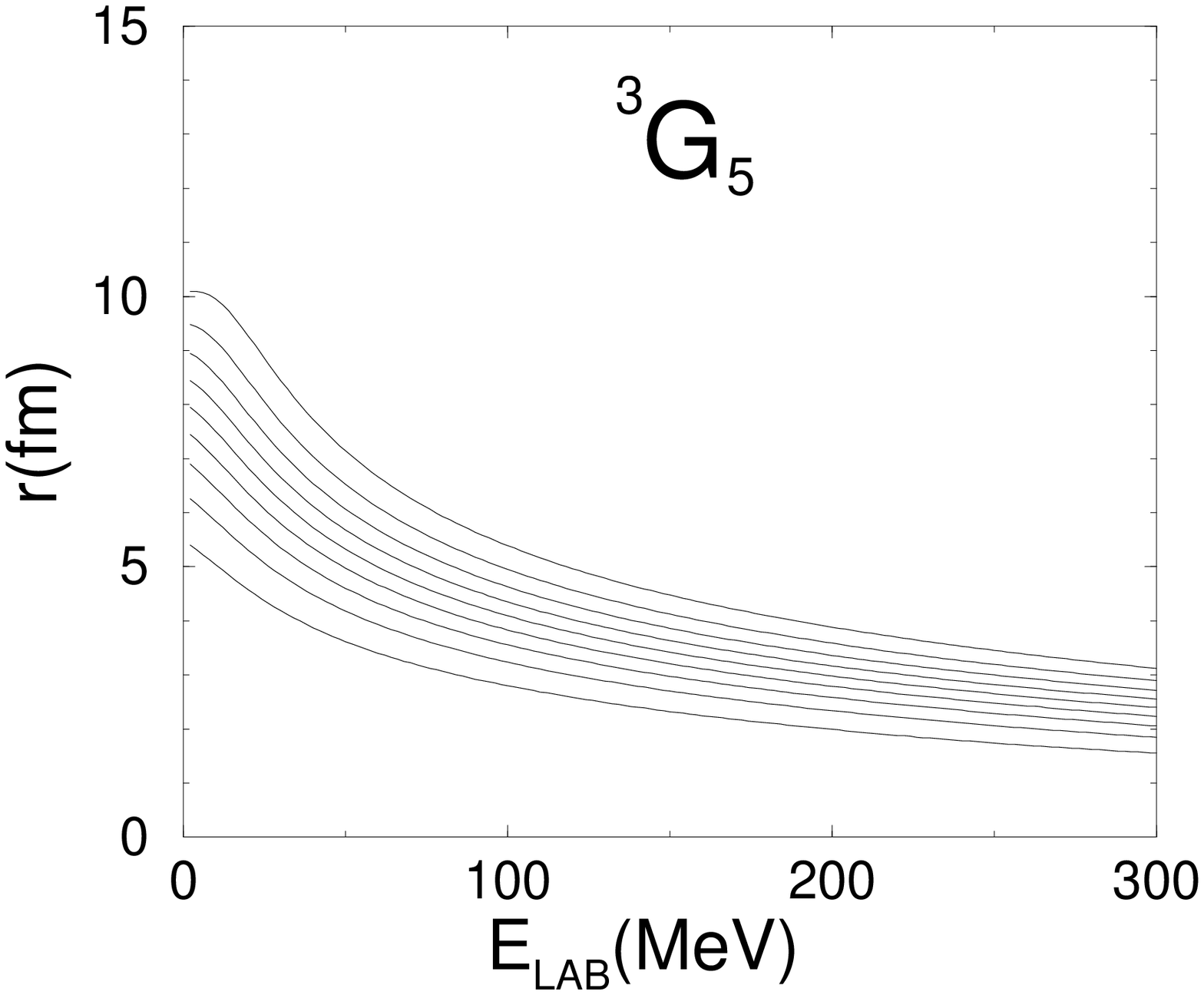, height=1.65in} \hspace{0.3in} &
\end{tabular}
\end{center}
\caption{Same as Fig.~\ref{figPF-F}, for $G$ waves.}
\label{figPF-G}
\end{figure}

\begin{figure}[hb]
\begin{center}
\begin{tabular}{ccc}
\epsfig{figure=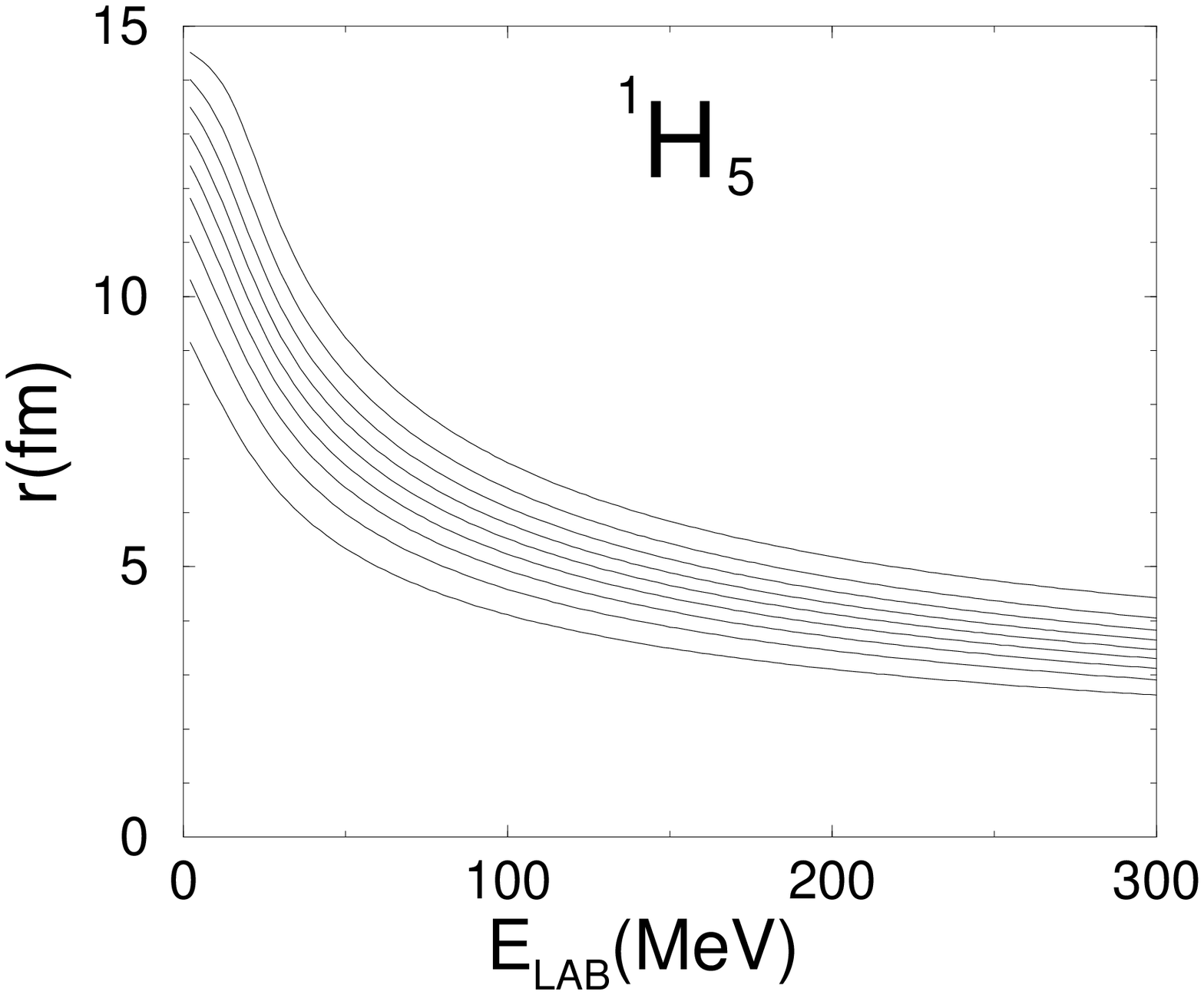, height=1.65in} \hspace{0.1in} &
\epsfig{figure=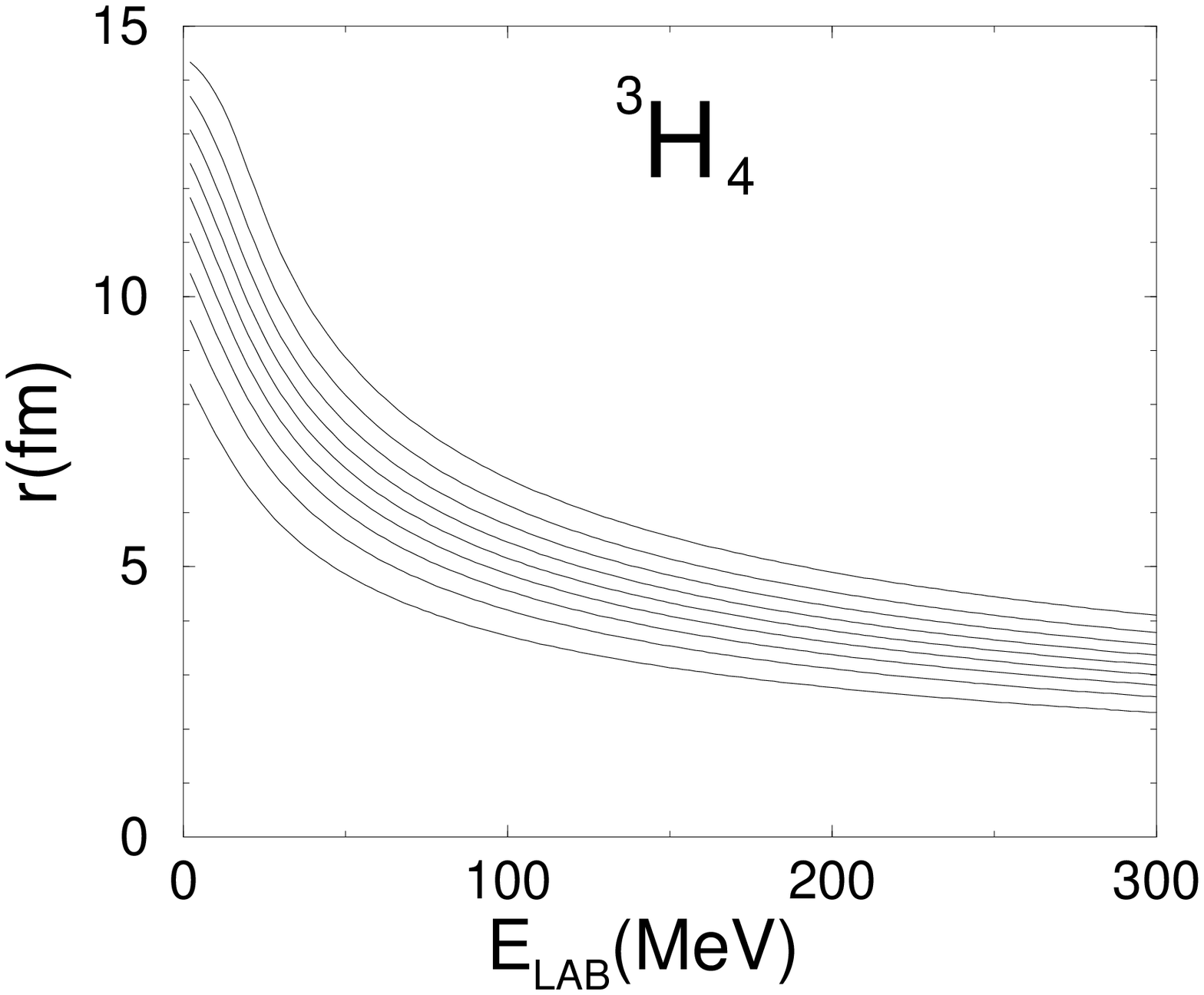, height=1.65in} \hspace{0.3in} &
\epsfig{figure=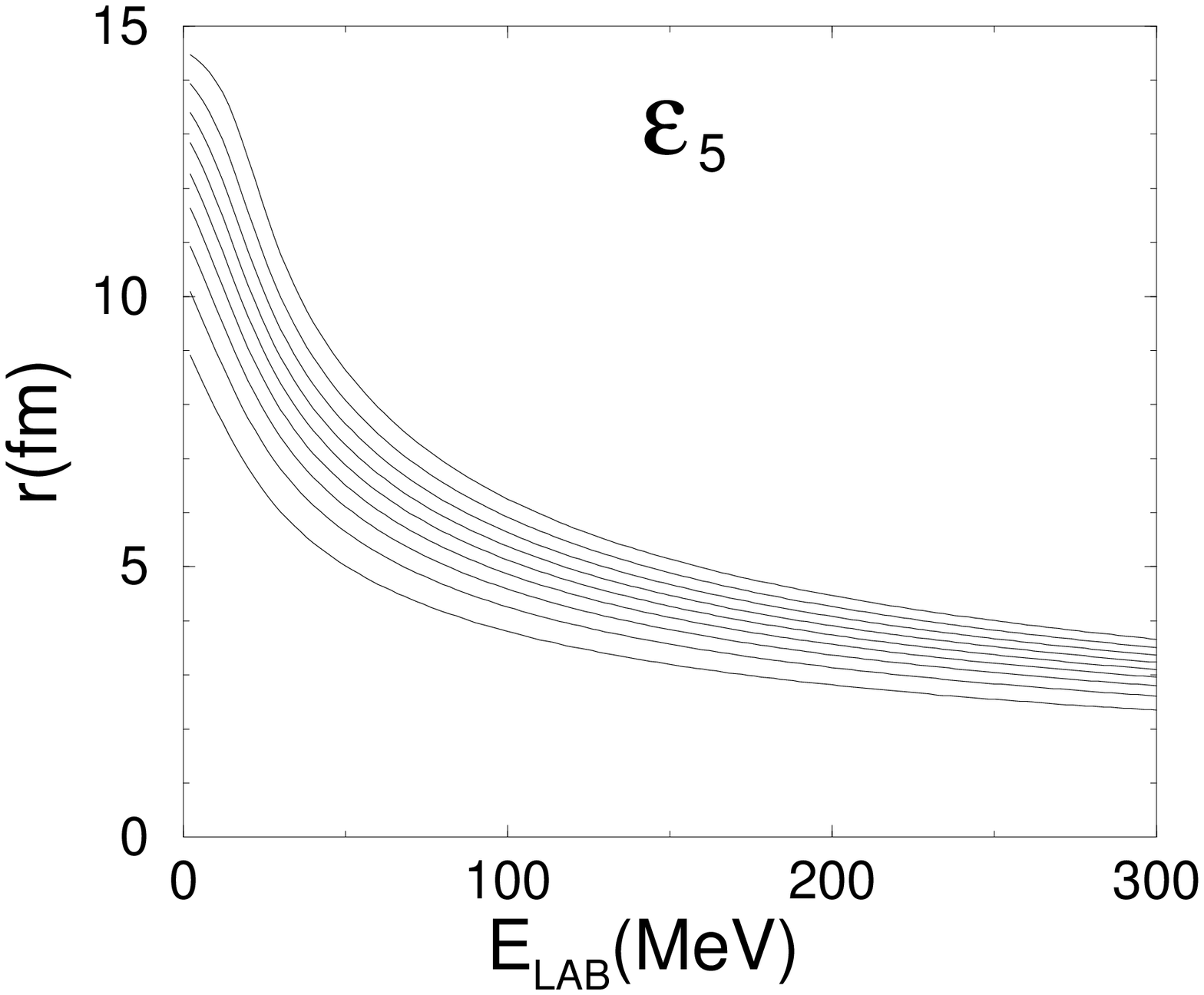, height=1.65in}\\[2mm]
\epsfig{figure=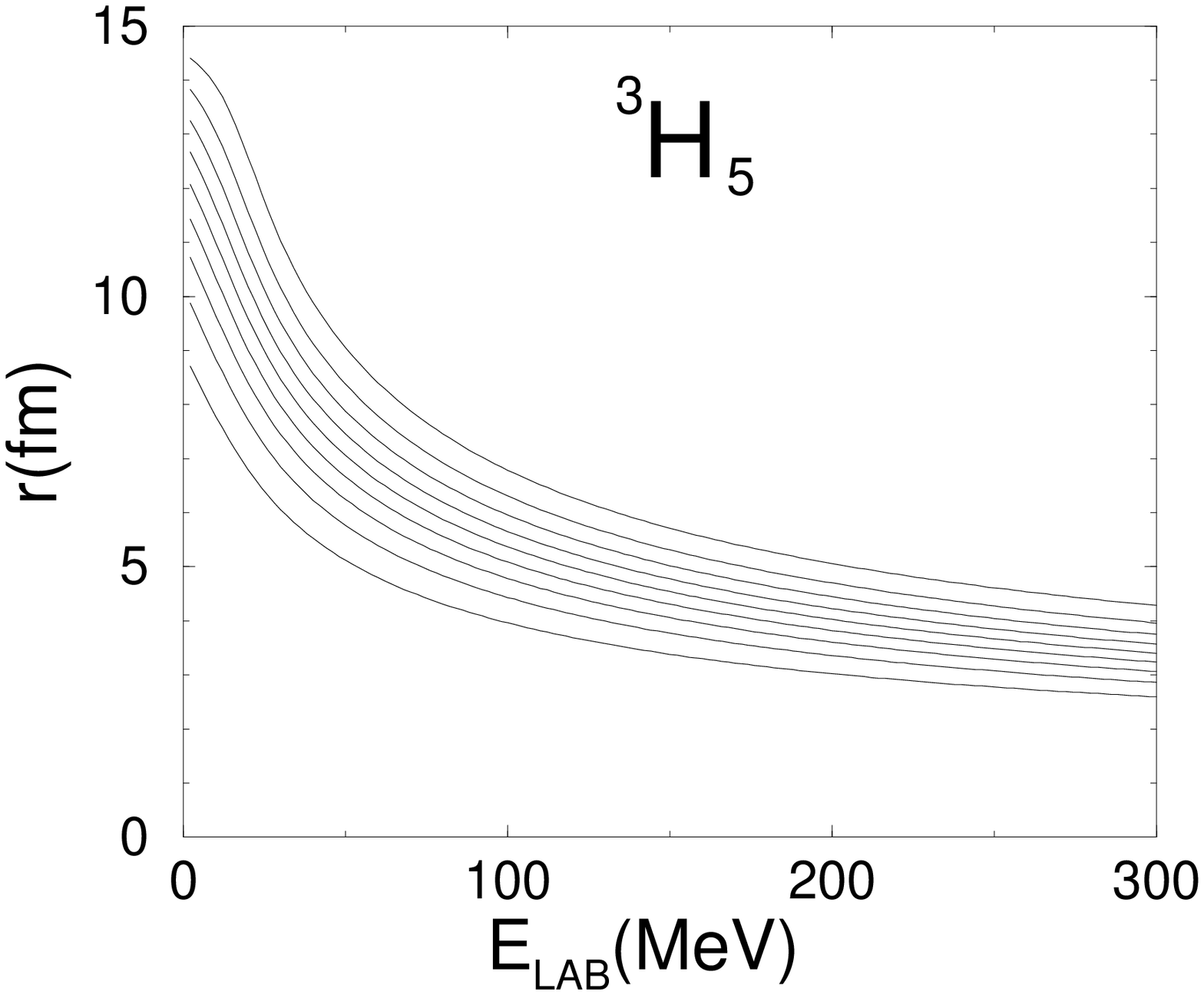, height=1.65in} \hspace{0.1in} &
\epsfig{figure=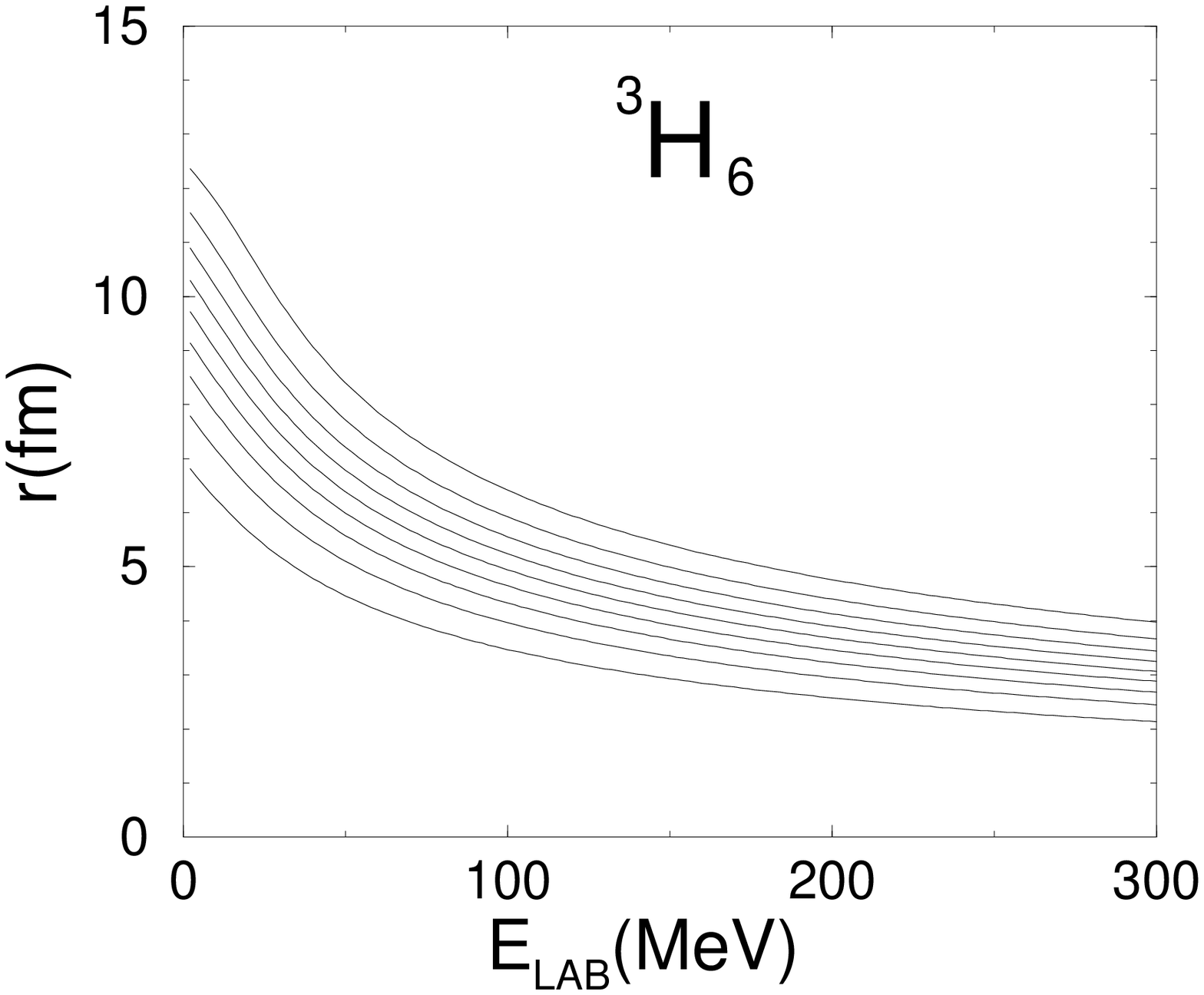, height=1.65in} \hspace{0.3in} &
\end{tabular}
\end{center}
\caption{Same as Fig.~\ref{figPF-F}, for $H$ waves.}
\label{figPF-H}
\end{figure}

\noindent Before addressing their dependence on different sets 
of LECs it is interesting to take advantage of the phase function 
formalism (Sec.~\ref{secIII}) to identify the most important ranges 
of the potential in the determination of these quantities. In 
Figs.~\ref{figPF-F}, \ref{figPF-G}, and \ref{figPF-H} we plot the 
distances where the phase function $\delta_L(r)$ reaches 10\%, 20\%, 
and so on, up to 90\%, of the final value, $\delta_L(\infty)$, as 
functions of the laboratory energy, $E_{\rm LAB}$. As in 
Sec.~\ref{secII}, we use here a potential with LECs from Entem 
and Machleidt (last column of table \ref{tab3}) and similar curves 
are obtained for the other sets. These figures show, quantitatively, 
the sensitivity of phase shifts on large distances at sufficiently low 
energies. We also notice that the range, delimited by the points where 
the phase function reaches 10\% and 90\% of the final value, gets 
narrower and closer to the inner region of the potential as we go higher 
in energies, but not smaller than 1fm (1.5fm) for $F$ ($G$ and $H$) 
waves, a clear manifestation of the centrifugal barrier. 


\newpage

\begin{figure}[hb]
\begin{center}
\begin{tabular}{ccc}
\epsfig{figure=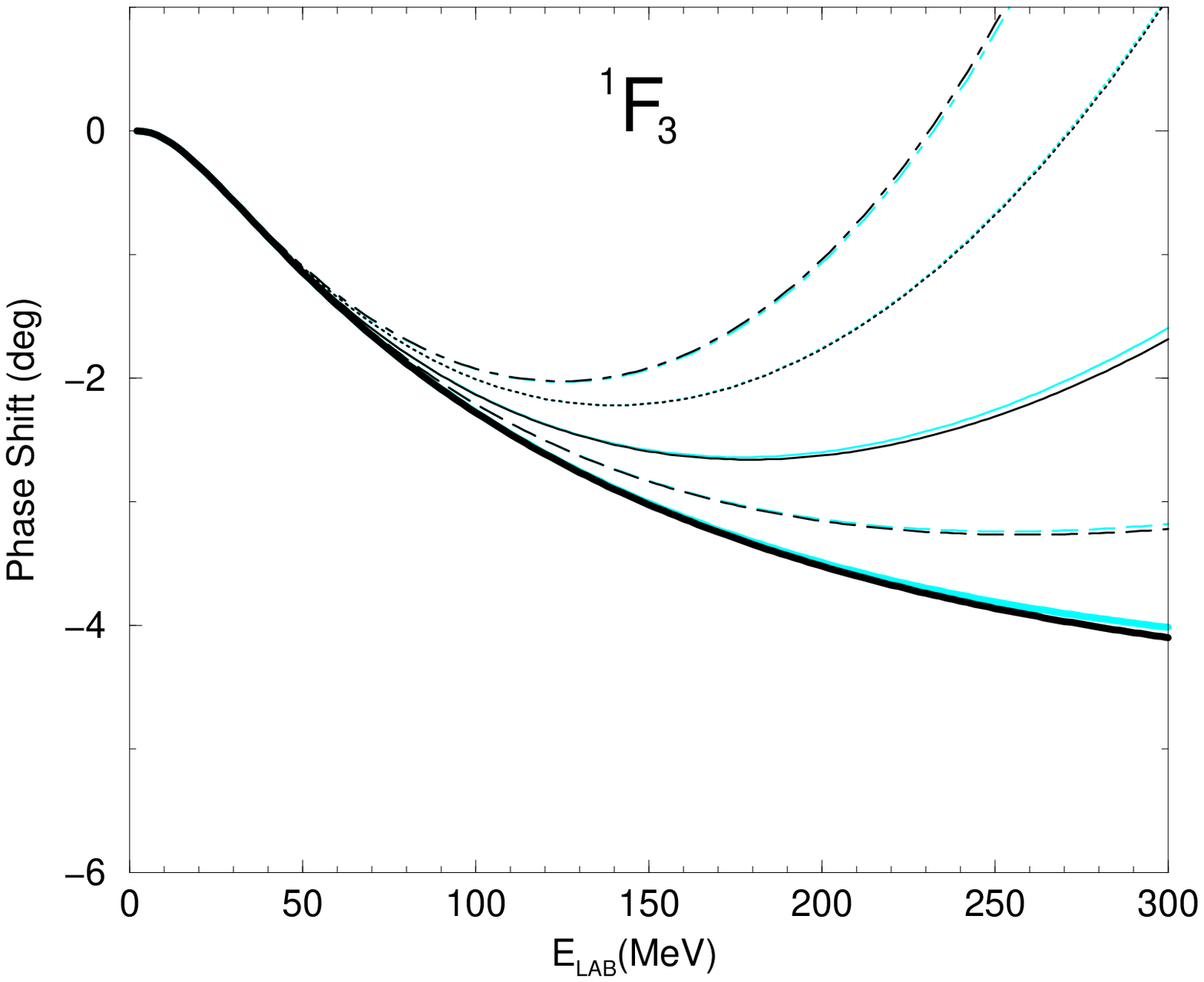, height=2.3in} & \hspace{5mm} &
\epsfig{figure=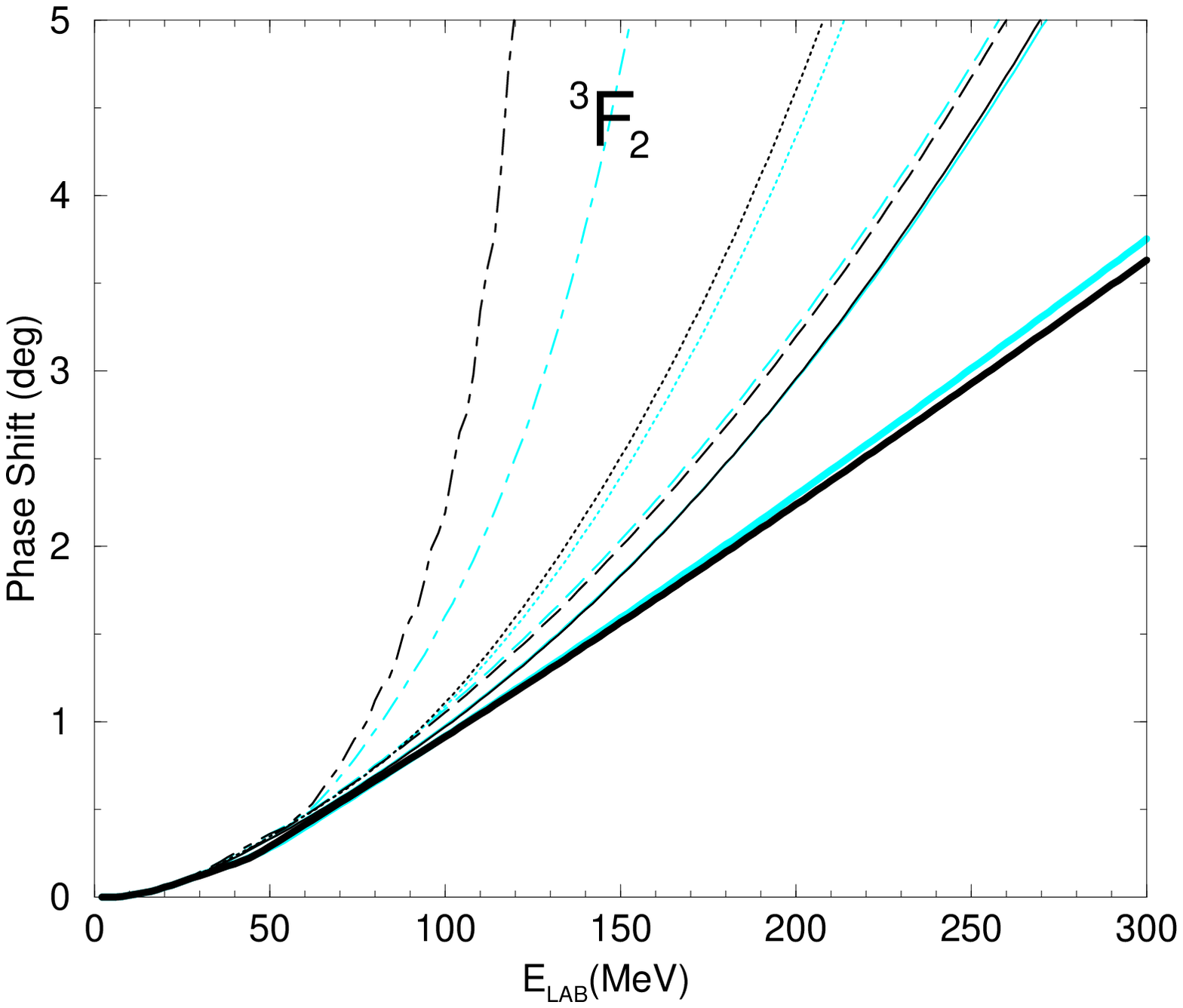, height=2.3in}\\[5mm]
\epsfig{figure=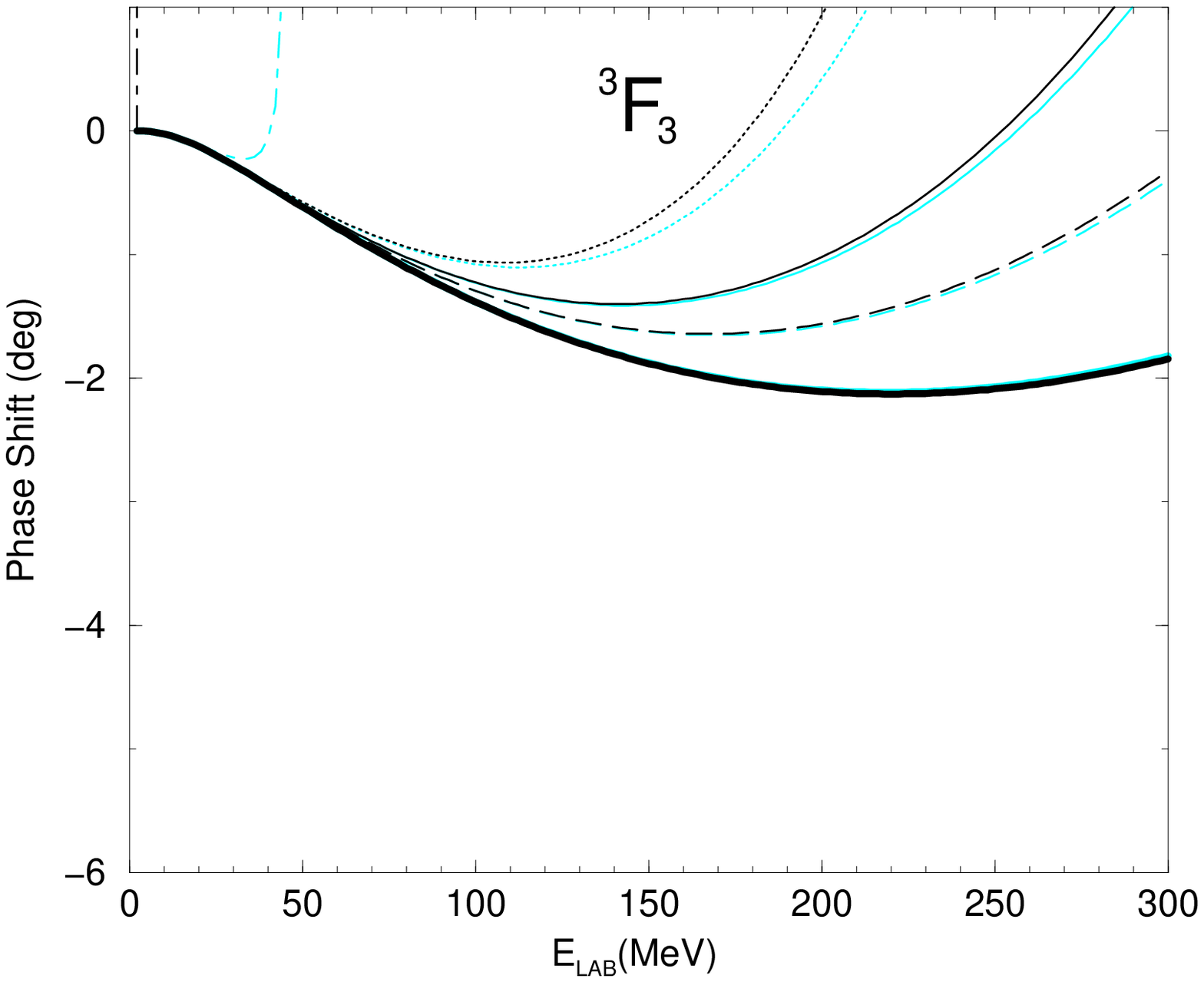, height=2.3in} & \hspace{5mm} &
\epsfig{figure=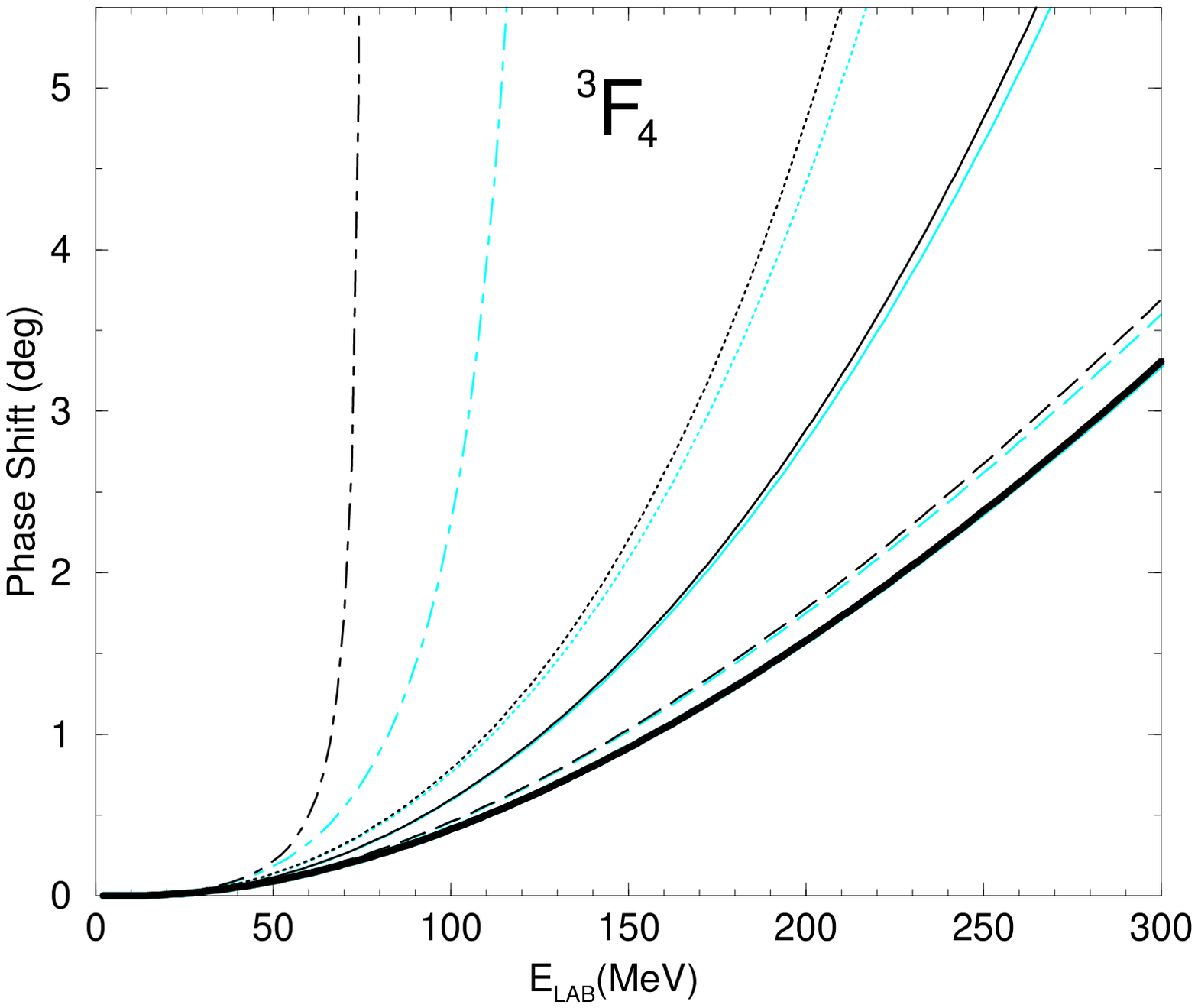, height=2.3in}\\[5mm]
\end{tabular}
\epsfig{figure=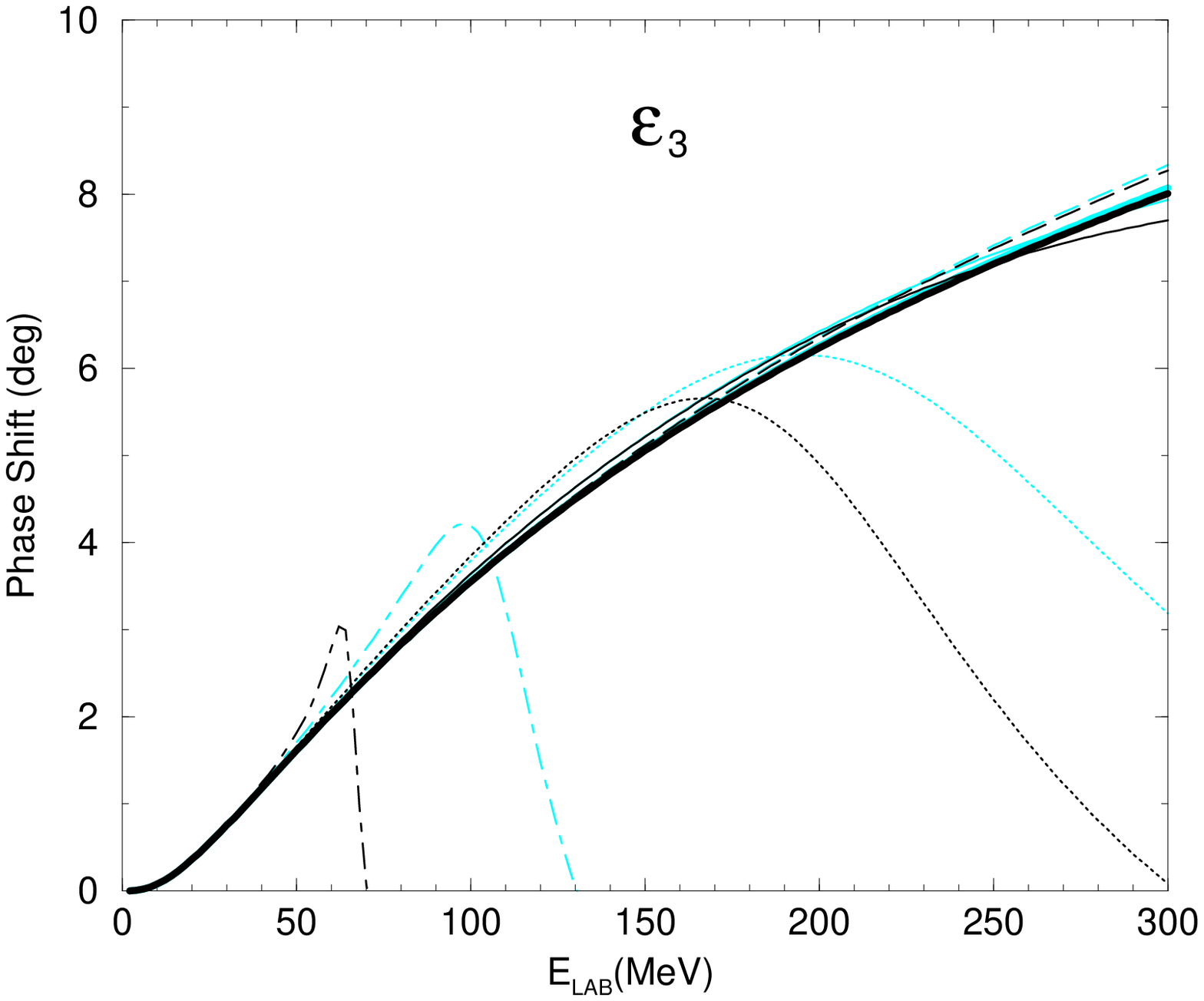, height=2.3in}
\end{center}
\caption{Phase shifts and mixing parameter for $F$ waves, as functions of 
the laboratory energy. The dotted, dot-dashed, dashed, solid, and 
thick-solid lines correspond to LECs from table~\ref{tab3}: 
columns 1, 2, 3, 4, and 5, respectively. The dark curves are results from 
the RB formalism, while the light ones, from the HB expressions.}
\label{figpsregF}
\end{figure}

\begin{figure}[ht]
\begin{center}
\begin{tabular}{ccc}
\epsfig{figure=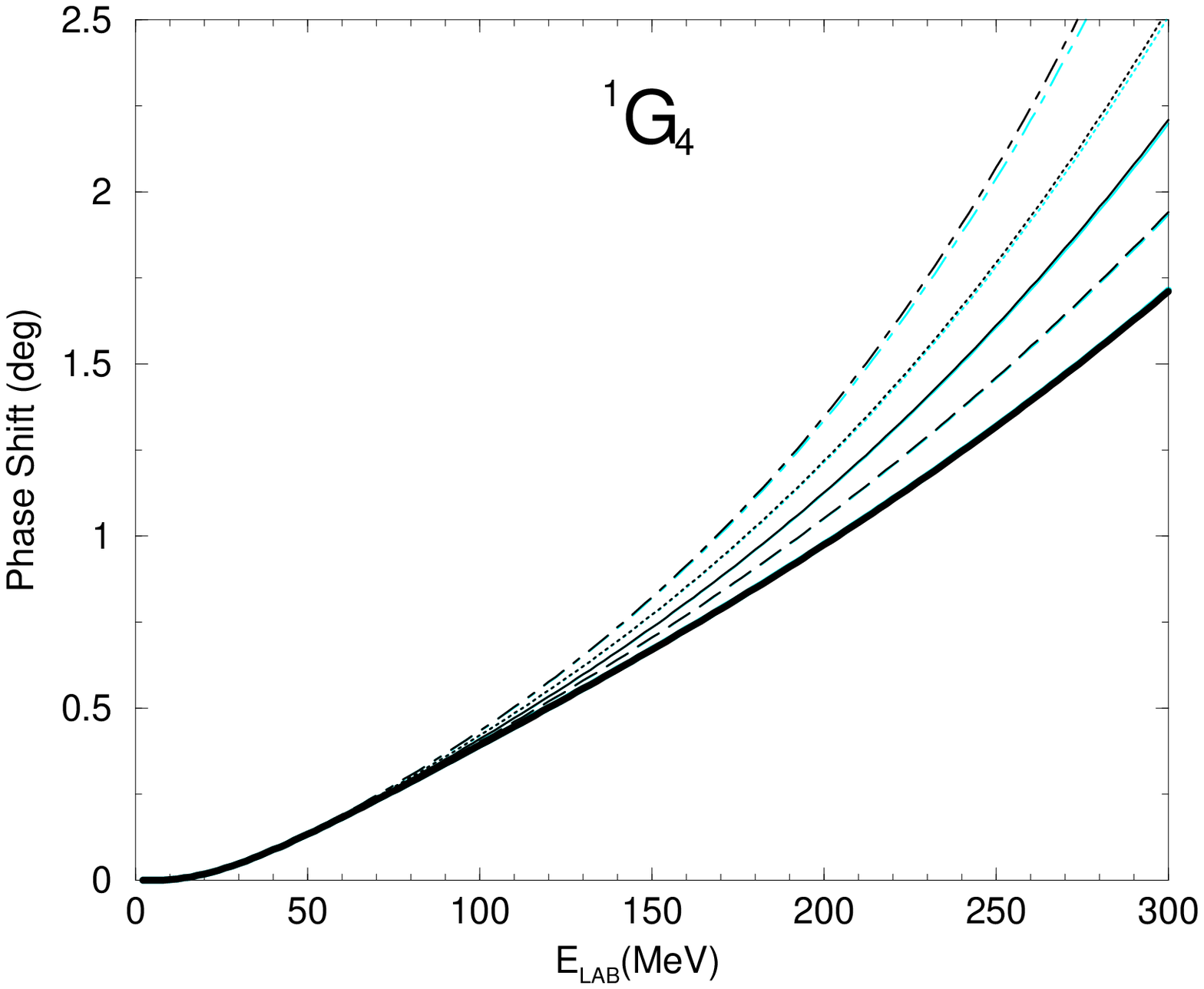, height=2.3in} & \hspace{5mm} &
\epsfig{figure=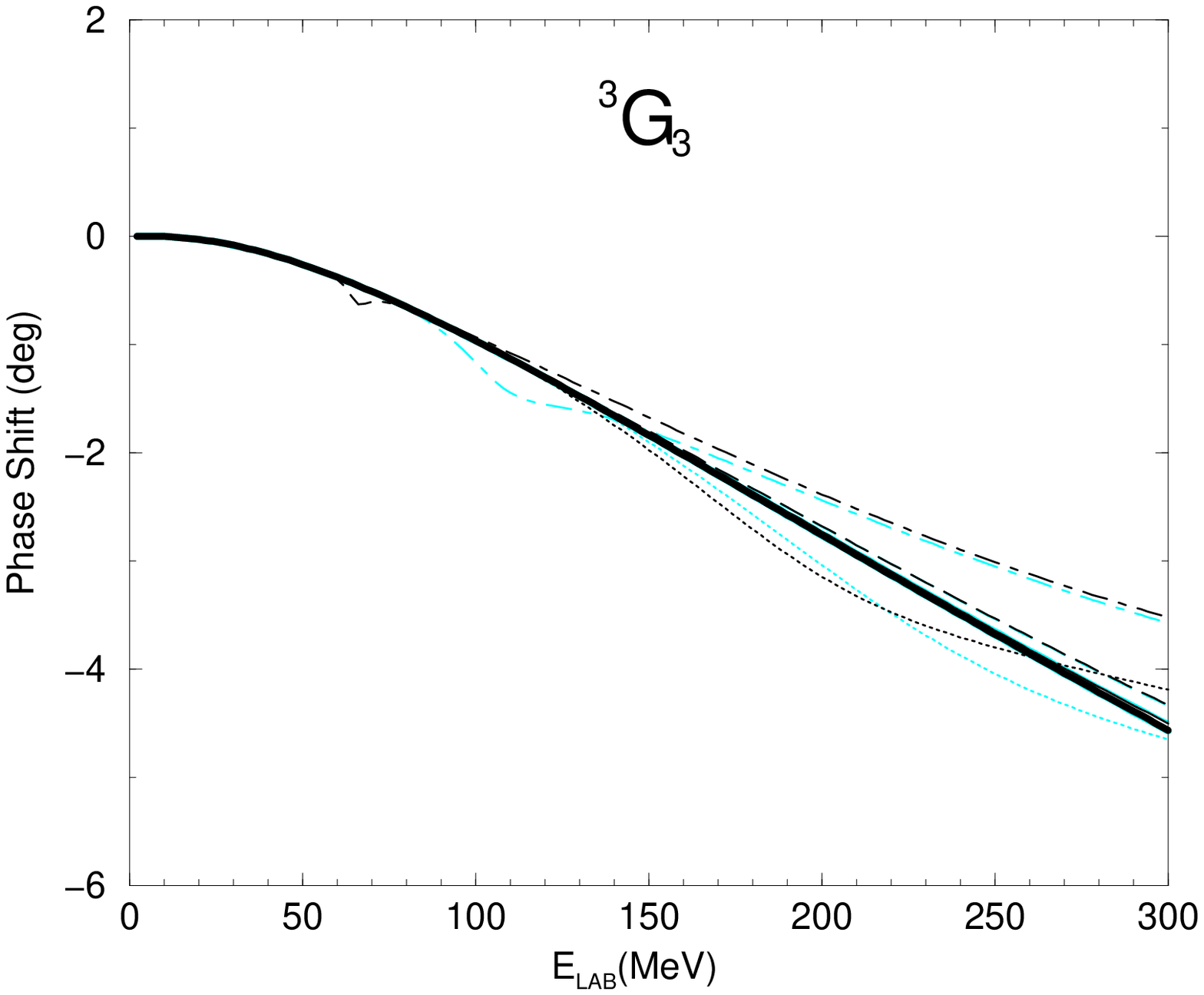, height=2.3in}\\[5mm]
\epsfig{figure=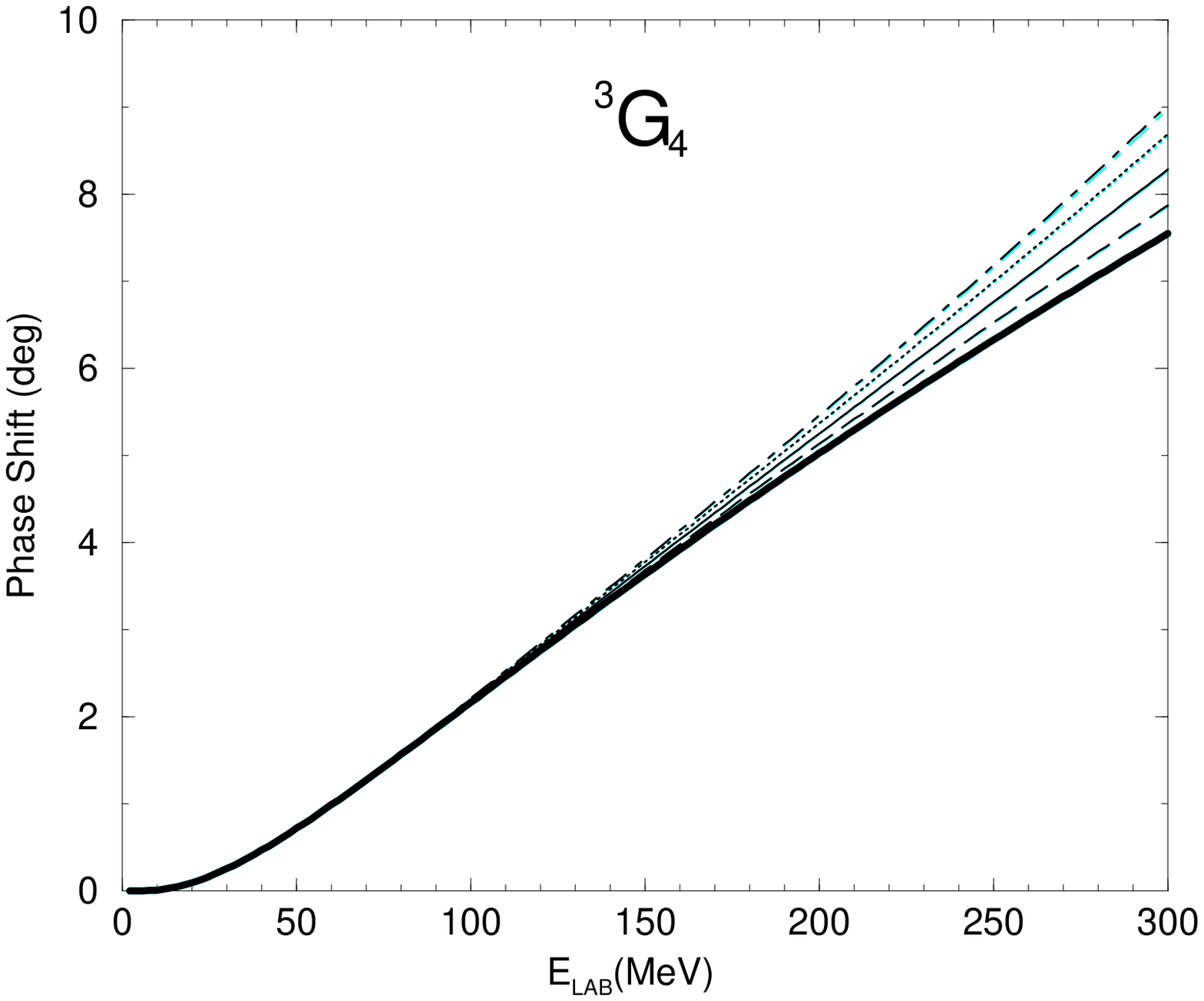, height=2.3in} & \hspace{5mm} &
\epsfig{figure=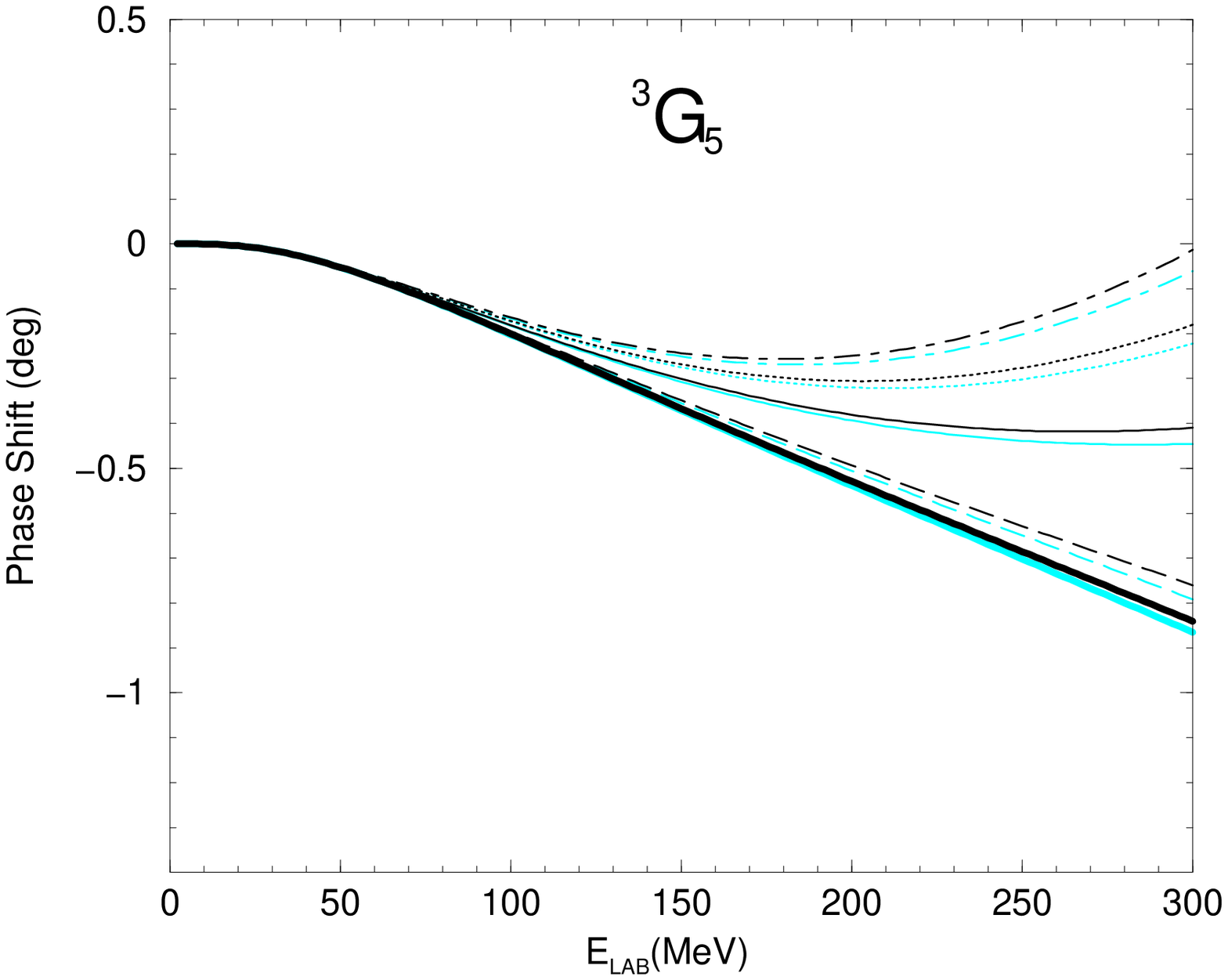, height=2.3in}\\[5mm]
\end{tabular}
\epsfig{figure=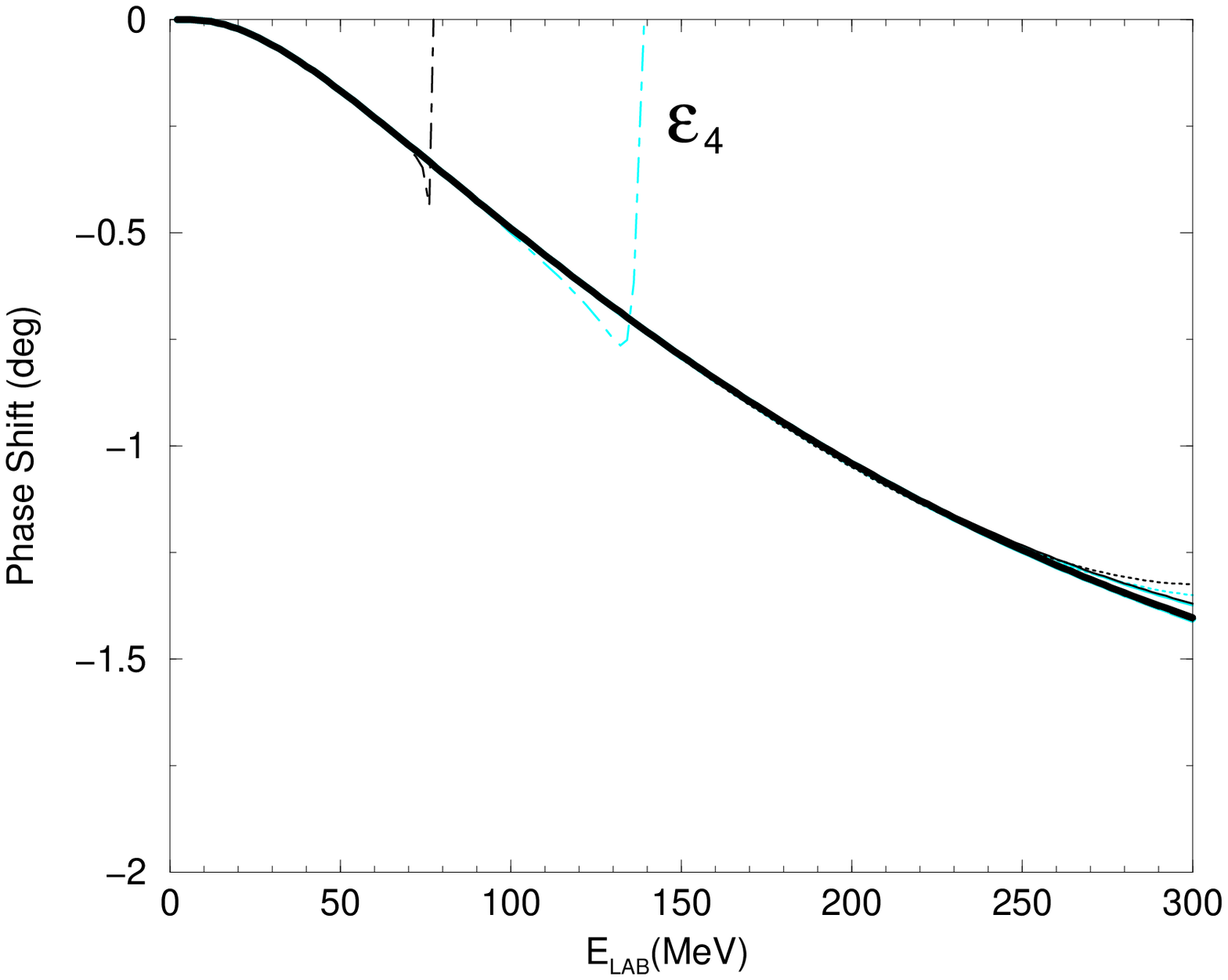, height=2.3in}
\end{center}
\caption{Phase shifts and mixing parameter for $G$ waves, as functions of 
the laboratory energy. Notation is the same used in Fig.~\ref{figpsregF}.}
\label{figpsregG}
\end{figure}

\begin{figure}[ht]
\begin{center}
\begin{tabular}{ccc}
\epsfig{figure=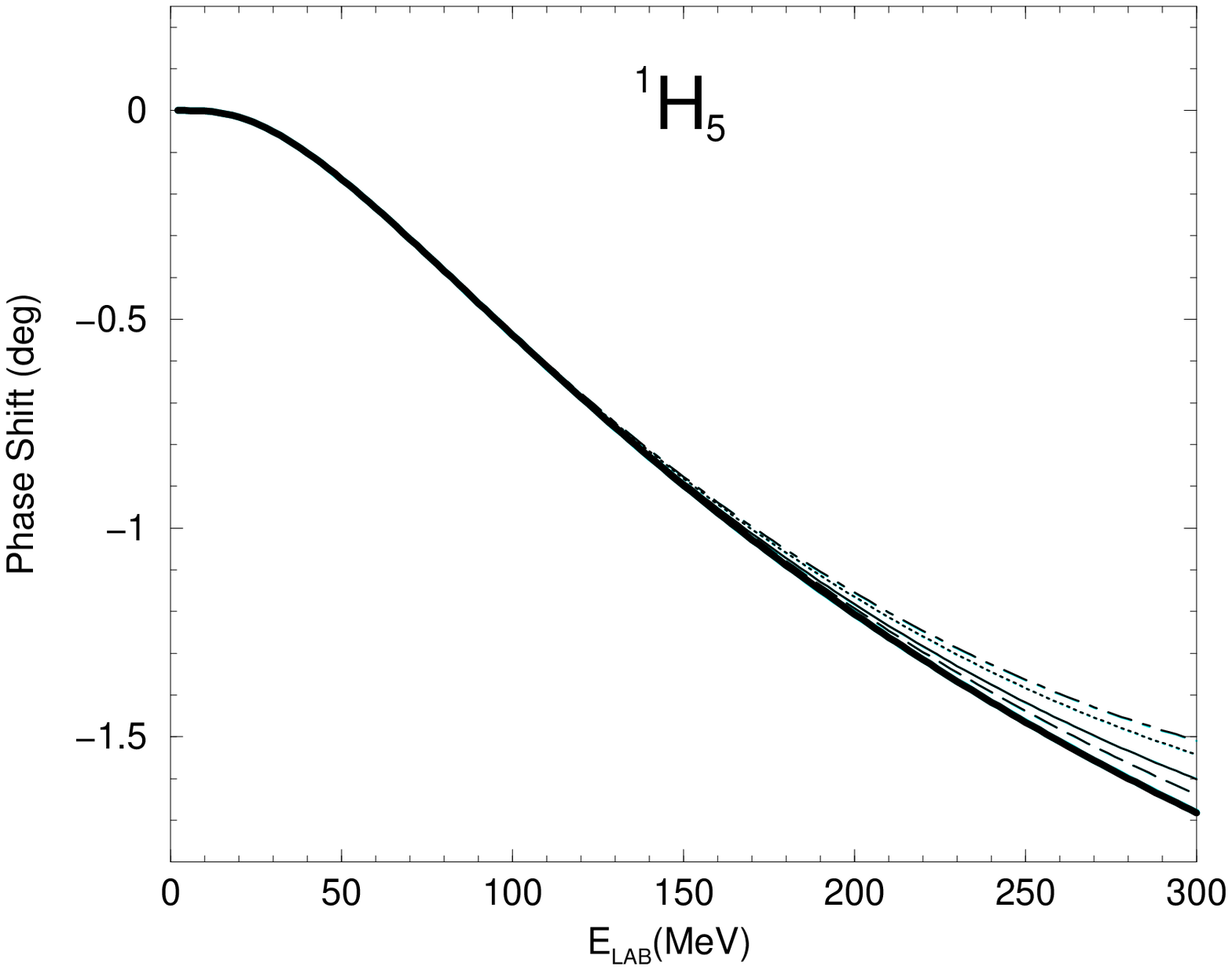, height=2.3in} & \hspace{5mm} &
\epsfig{figure=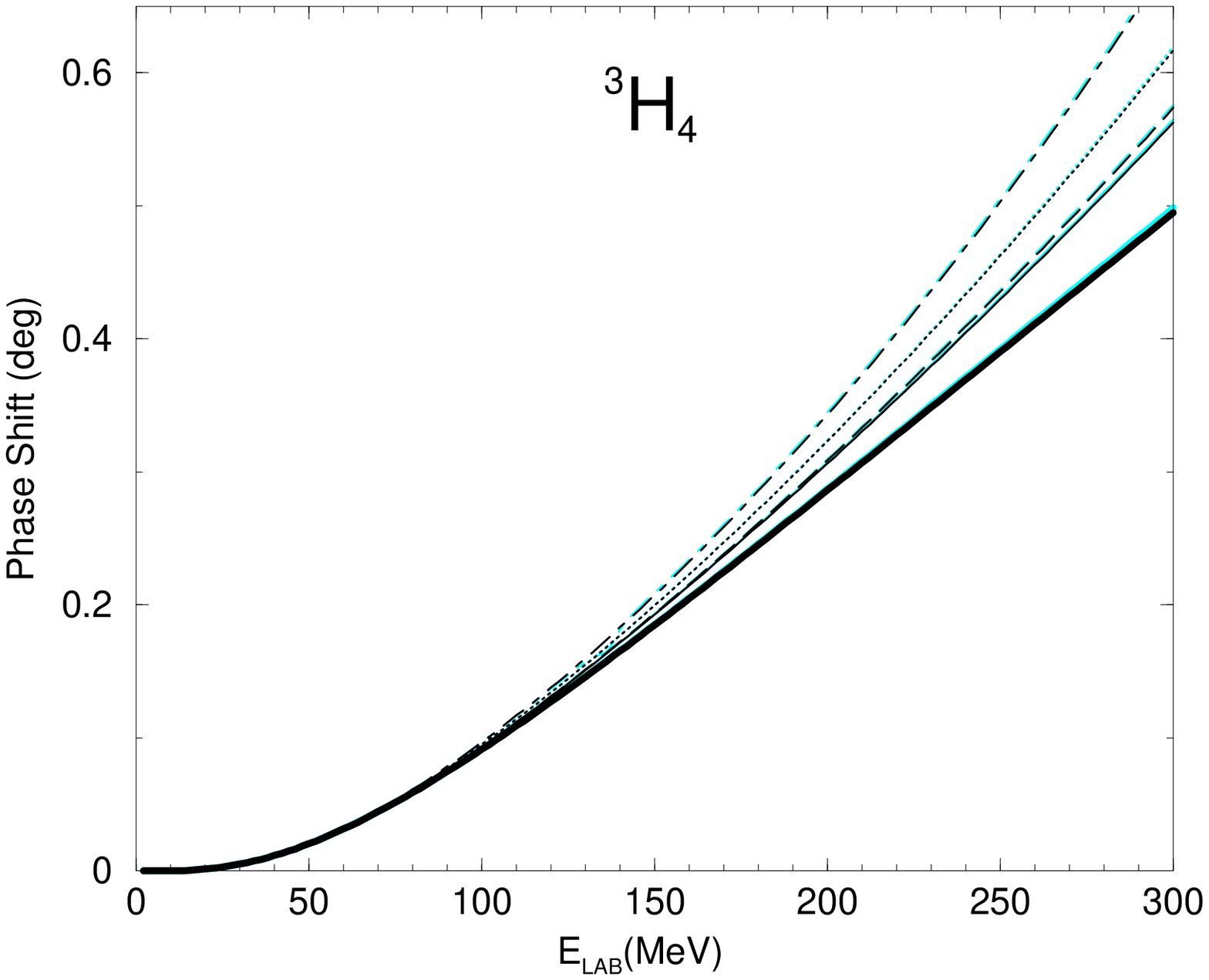, height=2.3in}\\[5mm]
\epsfig{figure=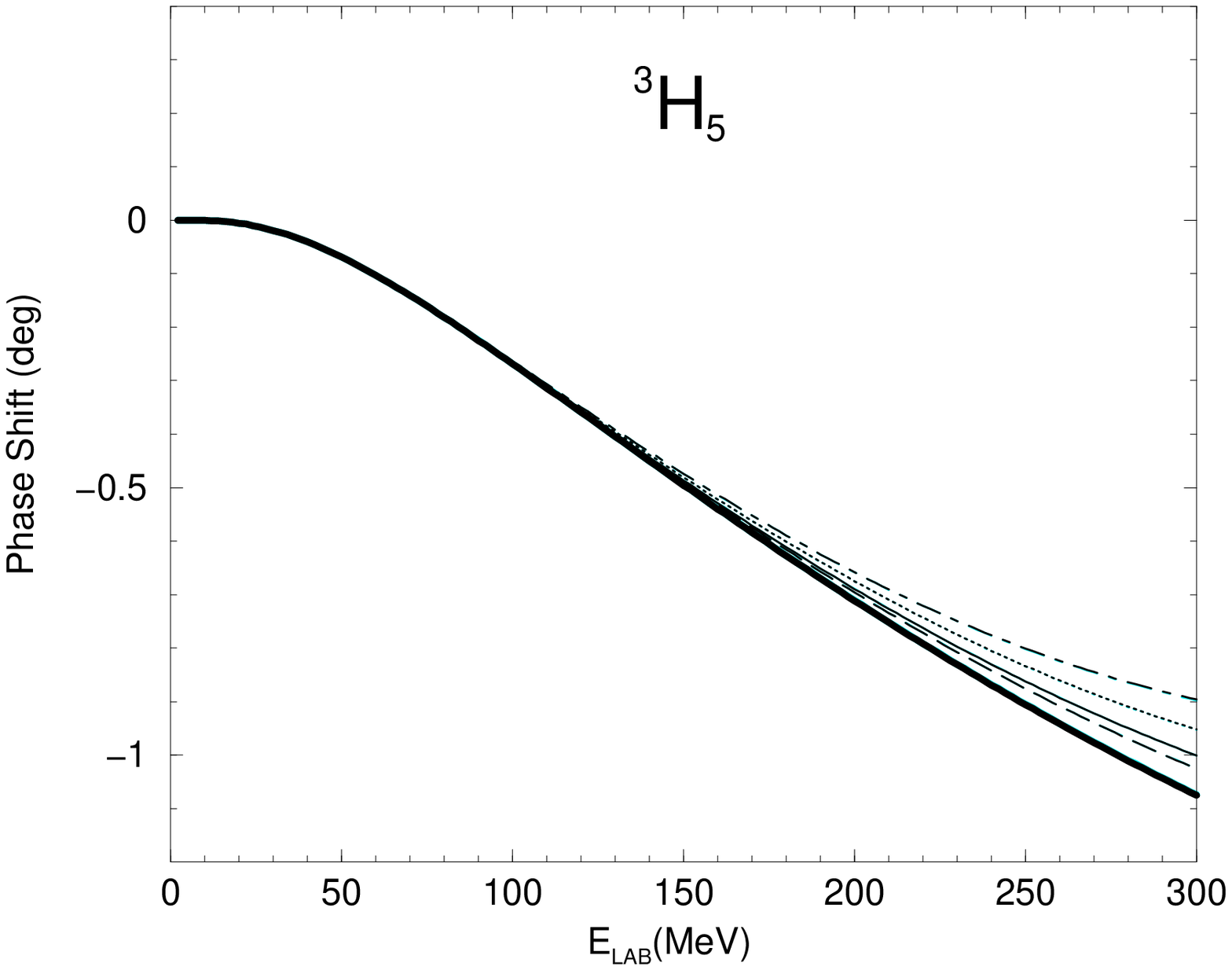, height=2.3in} & \hspace{5mm} &
\epsfig{figure=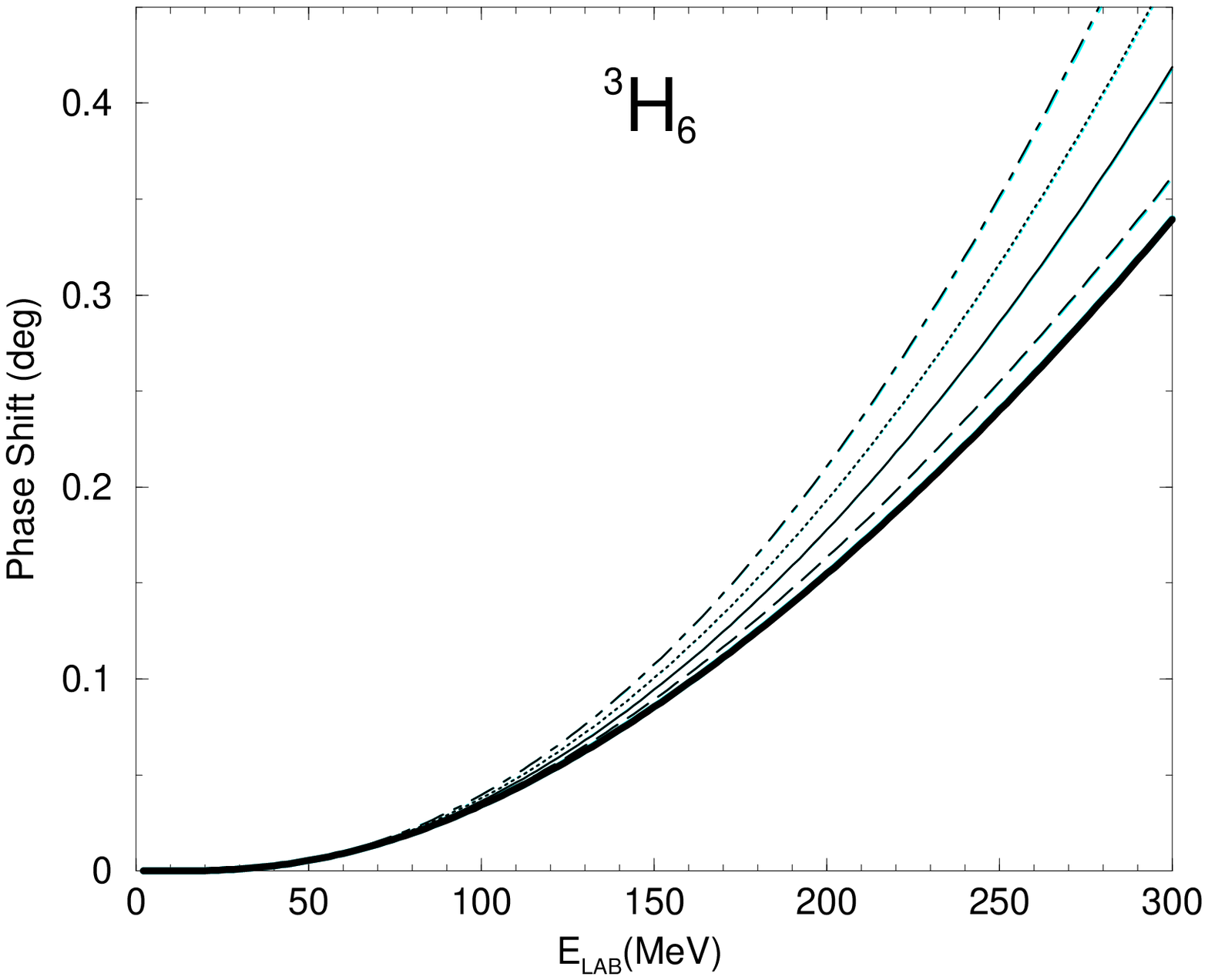, height=2.3in}\\[5mm]
\end{tabular}
\epsfig{figure=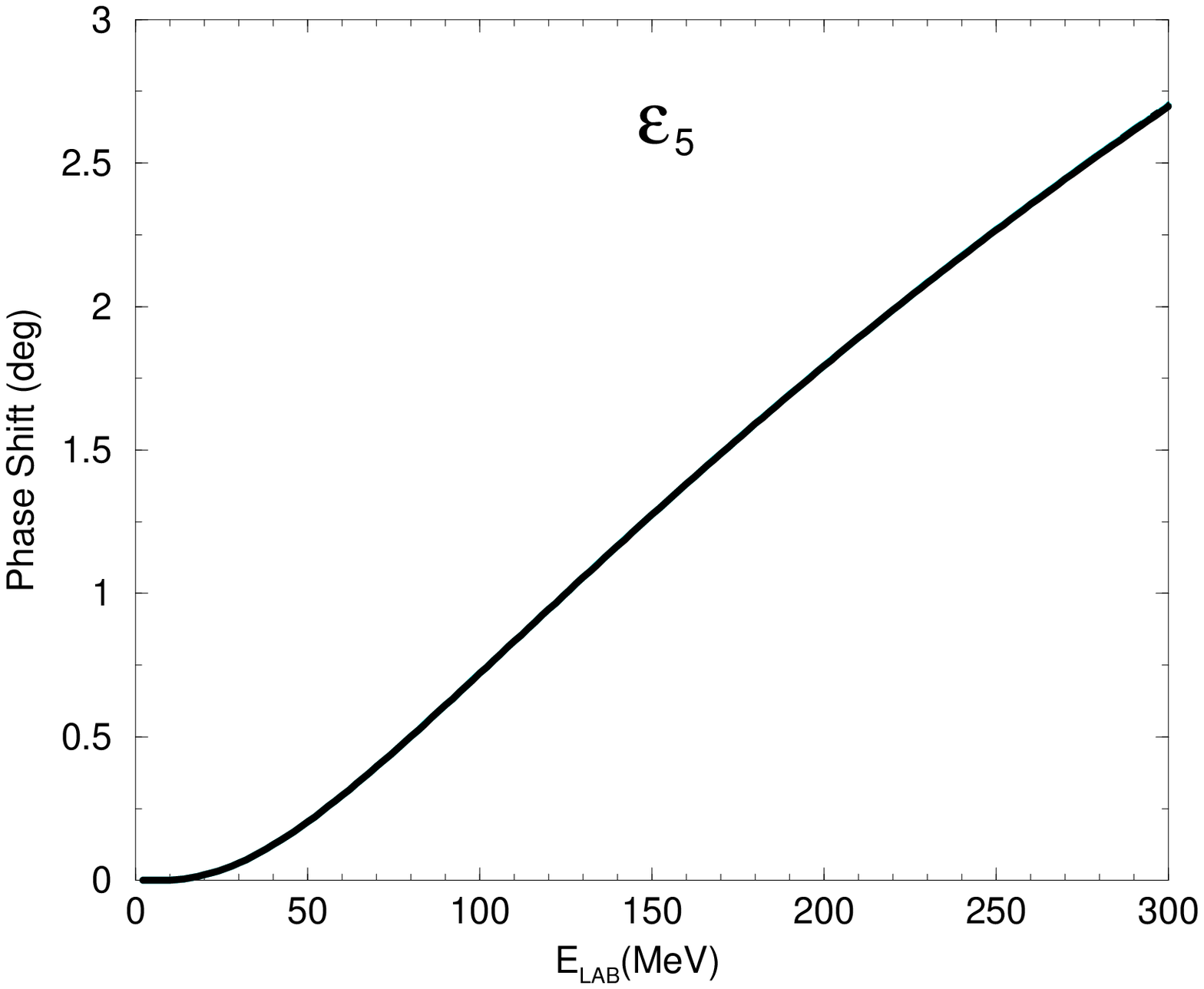, height=2.3in}
\end{center}
\caption{Phase shifts and mixing parameter for $H$ waves, as functions of 
the laboratory energy. Notation is the same used in Fig.~\ref{figpsregF}.}
\label{figpsregH}
\end{figure}

Figs.~\ref{figpsregF}, \ref{figpsregG}, and \ref{figpsregH} show, 
respectively, the phase shifts associated to $F$, $G$, and $H$ waves, as 
functions of the laboratory energy. \noindent The dotted 
lines correspond to LECs from Moj\v zi\v s (column 1), dot-dashed, from 
Fettes {\em et.al.} (column 2), dashed, from B\"uttiker and Mei\ss ner 
(column 3), solid, from the Nijmegen group (column 4), and thick solid, 
from Entem and Machleidt (column 5). The dark and light lines represent, 
respectively, calculations using the RB and HB formalism. 
As it is clear, from $NN$ scattering phase shifts one cannot observe 
any significant difference between RB and HB results --- despite the 
contrast between them, shown in Sec.~\ref{secIIcomp}, in $NN$ scattering 
one also has to add the OPEP, which gives a large contribution and makes 
such discrepancies harder to observe. Exeptions are the $F$ wave results 
using the LECs from Fettes {\em et.al.}, but their predictions are not 
very close from what one should expect. 

\begin{figure}[ht]
\begin{center}
\begin{tabular}{ccc}
\epsfig{figure=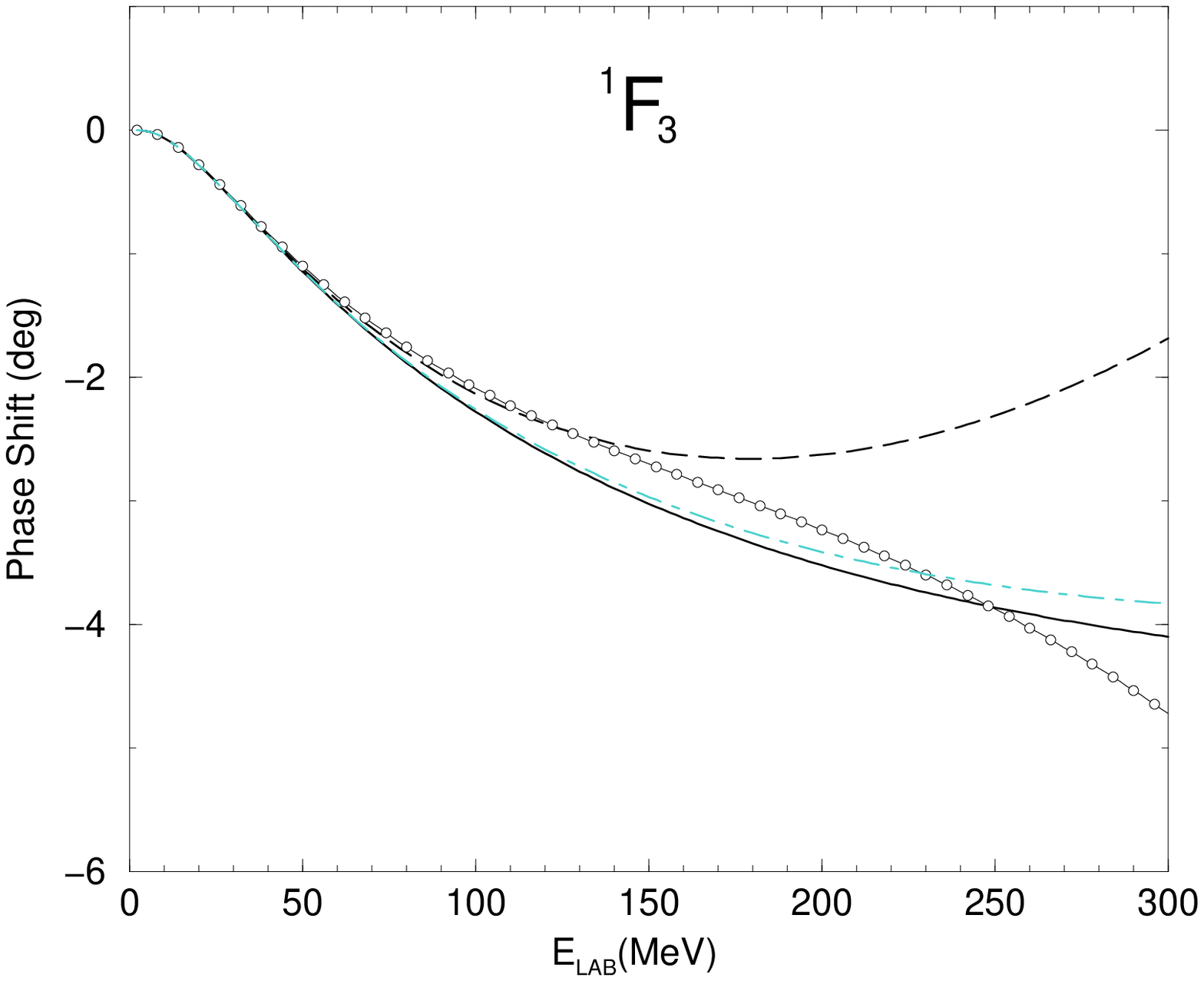, height=2.3in} & \hspace{5mm} &
\epsfig{figure=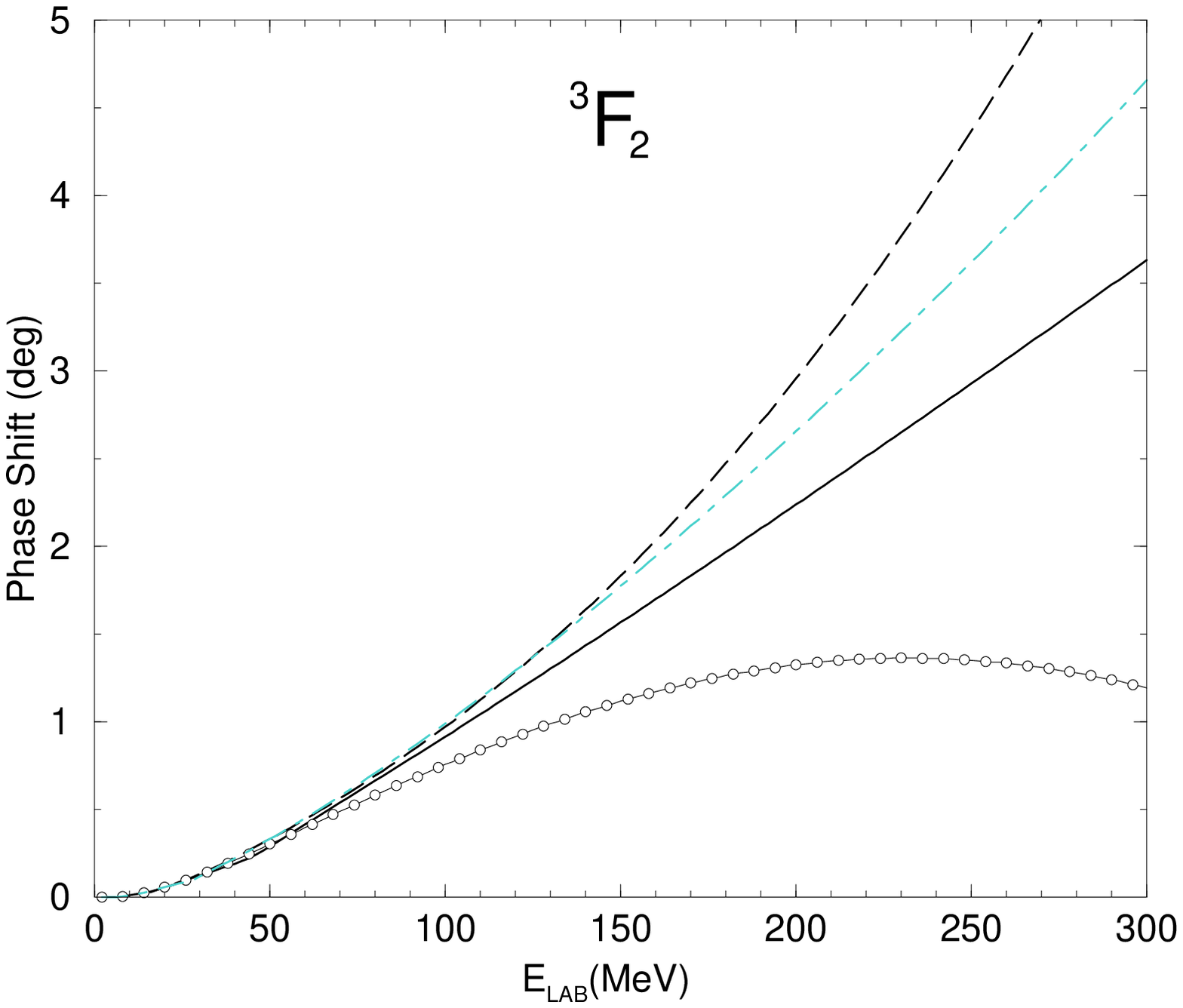, height=2.3in}\\[3mm]
\epsfig{figure=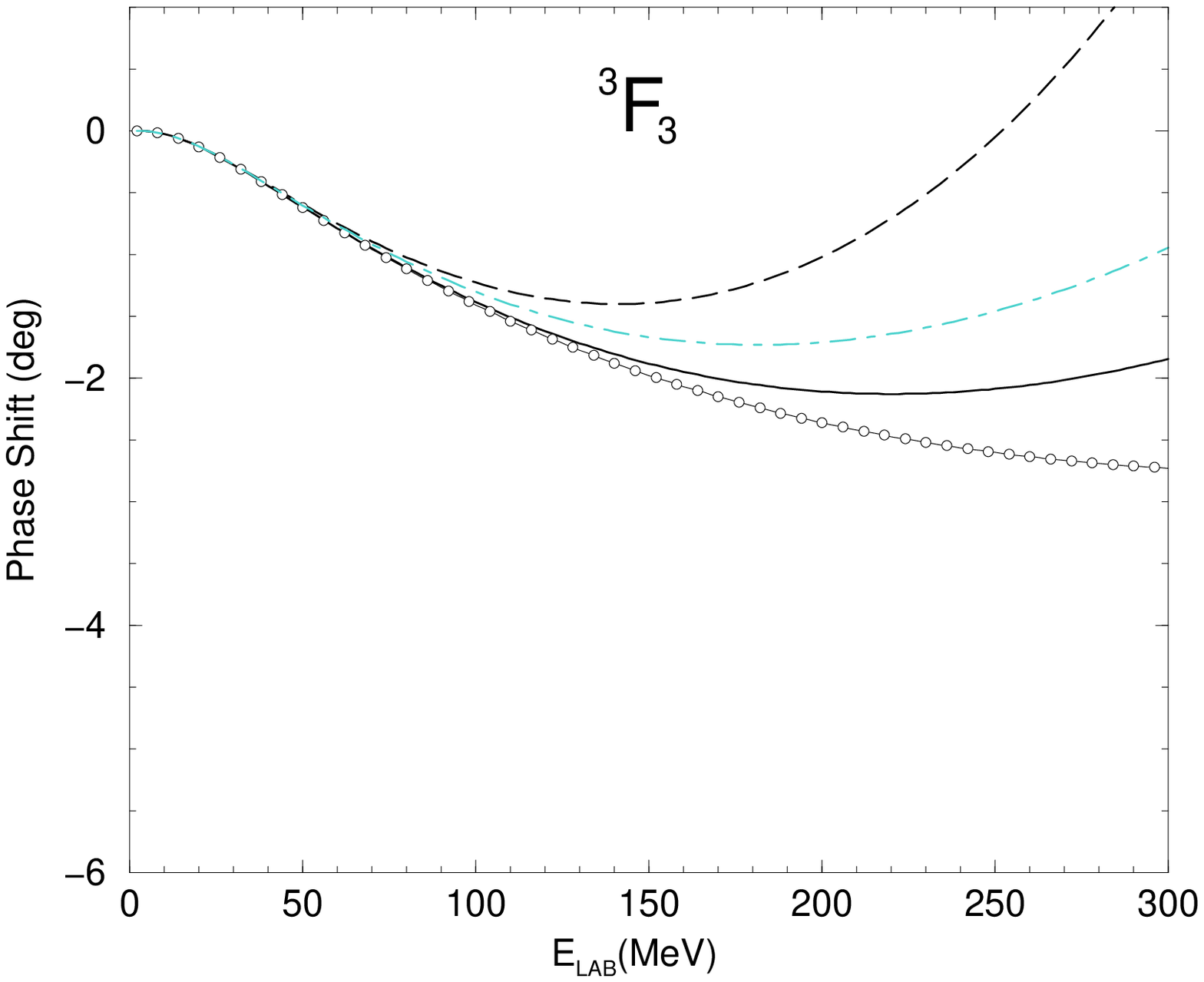, height=2.3in} & \hspace{5mm} &
\epsfig{figure=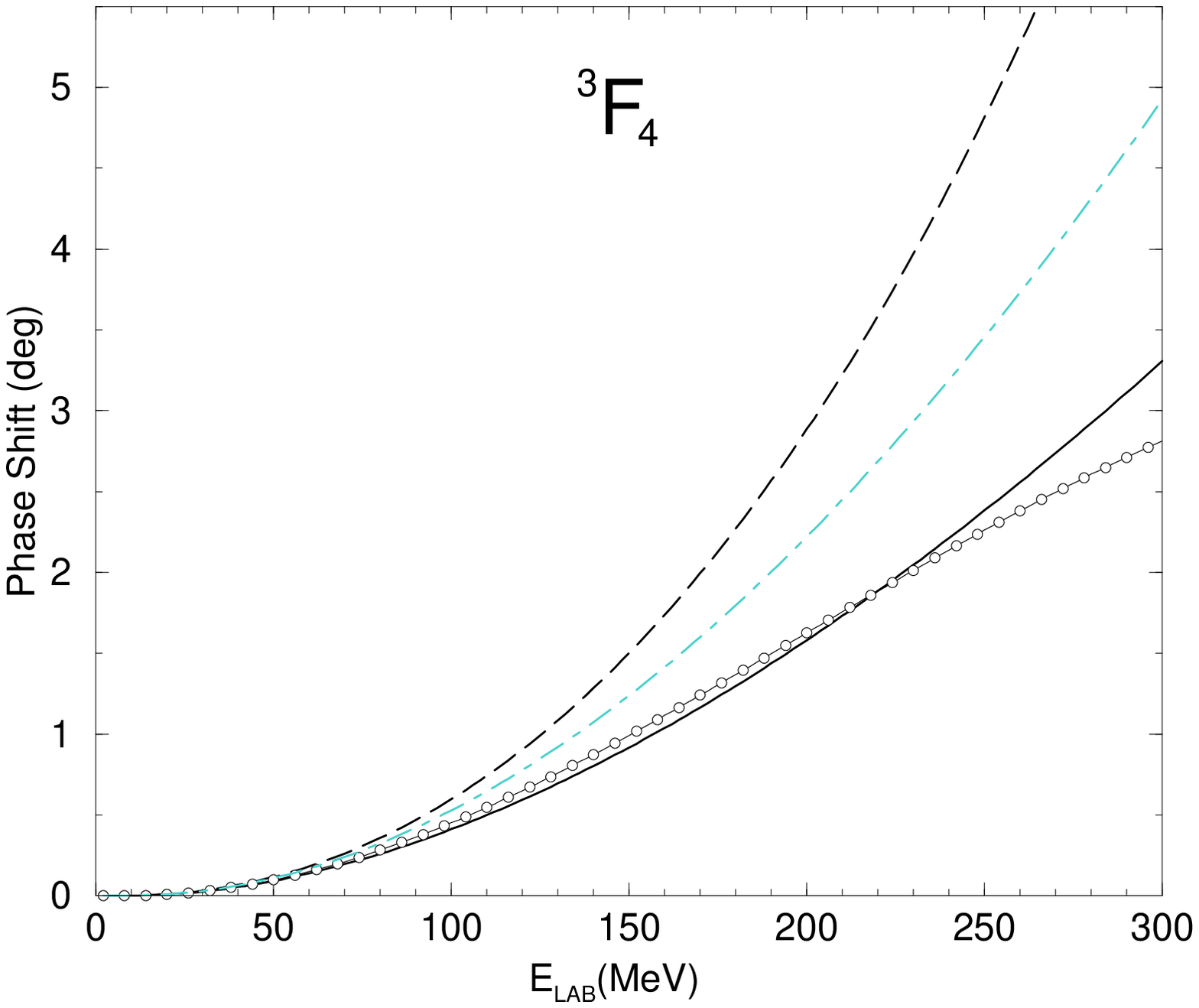, height=2.3in}\\[3mm]
\end{tabular}
\epsfig{figure=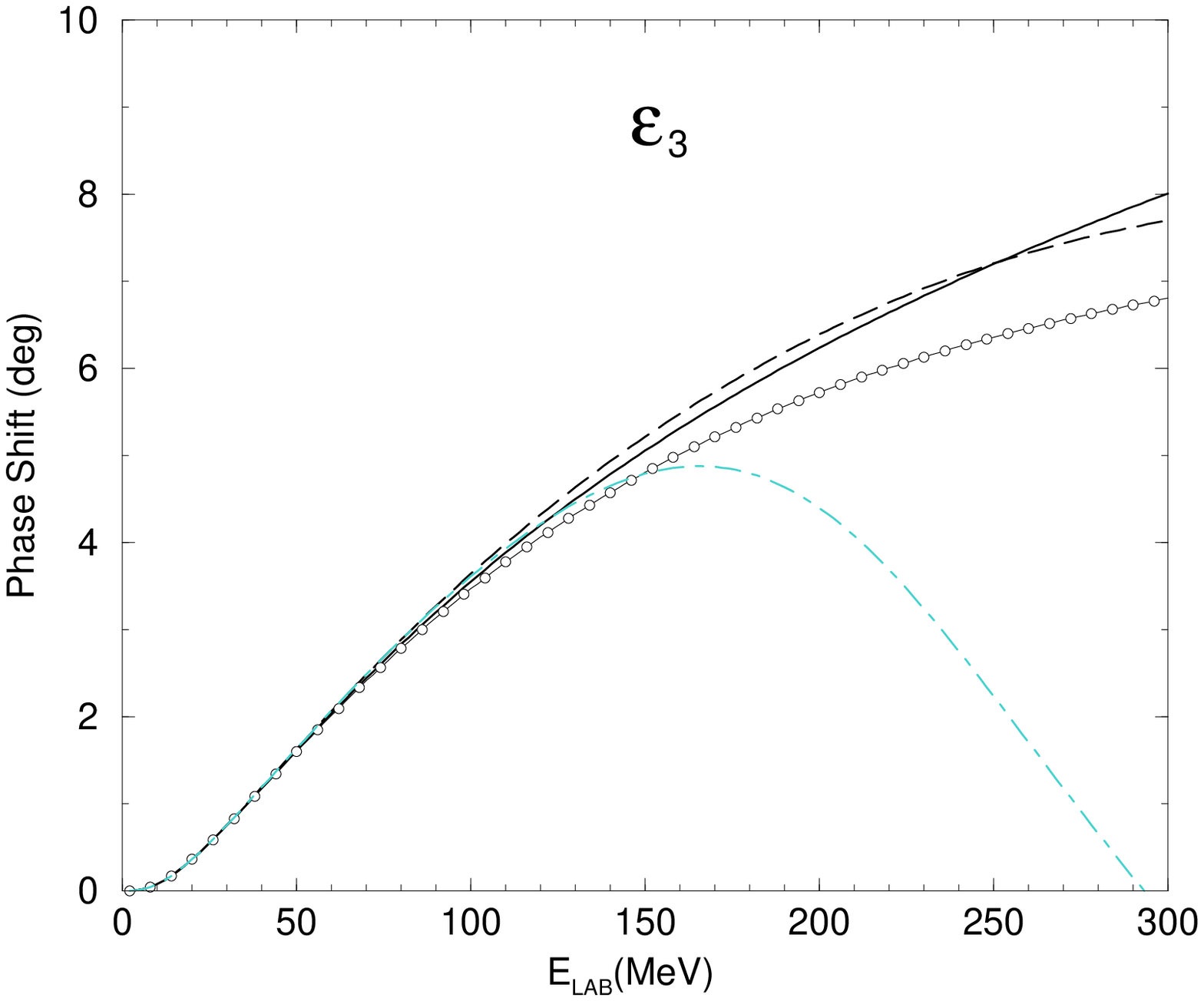, height=2.3in}
\end{center}
\caption{Phase shifts and mixing parameter for $F$ waves, as functions of 
the laboratory energy, compared with the Nijmegen partial wave analysis 
(circled line) \cite{nnonline}. The solid and dashed (dark) lines 
correspond to OPE plus TPE up to $O(q^4)$ using, respectively, the LECs 
from Nijmegen (column 4) and Entem and Machleidt (column 5), while the 
dot-dashed (light) curves are results from OPE+TPE up to $O(q^3)$, using 
the LECs from Nijmegen.}
\label{prepsF}
\end{figure}

\begin{figure}[ht]
\begin{center}
\begin{tabular}{ccc}
\epsfig{figure=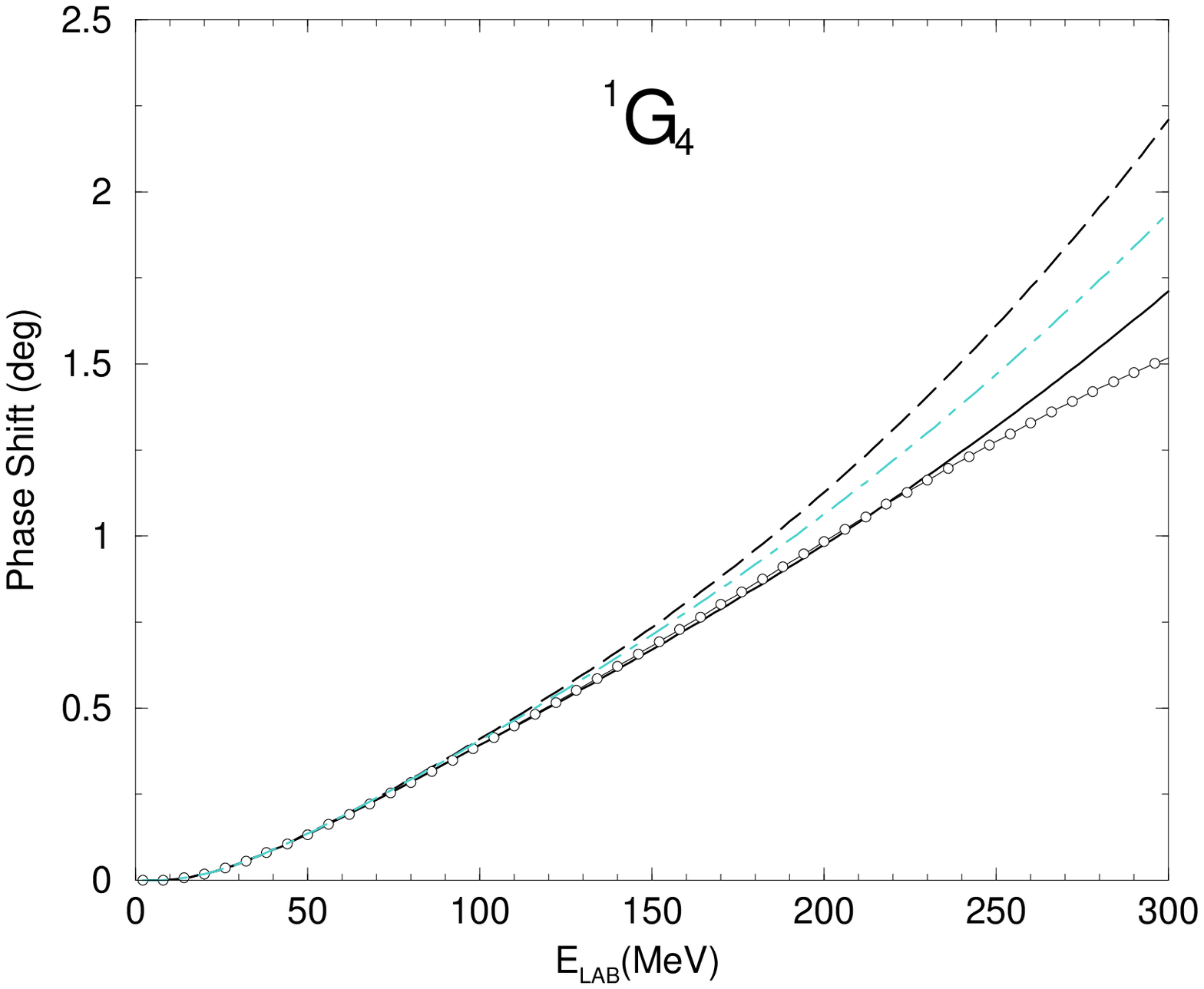, height=2.3in} & \hspace{5mm} &
\epsfig{figure=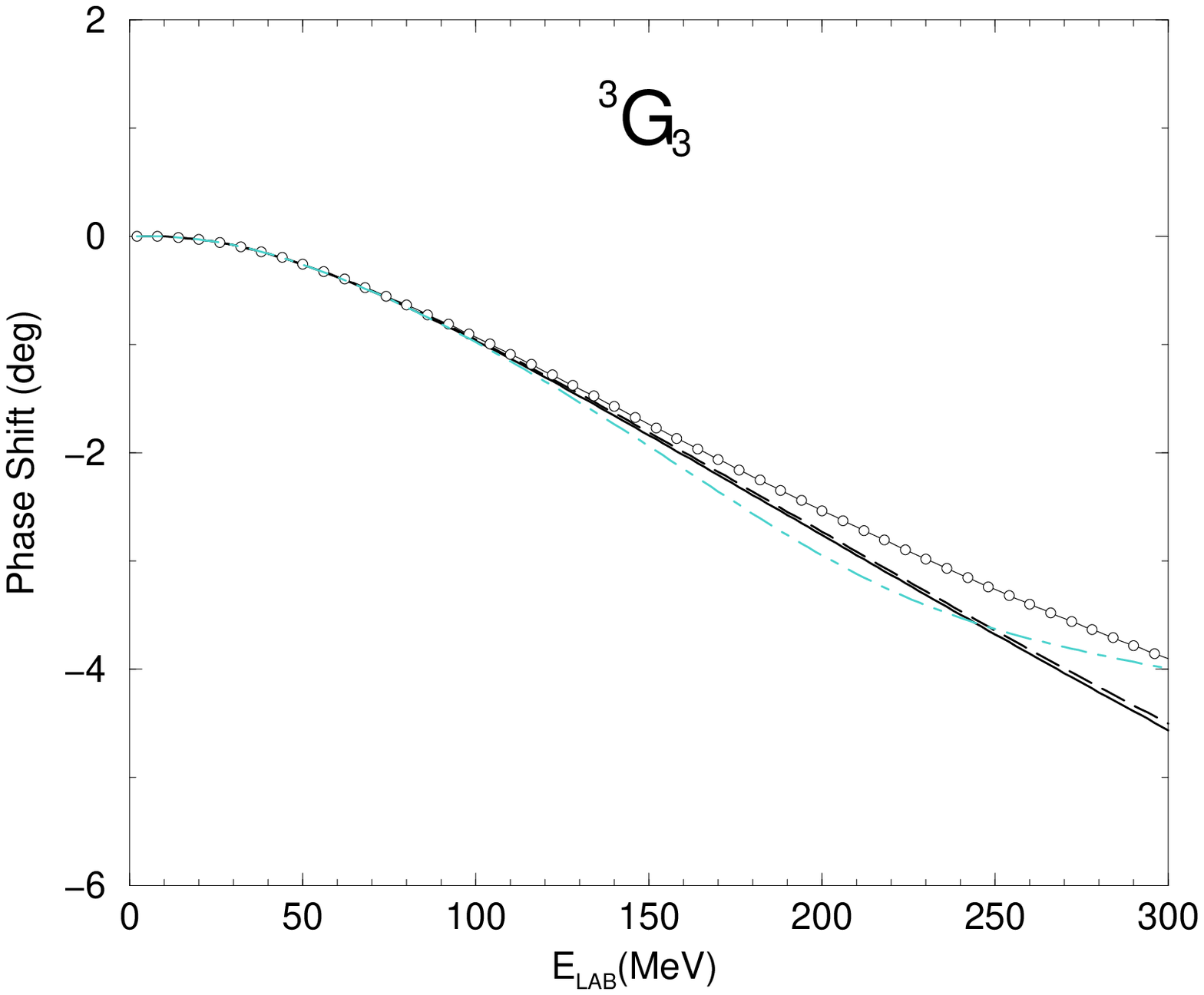, height=2.3in}\\[5mm]
\epsfig{figure=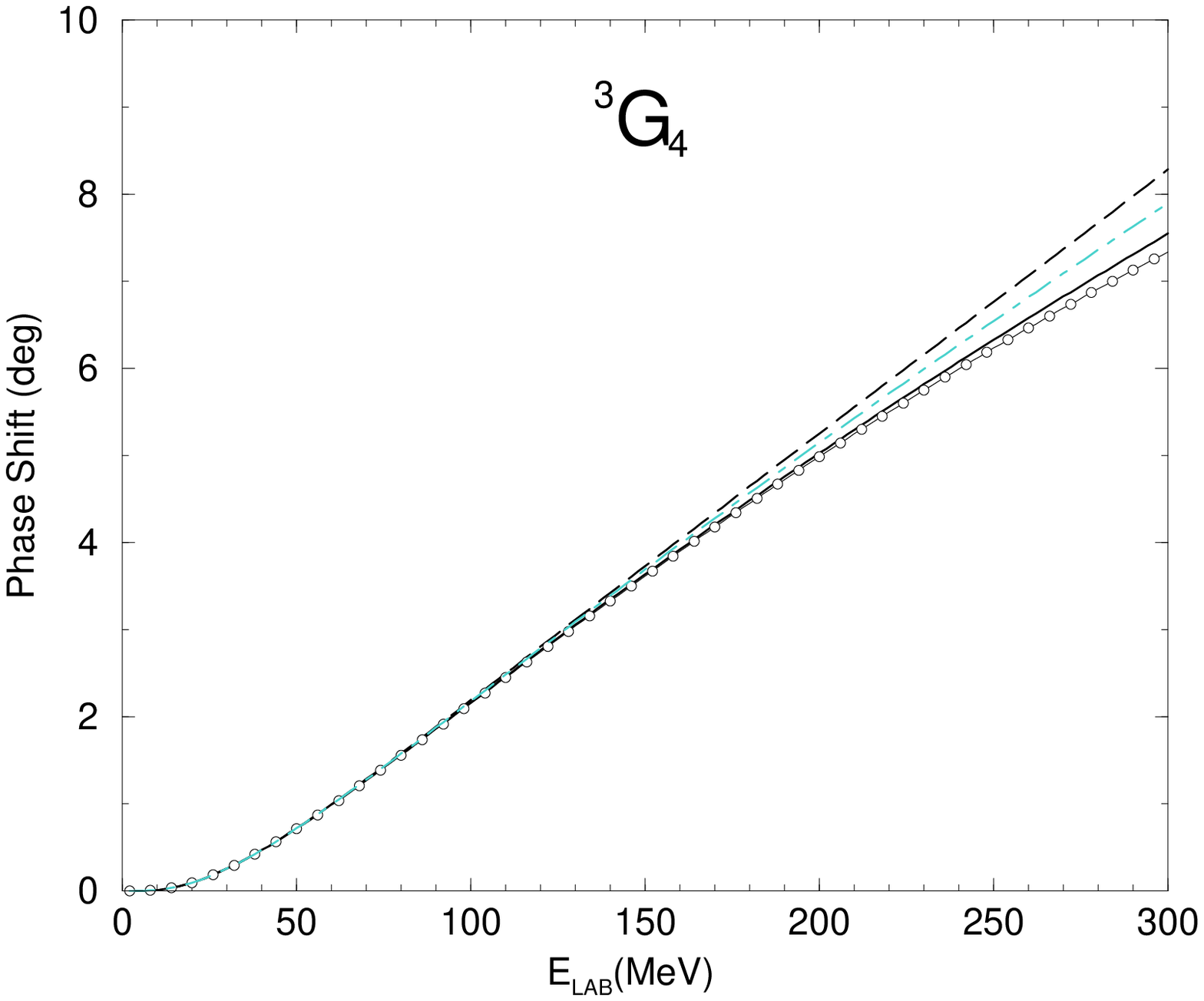, height=2.3in} & \hspace{5mm} &
\epsfig{figure=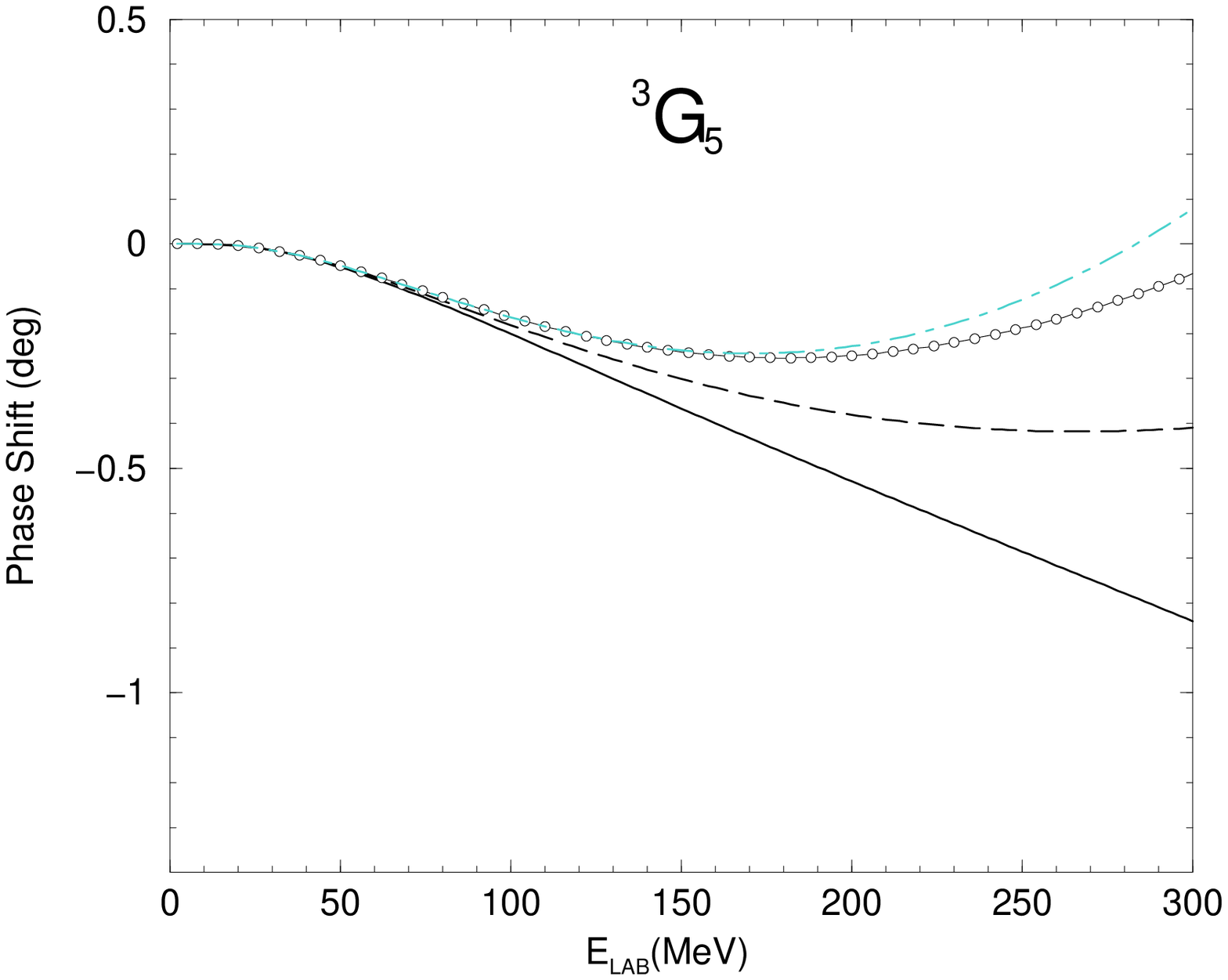, height=2.3in}\\[5mm]
\end{tabular}
\epsfig{figure=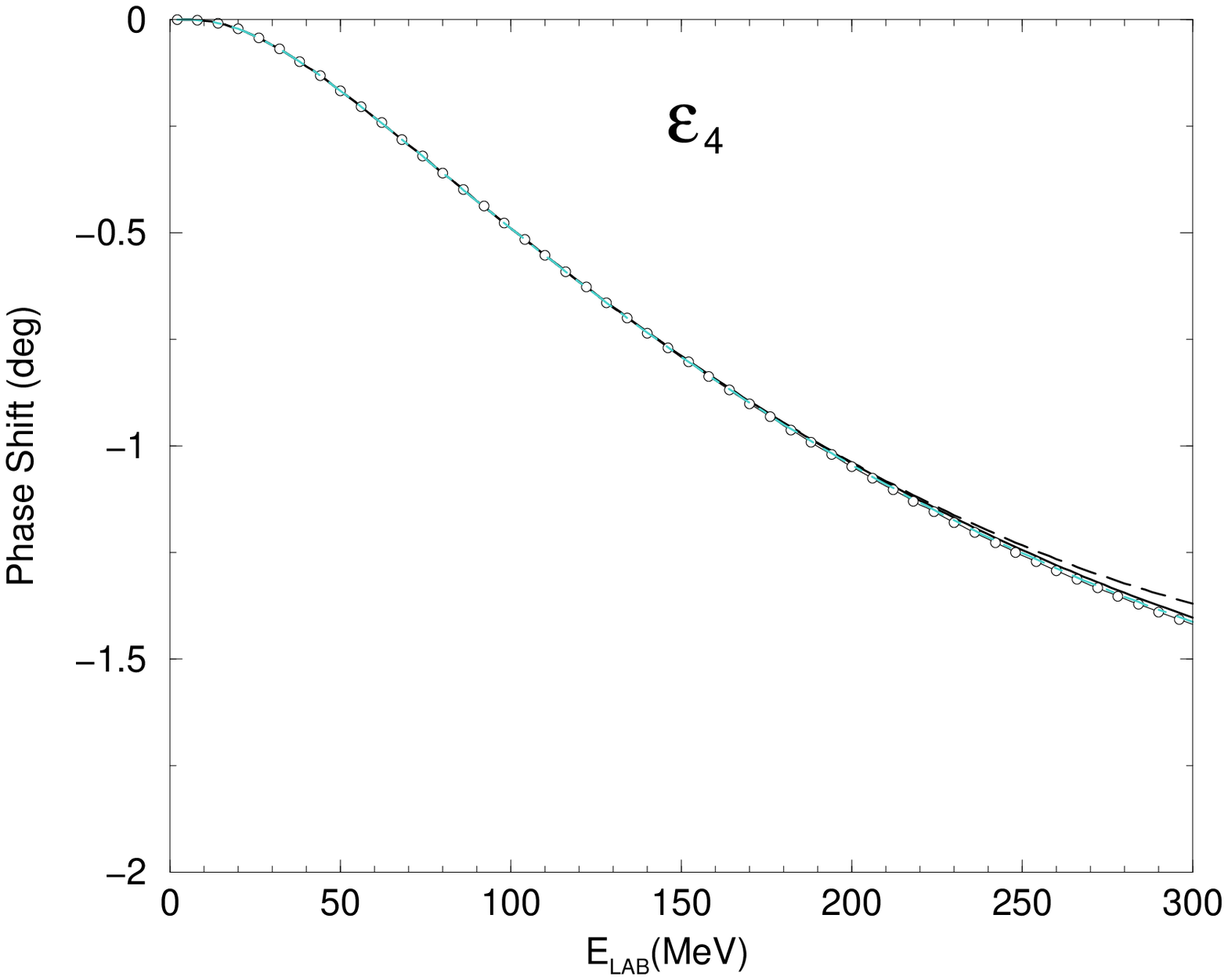, height=2.3in}
\end{center}
\caption{Phase shifts and mixing parameter for $G$ waves, as a function of 
the laboratory energy. Notation is the same used in Fig.~\ref{prepsF}.}
\label{prepsG}
\end{figure}

\begin{figure}[ht]
\begin{center}
\begin{tabular}{ccc}
\epsfig{figure=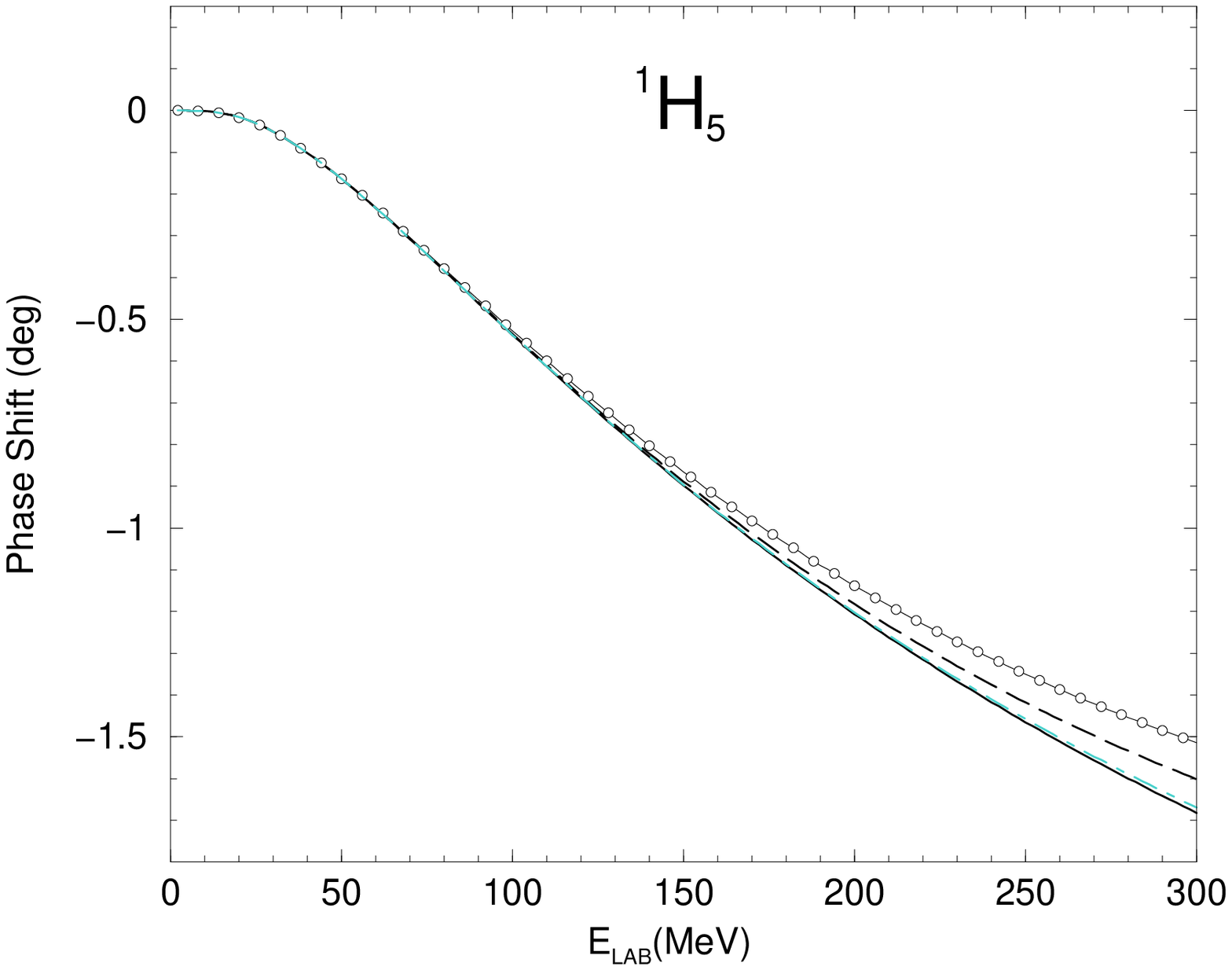, height=2.3in} & \hspace{5mm} &
\epsfig{figure=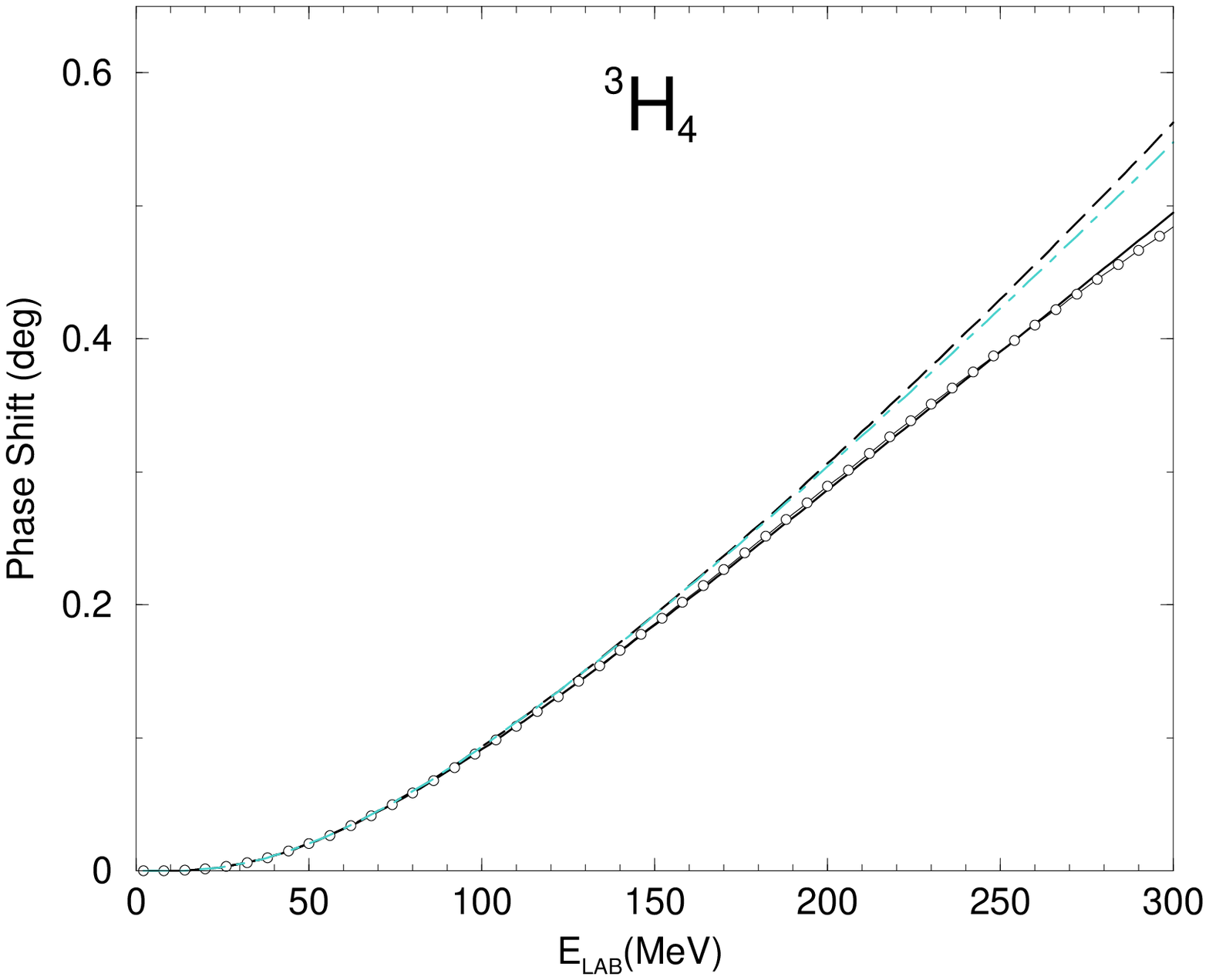, height=2.3in}\\[5mm]
\epsfig{figure=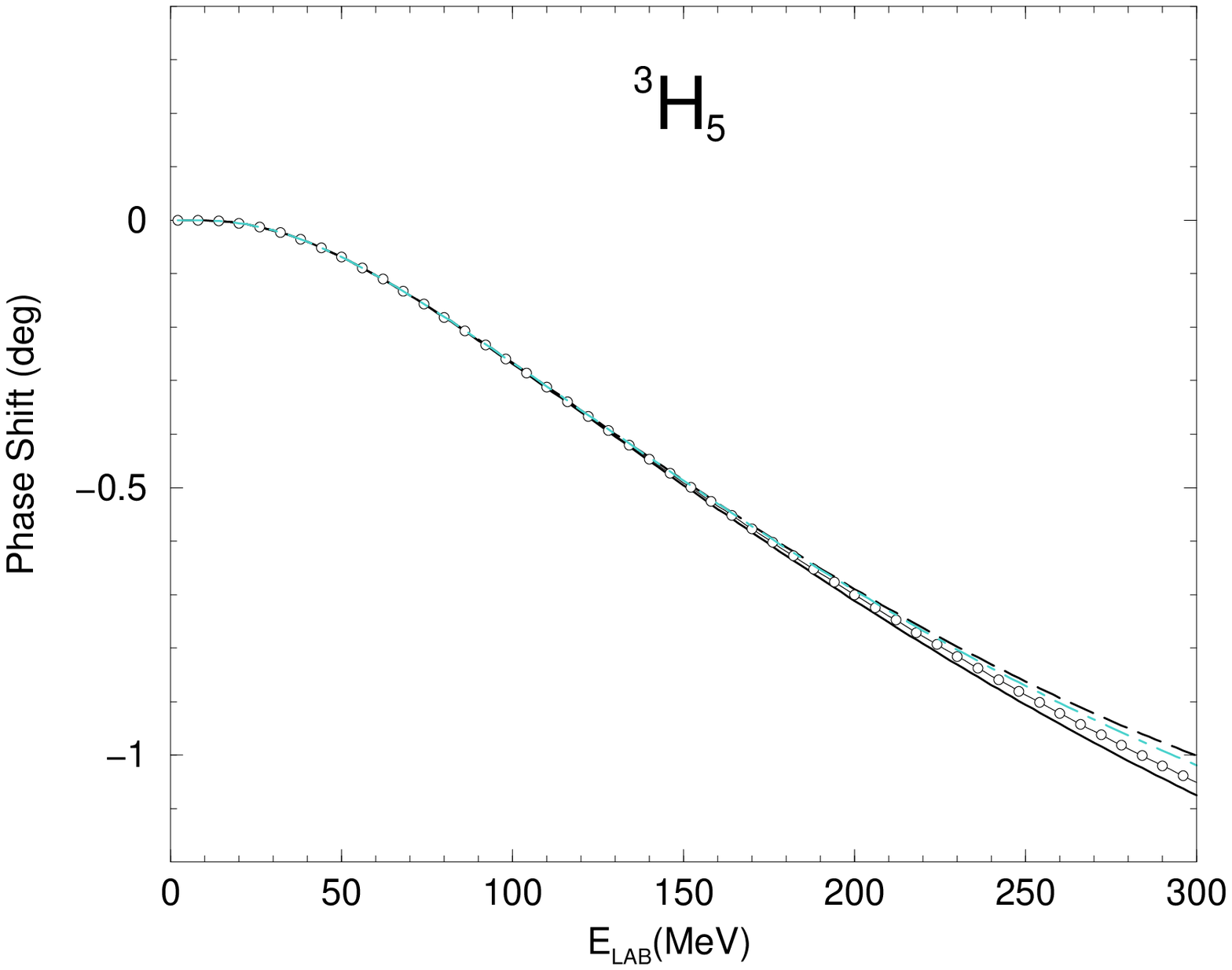, height=2.3in} & \hspace{5mm} &
\epsfig{figure=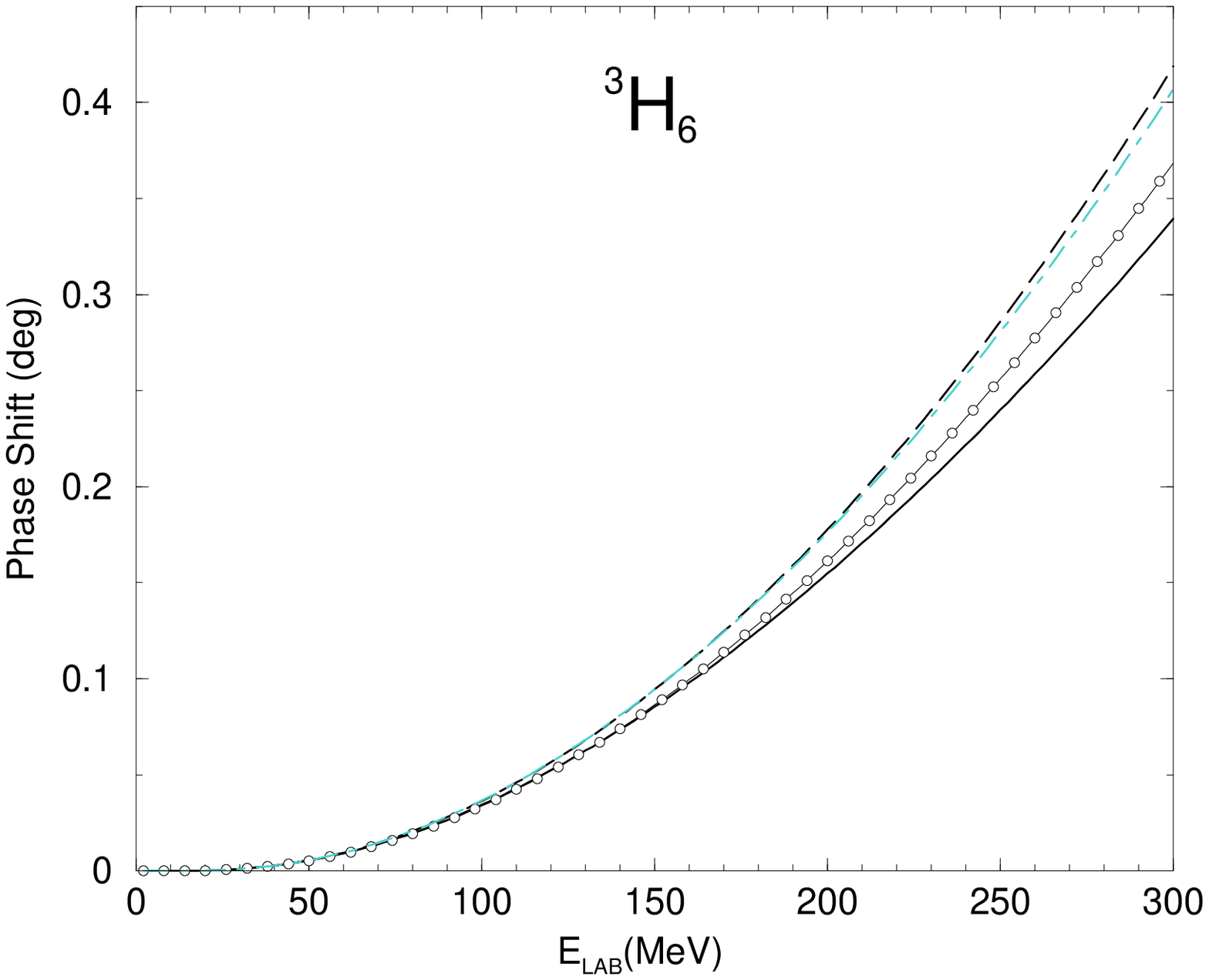, height=2.3in}\\[5mm]
\end{tabular}
\epsfig{figure=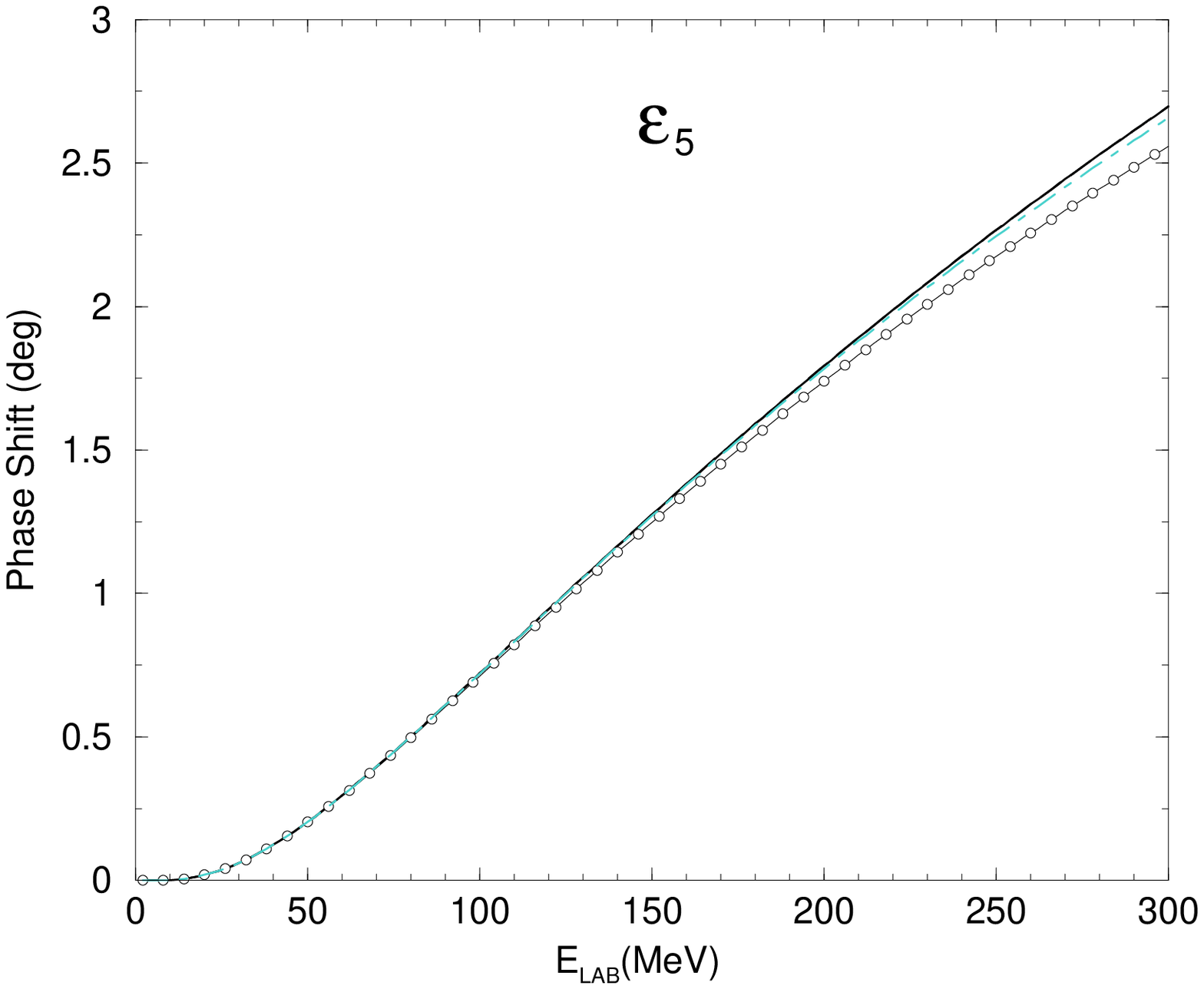, height=2.3in}
\end{center}
\caption{Phase shifts and mixing parameter for $H$ waves (as a function of 
the laboratory energy). Notation is the same used in Fig.~\ref{prepsF}.}
\label{prepsH}
\end{figure}

It is noticeable the large variations of the phase shifts with different 
sets of LECs, predominantly due to $c_3$ (the leading order term of the 
nucleon axial polarizability \cite{tarer}). As pointed out by Entem and 
Machleidt \cite{EM}, consistency with the experimental analysis 
(Figs.~\ref{prepsF}, \ref{prepsG}, and \ref{prepsH}) favors a smaller 
value for this particular LEC, $c_3\sim -3.4$. We will discuss this 
issue later in Sec.~\ref{secV}. 


\newpage

\newpage

\newpage

In Figs.~\ref{prepsF}, \ref{prepsG}, and \ref{prepsH} we compare 
the phase shifts associated to the LECs from column 4 (dark, dashed 
curves) and column 5 (dark, solid lines) of table~\ref{tab3} with 
the Nijmegen partial wave analysis (circled line) \cite{nnonline}. 
It is important to emphasize that, in Refs.~\cite{nij99,nij03}, the 
Nijmegen group employ the TPEP up to $O(q^3)$ to determine the 
dimension two LECs $c_1$, $c_3$, and $c_4$. For this reason, we 
also plot the phase shifts using the TPEP up to this order with 
their LECs, indicated by the (light) dot-dashed curves. 


\newpage

\newpage

\newpage


\section{comments and summary} \label{secV}

As mentioned in Sec.~\ref{secIIcomp}, the calculation of phase shifts 
requires the regularization of the potential near the origin and, as a 
consequence, makes our calculation dependent on the value of the 
regulator. 
This procedure is analogous to, in momentum space, introduce 
a cutoff in the dynamical equation, and seems to be unavoidable when 
dealing with two or more nucleon systems. There are, though, attempts 
to reformulate the problem using dimensional regularization \cite{PAH} 
or variants \cite{fuchs,goity} in the literature. 
One should emphasize that even 
if the chiral potential is finite in momentum space, its representation 
in configuration space shows a divergence at the origin of the type 
$r^{-n}$, where $n$ is some integer. It simply maps the fact that a 
chiral potential has a limited region of applicability, increasing 
as a polinomial in $q$ (in contrast, for instance, with phenomenological 
one-boson exchange potentials). The advantage to work in configuration 
space is that statements about ranges and nucleon sizes can be made more 
transparently, and also it is more suitable to be extended 
to few body calculations for nuclei up to $A=10$, {\em e.g.}, by means 
of the Variational or Green Function Monte Carlo techniques \cite{gfmc}. 

\begin{figure}[!ht]
  \begin{minipage}{3in}
  \epsfig{figure=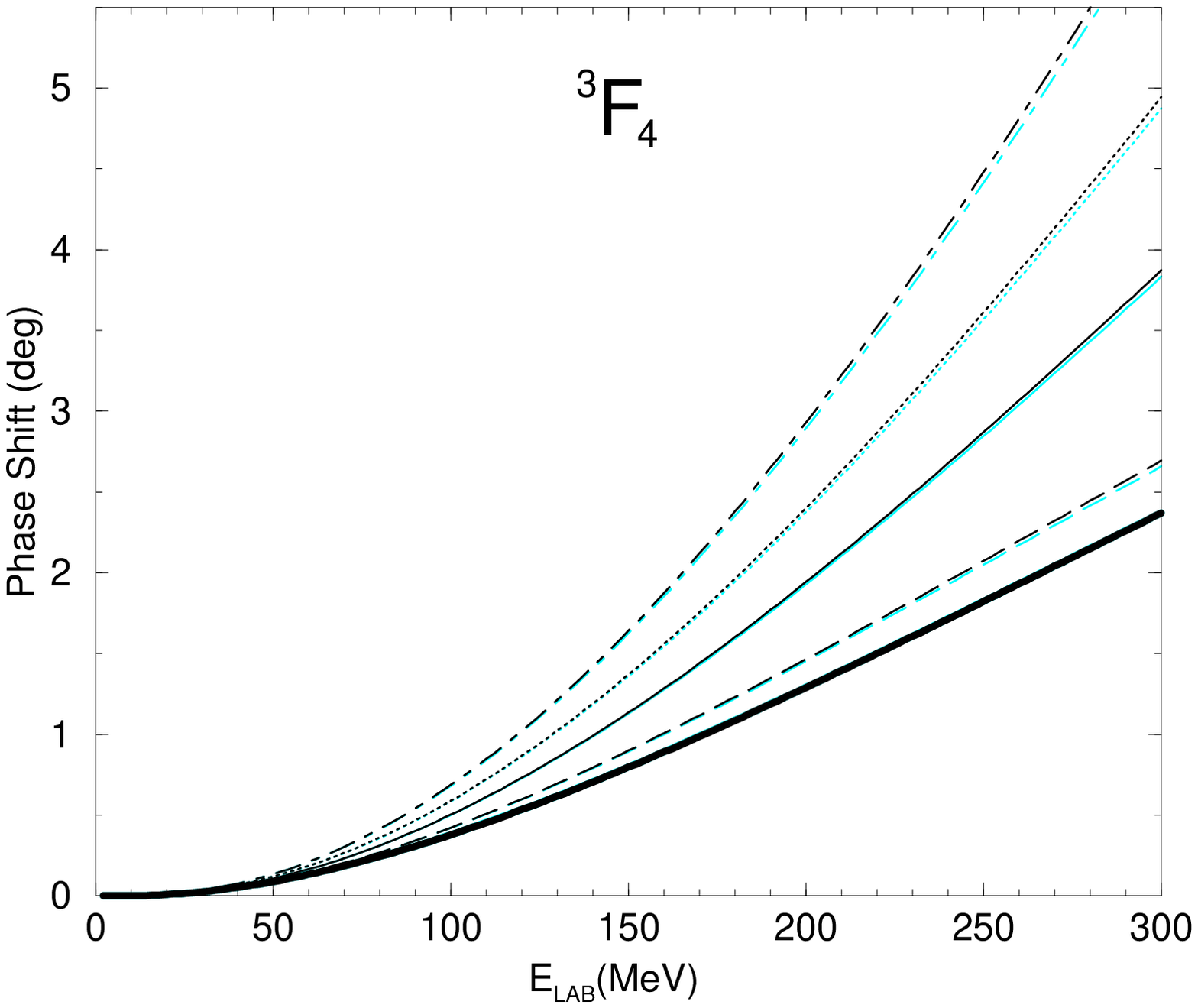, height=50mm}
  \caption{${}^3F_4$ phase shifts, using $c=1.0\mbox{\rm fm}^{-2}$ for 
the regulator. Notation is the same used in Fig.~\ref{figpsregF}.}
  \label{figreg3F4}
  \end{minipage}\hspace{10mm}
  \begin{minipage}{3in}
  \epsfig{figure=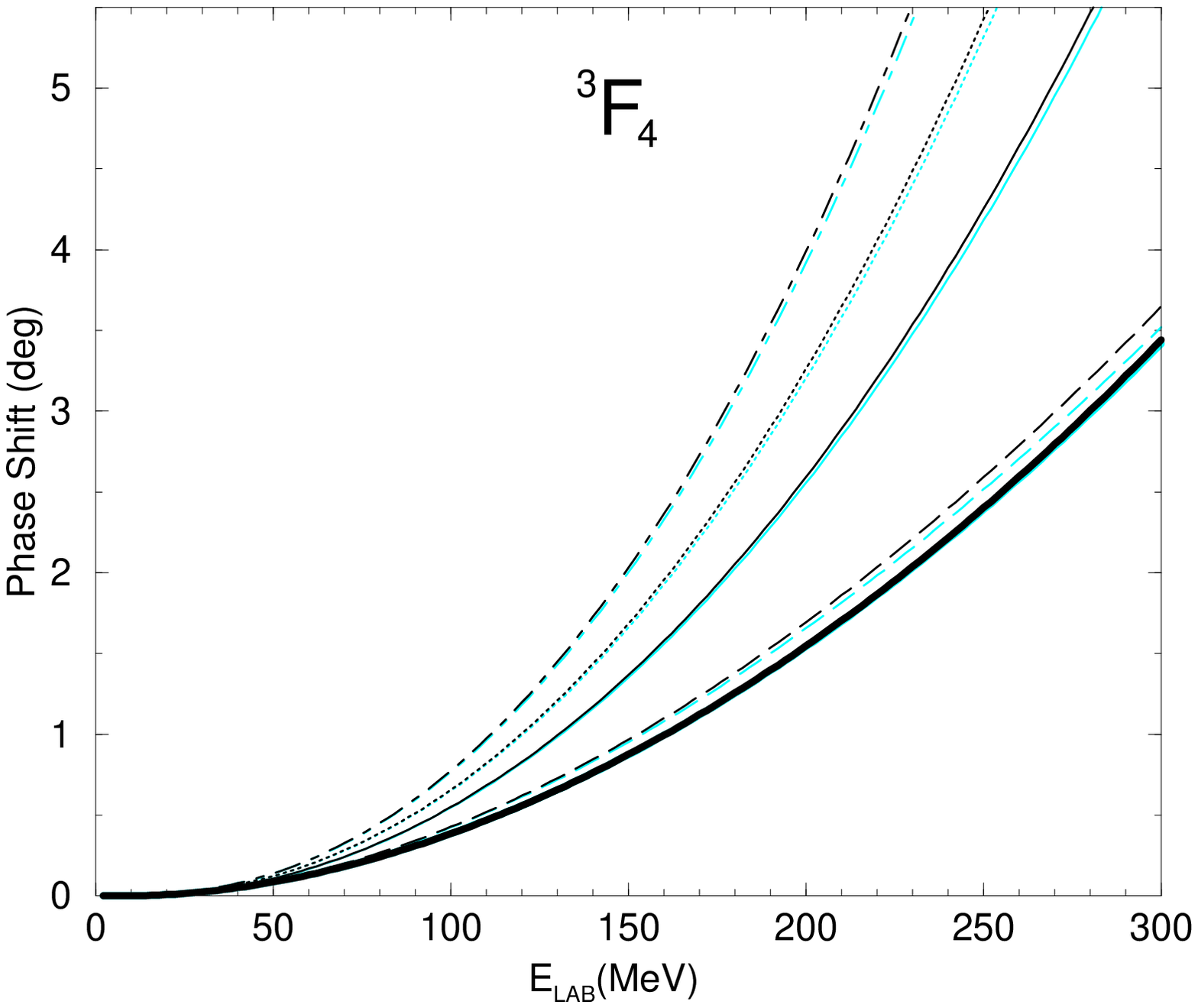, height=50mm}
  \caption{Born approximation for ${}^3F_4$ phase shifts. 
Notation is the same used in Fig.~\ref{figpsregF}.}
  \label{figborn}
  \end{minipage}
\end{figure}

We introduced a regulator given by 
Eq.(\ref{eq:regul}), with $c=2.0$fm${}^{-2}$, in our calculations. In 
principle there is not a real compelling justification to use such a 
value. With higher values, however, the phase shifts using the LECs from 
Fettes {\em et.al.}, or from Moj\v zi\v s, become even more unrealistic. 
For lower values the situation interchanges, as ilustrated in 
Fig.~\ref{figreg3F4}, but at the expense of 
large modifications in the strength of the potential. For example, using 
$c=1.0\mbox{\rm fm}^{-2}$ one has a reduction of almost half of the 
``bare" value of the  
potential at $r=1.4$fm. With $c=2.0\mbox{\rm fm}^{-2}$ at the same 
point, it reduces roughly 8\%. From the same figure we see that, 
even with this regulator, the sets with larger absolute values of $c_3$ 
are still far from PWA results (Figs.~\ref{prepsF}--\ref{prepsH}). There 
might be two explanations for this 
to happen. One, that $F$ waves are not peripheric enough to avoid the 
dependence on short distances of the potential, but we check that similar 
behavior holds for ${}^1G_4$ wave as well. Also, one can rule out this 
option by calculating the phase shifts in Born approximation, as shown in 
Fig.~\ref{figborn}, where there is no need to regulate the potential 
near the origin. 

\noindent The other possible explanation rely on EFT ideas to deal with 
divergences in the dynamical equation, implemented by the introduction 
of a cutoff. To be consistent, the same cutoff should also regularize 
loop integrals for the irreducible two-nucleon diagrams that constitute 
the kernel (or the potential) of the dynamical equation \footnote{In 
Ref.~\cite{egm04-2}, however, the authors performed a study using 
distinct cutoffs for the dynamical equation and for the irreducible 
kernel.}. This procedure can weaken the contribution of pion loops at 
short and intermediate distances, enough to allow larger absolute values 
for $c_3$ \cite{egm04-1}. We do not address this alternative in the 
present work, as we keep the possibility of renormalize the dynamical 
equation by other means, as the ones mentioned before. 

A further remark on the regulator is that our calculations, using the 
same LECs from Entem and Machleidt, are not very sensitive to its 
variations in the range 
$1.0\mbox{\rm fm}^{-2}\leq c \leq 3.0\mbox{\rm fm}^{-2}$, except for 
${}^3F_3$ and ${}^3F_4$ waves. Using $c=2.5\mbox{\rm fm}^{-2}$, we 
have all of our results very close to theirs~\cite{EM}. 

\begin{figure}[!ht]
  \epsfig{figure=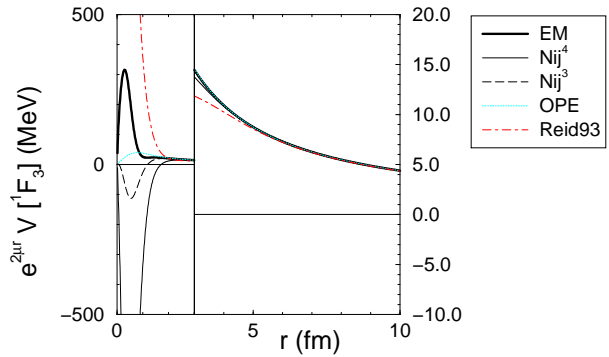, height=50mm}
  \caption{${}^1F_3$ potential. Notation is given in the text.}
  \label{compotF3}
\end{figure}

One comment on Figs.~\ref{prepsF}--\ref{prepsH} is 
in order. We can see that using the LECs from the Nijmegen group on 
the expressions of the $O(q^4)$ TPEP does not give a good description 
of results from PWA, but this procedure does not make sense, as 
these LECs were determined using a potential to $O(q^3)$ 
\cite{nij99,nij03}. It is well known that the LECs pinned down by any 
particular process depend on the order one is working on. Formally the 
difference between using an $O(q^3)$ and $O(q^4)$ TPEP should be of 
higher order, but in practice it is sizeable (see, for instance, the 
$\pi N$ scattering analysis from Ref.~\cite{fet04}). We can check that 
the too strong attraction, which arises if one uses the LECs from the 
Nijmegen group in the $O(q^4)$ potential (Nij${}^4$), is actually not 
that attractive if one considers it up to $O(q^3)$ (Nij${}^3$), bringing 
their phase shifts closer to the ones from PWA. 
Fig.~\ref{compotF3} ilustrates this point for ${}^1F_3$ 
wave. The thick-solid curve is the potential using the LECs from Entem 
and Machleidt, while the dashed and solid curves are the Nij${}^3$ and 
Nij${}^4$ potentials, respectively. For comparison purposes, we also 
include the OPEP (light, dotted curve) and the phenomenological Reid 93 
potential \cite{nnonline,reid93} (dot-dashed curve). 

From this figure and what we discussed above one can say that the 
argument given by Entem and Machleidt \cite{EM,EMach-com}, that the 
LEC $c_3$ from the Nijmegen group leads to a much stronger potential 
as a result of a cutoff in the potential at 1.4fm or 1.6fm (and set 
to zero until the origin) could have some meaning {\em only if} such 
a value were extracted from an $O(q^4)$ TPEP. Evidently this is not 
the case, primarily because the energy-dependent boundary conditions 
of Nijmegen PWA at $r=1.6$fm does not mean that the potential is zero 
inside this range. Besides that, the Nij${}^3$ potential shows a 
repulsion beyond $r\sim 1.2$fm, as can be seen in Fig.~\ref{compotF3}. 
The (now, not too large) disagreement of Nij${}^3$ must have other 
explanations. From our point of view, the reason is that the LECs 
are parameters in the TPEP beyond 1.6fm which are determined when the 
fit to scattering data gives the best $\chi^2$ with a minimum number 
of boundary conditions. This involves sums also over lower partial 
waves, where the nonperturbative aspect of the nucleon-nucleon 
interaction is more pronounced. Therefore, the way short distances 
are parametrized can contaminate the actual value of these LECs, 
contributing as a systematic error. Moreover, contact terms can start 
contributing at $O(q^5)$ or $O(q^6)$ to $F$ waves, which can change 
the the behavior of phase shifts even at not so large energies 
\cite{comun}. When fitting to data, these terms are inherently taken 
into account. 

To summarize, in this work we address the discrepancies we found 
when comparing our $1/m$ {\em expanded} results with the expressions 
from the HB formalism. It forced us to revise not only our $1/m$ 
expansions, but also our two loop diagrams, and now we have a better 
understanding about the origin of the remaining discrepant terms. 
There are, however, two terms in the central isovector component, 
originated from two loop calculations, which could not be understood. 
It will be necessary to solve them if claims for higher order 
calculations of the chiral $NN$ potential becomes imminent. In a 
second stage we calculate the phase shifts for peripheral waves 
($F$, $G$, and $H$), where short distances are supposed to play a 
minor role. This was shown in Figs.~\ref{figPF-F}--\ref{figPF-H}, 
where the ranges relevant to phase shifts where determined using 
the method of the phase function. Our results for phase shifts 
reveals that, due to the overwhelming contribution of the OPEP, any 
difference between the RB- and HB- TPEP is barely noticed. It is 
possible that these differences might be relevant in processes where 
the OPEP is cancelled, and the dominant tail of the interaction, 
given by the TPEP, for instance, in low-energy peripheral scattering 
of nucleons off isoscalar targets \cite{nisoscat}. A more relevant 
question, there are large variations of the phase shifts with the 
existing values of LECs, predominantly due to $c_3$, which dictates 
the strength of the attraction of the potential. Consistency with PWA 
favors smaller values, in magnitude, than the existent ones in the 
literature. There are, however, conceptual details mentioned before 
which prevent, at the moment, to constrain these LECs with a 
satisfactory precision. They should be well understood and quantified 
if one aims to extract values with good precision from $NN$ 
scattering.


\section*{acknowledgements}

I was greatly profited by enlightening discussions with E. Epelbaum 
during the preparation of this work. I also acknowledge helpful 
communications with R. G. E. Timmermans, M. R. Robilotta, 
C. A. da Rocha, F. Gross, 
and the hospitality of the Institute 
for Nuclear Theory at the University of Washington, where part of 
this work was developed. This work was supported by DOE Contract 
No.DE-AC05-84ER40150 under which SURA operates the 
Thomas Jefferson National Accelerator Facility. 


\end{document}